\documentclass[%
 reprint,
 amsmath,amssymb,
 aps,
floatfix,
]{revtex4-2}

\usepackage{graphicx}
\usepackage{dcolumn}
\usepackage{bm}
\usepackage{booktabs}
\usepackage{blkarray}
\usepackage{hyperref}
\hypersetup{colorlinks=true}
\usepackage[all]{hypcap}
\usepackage{lettrine}

\begin{document}

\preprint{APS/123-QED}

\title{Implementing High-fidelity Two-Qubit Gates in Superconducting Coupler Architecture with Novel Parameter Regions}

\author{Lijing Jin}
\email{jinlijing@baidu.com}
 \affiliation{Institute for Quantum Computing, Baidu Research}

\date{\today}

\begin{abstract}
Superconducting circuits with coupler architecture receive considerable attention due to their advantages in tunability and scalability. 
Although single-qubit gates with low error have been achieved, high-fidelity two-qubit gates in coupler architecture are still challenging. This paper pays special attention to examining the gate error sources and primarily concentrates on the related physical mechanism of ZZ parasitic couplings using a systematic  effective Hamiltonian approach.
Benefiting from the effective Hamiltonian, we provide simple and straightforward insight into the ZZ parasitic couplings that were investigated previously from numerical and experimental perspectives.
The analytical results obtained provide exact quantitative conditions for eliminating ZZ parasitic couplings, and trigger four novel realizable parameter regions in which higher fidelity two-qubit gates are expected. 
Beyond the numerical simulation, we also successfully drive a simple analytical result of the two-qubit gate error from which the trade-off effect between qubit energy relaxation effects and ZZ parasitic couplings is understood, and the resulting two-qubit gate error can be estimated straightforwardly. Our study opens up new opportunities to implement high-fidelity two-qubit gates in superconducting coupler architecture.  
\end{abstract}

\maketitle


\section{Background and Motivation}

Benefiting from the development of nano-technology and mature complementary metal-oxide-semiconductor technology, superconducting circuits become a promising hardware candidate for quantum computing  \cite{gambetta2017building,barends2016digitized,wendin2017quantum}. 
In the past few years, significant progress has been achieved in this field, including not only the remarkable improvements of qubits' quantity and quality \cite{arute2019quantum,kjaergaard2020superconducting,gong2021quantum,jurcevic2021demonstration}, but also the realizations of some meaningful noisy intermediate-scale quantum applications \cite{o2016scalable,kandala2017hardware,havlivcek2019supervised,yordanov2020efficient,harrigan2021quantum}. 

To execute various and complex quantum tasks, quantum hardware with many superconducting qubits has to be developed. In addition to qubits' quality,  a natural problem followed is qubit architecture, namely the way of connecting different qubits. According to different purposes, various types of qubit architecture were designed and studied. The simplest one is to connect two adjacent qubits directly via either a capacitor \cite{barends2013coherent} or an inductor \cite{johnson2011quantum,niskanen2007quantum}. The corresponding qubits' topological structure can be one dimensional or two-dimensional grid \cite{barends2014superconducting, kelly2015state}. Such kinds of architectures were used frequently to explore various exciting problems, e.g., demonstration of Greenberger–Horne–Zeilinger state using the full set of gates \cite{barends2014superconducting}, verification of surface codes for quantum error correction  \cite{kelly2015state}, and so on. An alternative way to connect qubits is to couple different qubits via a common resonator (named as a ``quantum bus") \cite{majer2007coupling,chow2014implementing,song2019generation,xu2018emulating,guo2021observation}.  In such indirect coupling architecture, multicomponent atomic Schrödinger cat states were realized \cite{song2019generation}, and intriguing physics of quantum many-body systems were simulated  \cite{xu2018emulating,guo2021observation}.

In the typical architectures mentioned above, the unavoidable problem is that the neighboring qubits will suffer from crosstalk. Moreover, even if the qubit's frequency is tuned away from others (so that the crosstalk can be suppressed), we will still encounter the frequency crowing problem. To overcome this difficulty, some novel qubit architectures were designed and studied. 
In 2014, Gmon was firstly proposed in Ref.~\cite{chen2014qubit}. It overcomes the challenge of incorporating tunable coupling with high coherence devices \cite{neill2018blueprint}. 
Very recently, one simple and generic
architecture with an additional qubit (named as ``coupler") attract wide attention and become the research forefront of superconducting circuits \cite{yan2018tunable, mundada2019suppression, li2020tunable, zhao2020high, xu2020high, zhao2020switchable, sung2020realization, collodo2020implementation,xu2020zz,zhao2020suppression,han2020error,xu2021realization,cai2021perturbation,sete2021parametric}. 
The impressive achievement is that such architecture made great success in Google's quantum supremacy experiment \cite{arute2019quantum}. In the quantum processor, each qubit is connected to its neighboring qubits using an adjustable coupler. 
Tunable coupling and relatively higher fidelity quantum gates were realized. 
In particular, the coupling strengths were able to be tuned continuously from $-40~ \rm MHz$ to $5 ~\rm MHz$, and the average single-qubit gate error can be reached as lower as 0.15$\%$.  However,  realizing fault-tolerant quantum computing \cite{campbell2017roads} with coupler architecture  is still out of reach because of the overhead needed for error-correction with state-of-the-art two-qubit gate performance. 
One of the main reasons for the slow progress in improving two-qubit gate fidelity could be an incomplete understanding of the gate error mechanism.
While some previous work mainly concentrates on the tunable coupling effects and ZZ coupling characteristics between computational qubits \cite{yan2018tunable,mundada2019suppression,li2020tunable,zhao2020high,xu2020zz,zhao2020suppression,han2020error}, other essential problems
are less explored. For instance, what are the error sources of two-qubit gates and the corresponding physical mechanism behind? How does the higher energy level of the coupler affect the resulting gate fidelity?  What is the optimized gate fidelity using this architecture? Could we find some alternative parameter regions or schemes whose gate performances are better than the traditional ones?
To better understand and solve these problems, we focus on studying ZZ parasitic coupling mechanism and exploring novel parameter regions, which may advance the technology of large-scale coupler architecture.

This paper concentrates on exploring the physical mechanism of two-qubit gate error sources from the effective Hamiltonian perspective. As the primary error source for the targeted gate, we pay special attention to the characteristics and physical mechanisms of ZZ parasitic couplings \cite{mundada2019suppression,barends2019diabatic,zhao2020switchable,zhao2020high, sung2020realization,foxen2020demonstrating}. Interestingly and surprisingly, we find some novel parameter regions in which high-fidelity two-qubit gates are expected. The main contributions and findings of this work are summarized as follows:
i) we provide clear and straightforward understandings to the physical mechanism of ZZ parametric coupling in coupler architecture. Using the effective Hamiltonian derived, the physical processes that describe different parametric coupling can be explained clearly.
To the author's knowledge, this is the first time to find the physical mechanism of ZZ parasitic couplings from the effective Hamiltonian perspective which usually contains richer physics than other methods;
ii) using the analytical results obtained in this paper, some impressive results of previous work \cite{li2020tunable,zhao2020high,zhao2020suppression,ku2020suppression} can be explained and the related physical mechanism can be understood. More importantly, four unexplored parameter regions are inspired for eliminating ZZ parasitic coupling. The physical mechanism for ZZ coupling elimination is: the coupler's high energy level can be used to neutralize the energy shift induced by computational qubits' high energy level;
iii) we demonstrate high-fidelity two-qubit gates are realizable using our suggested parameter regions. Beyond numerical simulations, an analytical expression is derived for the two-qubit gate error.  As applications, it can be applied to estimate the average gate error of superconducting quantum processor with coupler architecture conveniently. 

The remainder of this paper is organized as follows. We start from the system Hamiltonian in lab frame and derive the effective Hamiltonian using Schrieffer-Wolf transformation (SWT) \cite{bravyi2011schrieffer} in Sec.~\ref{sec:Coupler architecture and Hamiltonian}. With the help of the resulting effective Hamiltonian, the physical mechanisms of ZZ parasitic couplings are discussed and analyzed in Sec.~\ref{sec:Physical mechanisms of parasitic couplings}.  In Sec.~\ref{sec:New parameters regions}, inspired by the analytical results, we propose four novel parameter regions in which ZZ parasitic couplings are expected to be eliminated. As a further step, we also suggest some possible experimental realization to achieve  high-fidelity two-qubit gates.  Involving different types of noises, we study the gate error characteristics using the suggested parameter regions  in Sec.~\ref{sec:Gate error analysis}; moreover, the tradeoff effects between energy relaxation effect and parasitic couplings are discussed as well.  We conclude in Sec.~\ref{Sec:Summary and Perspectives} and give some technical details in Appendices.

\section{Coupler architecture and Effective Hamiltonian}
\label{sec:Coupler architecture and Hamiltonian}

\begin{figure}[h]
\centering
\includegraphics[width=\columnwidth]{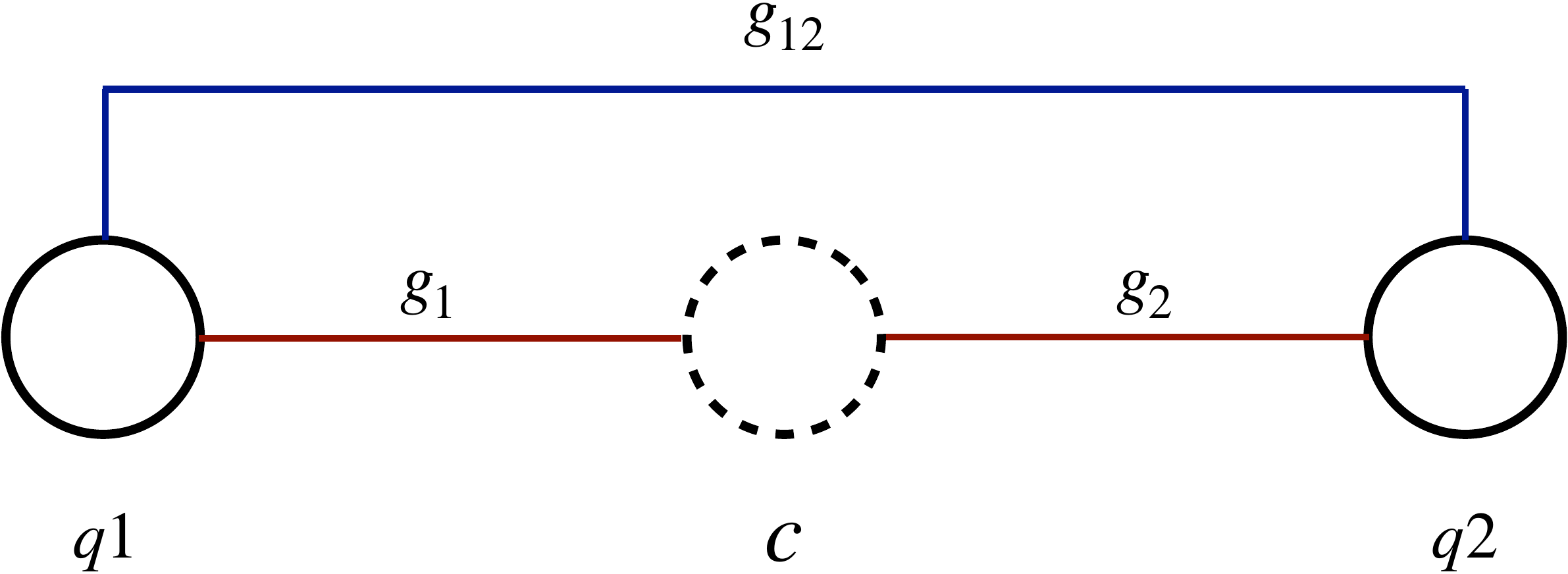}
\caption{Two computational qubits ($q1$ and $q2$, solid circles) are connected directly with a coupling strength $g_{12}$. Besides, a coupler ($c$, dashed circle), usually realized with a qubit, is introduced to connect the two qubits indirectly.  Benefiting from the interference effect between these two different coupling paths, the effective coupling $g_{\rm eff}$ between two computational qubits becomes tunable. More importantly, $g_{\rm eff}$ is allowed to be tuned off if necessary. }  \label{fig-circuit}
\end{figure}

As shown in Fig.~\ref{fig-circuit}, our studied architecture consists of two computational qubits ($q1$ and $q2$, solid circles), which has a direct coupling $g_{12}$. An auxiliary qubit is introduced as a coupler ($c$, dashed circle) to interact with each computational qubit, which will generate an effective indirect coupling. 
Both computational qubits and coupler are modelled by Duffing oscillators \cite{kovacic2011duffing},
the Hamiltonian in lab frame describes the coupler architecture consists of three parts:
\begin{equation}\label{Hamiltonian:lab frame}
\hat{H}_{\rm Lab} = \hat{H} _0 + \hat{H}_{qq} + \hat{H}_{qc}
\end{equation}
with
\begin{eqnarray}
 \hat{H} _0 &=&  \sum_{\lambda=q1,q2,c}  \omega_{\lambda} \hat{a}^{\dagger}_{\lambda} \hat{a}_{\lambda} + \frac{\alpha_{\lambda}}{2} \hat{a}^{\dagger}_{\lambda} \hat{a}^{\dagger}_{\lambda} \hat{a}_{\lambda} \hat{a}_{\lambda}, \label{Hamiltonian:0}\\ 
 \hat{H}_{qq} &=& g_{12} \left( \hat{a}^{\dagger}_{q1}  \hat{a}_{q2}  + \hat{a}_{q1}       \hat{a}^{\dagger}_{q2} - \hat{a}^{\dagger}_{q1}  \hat{a}^{\dagger}_{q2} - \hat{a}_{q1}  \hat{a}_{q2} \right), \\ 
  \hat{H}_{qc} &=& \sum_{k=1,2}  g_k \left( \hat{a}^{\dagger}_{qk}  \hat{a}_{c}  + \hat{a}_{qk}       \hat{a}^{\dagger}_{c} - \hat{a}^{\dagger}_{qk}  \hat{a}^{\dagger}_{c} - \hat{a}_{qk}  \hat{a}_{c}   \right), \label{Hamiltonian:qc} 
\end{eqnarray}
where $ \hat{H} _0$ describes the free energy of these three subsystems, $\omega_{\lambda}$ and $\alpha_{\lambda}$ ($\lambda={q1, q2, c}$) are the frequency and anharmonicity of the subsystem $\lambda$, respectively.
The operators $\hat{a}_{\lambda}$, $ \hat{a}^{\dagger}_{\lambda}$ are annihilation and creation operators for each qubit.
 $ \hat{H} _{qq}$ represents the direct coupling between two computational qubits, and $g_{12}$ is the coupling strength.  $ \hat{H} _{qc}$ means the couplings between computational qubits and coupler, and $g_{1}$, $g_{2}$ is the corresponding coupling strength. 
 It is noticeable that we keep not only the usual Jaynes-Cummings interaction terms but also the counter-rotating terms in $\hat{H}_{\rm Lab}$. This is because the couplings among the three subsystems are usually charge-charge couplings; all of these terms should be involved when one expends the charge operators in terms of annihilation and creation operators.
 
One of the primary purposes for inserting the coupler in superconducting circuits is to create destructive interference between the direct and indirect coupling of two computational qubits, so the first and foremost task is to derive the effective indirect coupling.
Since we  mainly concentrate on two-qubit gates realized between $q_1$ and $q_2$,   an approach to  decouple the coupler from the whole system is required. As a perturbed method,  SWT  is applied to
adiabatically eliminate qubit-coupler couplings and  work out the indirect coupling.  In particular,  the system Hamiltonian in lab frame is transformed to,   $\hat{{H}}^1_{\rm eff} = e^{\hat{s}_1} \hat{H}_{\rm Lab}  e^{-\hat{s}_1}$, $\hat{{H}}^2_{\rm eff} = e^{\hat{s}_2} \hat{{H}}^1_{\rm eff}  e^{-\hat{s}_2}$  with choosing
\begin{equation}\label{s operator}
\hat{s}_1 = \sum_{k=1,2}  \frac{g_k}{\Delta_k} \left(\hat{a}^{\dagger}_{qk}  \hat{a}_{c}  - \hat{a}_{qk}       \hat{a}^{\dagger}_{c} \right) - \frac{g_k}{\sum_{k}} \left(\hat{a}^{\dagger}_{qk}  \hat{a}^{\dagger}_{c} - \hat{a}_{qk}  \hat{a}_{c}  \right), 
\end{equation}
\begin{eqnarray}
\hat{s}_2 &=&     \sum_{k=1,2}   \frac{g_k \alpha_{qk}}{\Delta_k \left(\tilde{\Delta}_{k} + \tilde{\alpha}_{qk} \right) }  \left( \hat{a}^{\dagger}_{qk} \hat{a}_{qk} \hat{a}_{qk} \hat{a}^{\dagger}_{c} - \hat{a}^{\dagger}_{qk} \hat{a}^{\dagger}_{qk} \hat{a}_{qk} \hat{a}_{c} \right) \nonumber \\
&& +   \sum_{k=1,2}  \frac{g_k \alpha_c}{\Delta_k \left(\tilde{\Delta}_{k} - \tilde{\alpha}_c \right) } \left(  \hat{a}_{qk}  \hat{a}^{\dagger}_{c} \hat{a}^{\dagger}_{c} \hat{a}_{c} - \hat{a}^{\dagger}_{qk} \hat{a}^{\dagger}_{c} \hat{a}_{c} \hat{a}_{c}   \right) ,  \nonumber\\
\end{eqnarray}
where the detunings $\Delta_{k}=\omega_{{qk}}-\omega_c$ with $k=1,2$,  $\Sigma_{k}=\omega_{qk}+\omega_c$, and $\tilde{\Delta}_{k}=\tilde{\omega}_{{qk}}-\tilde{\omega}_c$. Here, the shifted qubits frequencies $\tilde{\omega}_{{qk}}$, $\tilde{\omega}_c$ and the shifted anharmonicities $ \tilde{\alpha}_{qk}$ will be given in Eqs.~\eqref{shifted qubit frequencies} and \eqref{shifted qubit anharmornicities}.


Applying two times SWT to the fourth order, and considering  dispersive regimes, i.e., $g_k \ll \vert \Delta_k \vert $, the Hamiltonian in new representation $\hat{{H}}^2_{\rm eff} $ is obtained as follows. 
More details concerning the cumbersome derivation are given in Appendix \ref{appendix: new representation}. 
This effective Hamiltonian will be the cornerstone of the following analysis and discussions. 
In addition, it maybe also helpful in exploring other problems in coupler architecture.

\begin{eqnarray} \label{Hamiltonian: new representation}
\hat{{H}}^2_{\rm eff}  &\approx & \sum_{\lambda = q1,q2,c}  \tilde{\omega}_{\lambda} \hat{a}^{\dagger}_{\lambda} \hat{a}_{\lambda} + \frac{\tilde{\alpha}_{\lambda}}{2} \hat{a}^{\dagger}_{\lambda} \hat{a}^{\dagger}_{\lambda} \hat{a}_{\lambda} \hat{a}_{\lambda} \\
&& +{g}_{\rm eff} \left( \hat{a}^{\dagger}_{q1}  \hat{a}_{q2}  - \hat{a}^{\dagger}_{q1}  \hat{a}^{\dagger}_{q2} + H.c. \right) 
\nonumber\\
  &&  - \frac{1}{2}  \frac{g_1 g_2 \alpha_{q1}}{\Delta_1 \Delta_2}
 \left( \hat{a}^{\dagger}_{q1} \hat{a}_{q1} \hat{a}_{q1} \hat{a}^{\dagger}_{q2}  + H.c. \right)\nonumber\\
    &&  - \frac{1}{2}  \frac{g_1 g_2 \alpha_{q2}}{\Delta_1 \Delta_2}
 \left(
  \hat{a}^{\dagger}_{q2} \hat{a}_{q2} \hat{a}_{q2} \hat{a}^{\dagger}_{q1} + H.c. \right)\nonumber\\
 &&
+ \frac{g_1 g_2 \alpha_c}{\Delta_1 \Delta_2}
\left( \hat{a}_{q1}\hat{a}_{q2} \hat{a}^{\dagger}_{c} \hat{a}^{\dagger}_{c} +H.c. \right)  \nonumber\\
&& +\frac{1}{2} \left(\frac{ g_1 g_2 }{\Delta_1 \Delta_2}\right)^2  \left(\alpha_{q1}+\alpha_{q2}+4\alpha_{c}\right)  \hat{a}^{\dagger}_{q1}   \hat{a}_{q1}  \hat{a}^{\dagger}_{q2} \hat{a}_{q2} , \nonumber
\end{eqnarray}
where the shifted qubit frequencies and anharmornicities  are obtained as
\begin{eqnarray} \label{shifted qubit frequencies}
\tilde{\omega}_{qk} &=& {\omega}_{qk} +  \frac{ g^2_k}{\Delta_k} - \frac{ g^2_k}{\Sigma_k} ~  , ~ k=1,2,  \nonumber\\
\tilde{\omega}_{c} &=& {\omega}_{c} -  \sum_{k=1}^2 \left( \frac{ g^2_k}{\Delta_k} + \frac{ g^2_k}{\Sigma_k}  \right),
\end{eqnarray}
\begin{eqnarray} \label{shifted qubit anharmornicities}
\tilde{\alpha}_{qk} &=& \alpha_{qk} \left[1 - \frac{2 g_k^2}{\Delta_k (\Delta_k + \alpha_{qk}) } \right] , ~  k=1,2,  \nonumber\\
 \tilde{\alpha}_{c} &=& \alpha_{c} \left[  1- \sum_{k=1}^2 \frac{2 g_k^2}{\Delta_k (\Delta_k - \alpha_c) }  \right],
\end{eqnarray}
respectively, and the effective coupling between two computational qubits is obtained as ${g}_{\rm eff} = {g}_{12}+ \tilde{g}_{12}$ with
\begin{equation} \label{coupling: q1 q2}
 \tilde{g}_{12} = \frac{g_1 g_2}{2} \sum_{k=1}^2 \left(\frac{1}{\Delta_k}-\frac{1}{\Sigma_k} \right).
\end{equation}

As seen obviously from the Hamiltonian $\hat{{H}}^2_{\rm eff}$,  the effective coupling $g_{\rm eff}$ becomes tunable through simply varying the coupler frequency $\omega_c$. Moreover, this coupling can be switched off, i.e., $g_{\rm eff}=0$,  if necessary.
Comparing with previous work \cite{yan2018tunable} which only give the effective coupling between computational qubits, we check carefully the additional second and fourth-order perturbative contributions, which contain more fruitful physics. In particular, we will see later that these terms induced by the nonlinear terms exactly correspond to gate error sources. Apart from these, we apply a second SWT to extend the analytical results  to a more general regime, i.e., $\alpha_{\lambda} \sim \vert \Delta_{k} \vert$ (in Ref.~\cite{yan2018tunable}, it was restricted to $\alpha_{\lambda} \ll \vert \Delta_{k} \vert$). Particularly, the second SWT results in the modification of qubits' anharmonicity. We will see that the second SWT becomes very important when the coupler architecture is studied in certain regimes.



As we know, the original idea for  coupler architecture is to make the coupling between computational qubits tunable, and more importantly to isolate one qubit from the neighboring qubits if necessary. As a further step, we specify the explicit parameter regions and conditions for realizing a switch. 
The first condition is  $g_k \ll \vert  \Delta_k \vert$ (dispersive couplings), meanwhile we have to take $\Delta_k <0$ which is used to generate negative indirect couplings between computational qubits. 
As obtained approximately from Eq.~\eqref{coupling: q1 q2}, the required coupler frequency for ${g}_{\rm eff}  =0$ is estimated  roughly as
\begin{equation} \label{omega C:switch condition}
\omega_c^{\rm off} \approx \omega_{q} + \frac{g_1 g_2} {g_{12}},
\end{equation}
where we assumed $\omega_{q1} \approx \omega_{q2} = \omega_q $ and $\sum_{k} \gg \Delta_{k}$, $k=1,2$.
To meet the dispersive  conditions $g_{k} /(\omega_c^{\rm off} - \omega_{qk}) \ll 1$, and zero effective coupling condition Eq.~\eqref{omega C:switch condition} simultaneously, it requires the direct coupling $g_{12} \ll g_{1,2}$. This is exactly the usual parameter regimes in realistic coupler type experiments. 

Once achieving ${g}_{\rm eff} =0$, one may think  qubits $q1$ and $q2$ become completely isolated from each other.   As a consequence, high-fidelity single-qubit gates are expected. Furthermore, if we consider the coupler architecture with many qubits  (e.g., \cite{arute2019quantum}), two-qubit gates can also avoid the crosstalk from other neighboring qubits, resulting in high-fidelity gates. Some previous work \cite{li2020tunable} indeed held similar arguments. However, our findings indicate it is not the case. As seen clearly from the effective Hamiltonian $\hat{H}^2_{\rm eff}$ [Eq.~\eqref{Hamiltonian: new representation}], even with ${g}_{\rm eff}=0 $, the parasitic couplings between computational qubits could still introduce unavoidable crosstalk. This will be discussed in the following sections.

\section{Characteristics and  Physical mechanisms of parasitic couplings}
\label{sec:Physical mechanisms of parasitic couplings}

To realize two-qubit native gates in superconducting quantum computing, for instance iSWAP gate, the XY type of coupling (i.e., $\hat{\sigma}^{q1}_{x} \hat{\sigma}^{q2}_{x} + \hat{\sigma}^{q1}_{y} \hat{\sigma}^{q2}_{y} $) between computational qubits is required \cite{krantz2019quantum}. Except for it, other couplings with different forms are counted as parasitic couplings, which will induce gate errors. In this section,  
we study the characteristics and  physical mechanisms of these parasitic couplings from the effective Hamiltonian perspective. 

First of all, the effective Hamiltonian $\hat{{H}}^2_{\rm eff}$ reduces approximately to
$ \sum_{\lambda = q1,q2,c}  \tilde{\omega}_{\lambda} \hat{a}^{\dagger}_{\lambda} \hat{a}_{\lambda} + ({\tilde{\alpha}_{\lambda}}/{2}) \hat{a}^{\dagger}_{\lambda} \hat{a}^{\dagger}_{\lambda} \hat{a}_{\lambda} \hat{a}_{\lambda} +{g}_{\rm eff} ( \hat{a}^{\dagger}_{q1}  \hat{a}_{q2}   - \hat{a}^{\dagger}_{q1}  \hat{a}^{\dagger}_{q2} + H.c. ) $ and 
high-order contributions  are neglected when we consider the regime $\vert \alpha_{\lambda}\vert \ll \vert \Delta_{k} \vert, \Sigma_{k}$ \cite{yan2018tunable}. 
Next, tuning the two computational qubits to be resonant, i.e., $\tilde{\omega}_{q1} = \tilde{\omega}_{q2} $, and reducing to computational basis (i.e., using Pauli representation),
 moreover transforming the resulting Hamiltonian into rotating representation with qubit frequency $\tilde{\omega}_{q1}$ and $\tilde{\omega}_{q2} $,  we ultimately get an effective Hamiltonian $ {g}_{\rm eff}  \left(\hat{\sigma}_{x}^{q1} \hat{\sigma}_{x}^{q2} + \hat{\sigma}_{y}^{q1} \hat{\sigma}_{y}^{q2}  \right)/2$ which could straightforward realize  perfect iSWAP gates with gate time $t_g= \pi/(2 g_{\rm eff})$ \cite{krantz2019quantum}. 
However, this is not the case in practice because the anharmonicities of the computational qubits and coupler do not always hold the condition $\vert \alpha_{\lambda} \vert \ll \vert \Delta_{k} \vert, \Sigma_{k}$.  Therefore,  the contributions of those terms originated from the nonlinear terms, which were neglected in the ideal case, have to be considered. More importantly, we will see that the physical mechanisms of parasitic couplings can be understood with the help of these terms. 

In superconducting circuits with coupler architecture, we mainly pay attention to the computational space of computational qubits while the coupler is assumed to stay in the ground state all the time. 
As a consequence, the computational space consists of the states $\vert 000 \rangle$, $\vert 100 \rangle$, $\vert 001 \rangle$ and $\vert 101 \rangle$ ($\vert q1,c,q2 \rangle$, represented in the Fock basis and labeled by the
approximate bare states when the coupler is far detuned; the corresponding eigenenergy denotes as $\omega_{q1,c,q2}$).  In addition to the  states mentioned above, those states (out of the computational space) that affect the states in computational space should be considered as well. To be able to explain clearly the physical mechanism of parasitic couplings, we rewrite the effective Hamiltonian \eqref{Hamiltonian: new representation} in terms of the basis $\vert q1,c,q2 \rangle$. In particular, we keep only the computational basis as well as those couple directly with computational basis. In the end, we obtain
\begin{widetext}
 \begin{eqnarray} \label{Hamiltonian: computational basis}
 {\hat{H}}'_{\rm eff} &=& \tilde{\omega}_{q1} \vert 100 \rangle \langle 100 \vert  +  \tilde{\omega}_{q2} \vert 001 \rangle \langle 001 \vert  +  (\tilde{\omega}_{q1} + \tilde{\omega}_{q2} )\vert 101 \rangle \langle 101\vert  + (2\tilde{\omega}_{q1} + \tilde{\alpha}_{q1} )\vert 200 \rangle \langle 200\vert   +(2\tilde{\omega}_{q2} + \tilde{\alpha}_{q2} ) \vert 002 \rangle \langle 002 \vert  \nonumber\\
 &&   +  (2\tilde{\omega}_{c} + \tilde{\alpha}_{c} )\vert 020 \rangle \langle 020 \vert   + g_{\rm eff} \left( \vert 100 \rangle \langle 001 \vert + \vert 001 \rangle \langle 100 \vert  \right)+ \tilde{g}_{200}  \left(\vert 200 \rangle \langle 101 \vert + \vert 101 \rangle \langle 200 \vert \right)    \nonumber\\
 && + \tilde{g}_{002}  \left( \vert 002 \rangle \langle 101 \vert +  \vert 101 \rangle \langle 002 \vert  \right) +  \tilde{g}_{020}  \left( \vert 020 \rangle \langle 101 \vert + \vert 101 \rangle \langle 020 \vert \right)  + \tilde{g}_{\rm cross-Kerr}   \left(\alpha_{q1}+\alpha_{q2}+4\alpha_{c}\right)  \vert 101 \rangle \langle 101 \vert,\nonumber\\
 \end{eqnarray}
where the coupling strengths for different physical processes are computed as
\begin{equation} \label{expressions: coupling strength}
\tilde{g}_{200} =  \sqrt{2} \left( g_{\rm eff} - \frac{1}{2}  \frac{g_1 g_2}{\Delta_1 \Delta_2} \alpha_{q1} \right), ~
 \tilde{g}_{002} = \sqrt{2} \left( g_{\rm eff}  - \frac{1}{2}  \frac{g_1 g_2}{\Delta_1 \Delta_2} \alpha_{q2} \right), ~
 \tilde{g}_{020} = \sqrt{2} \frac{g_1 g_2 }{\Delta_1 \Delta_2} \alpha_{c}, ~
 \tilde{g}_{\rm cross-Kerr} = \frac{1}{2}  \left(\frac{ g_1 g_2 }{\Delta_1 \Delta_2}\right)^2 .
  \end{equation}
\end{widetext}
 
 It is noticeable that we  consider only those states with excitation not more than two (neglect the states with larger excitation) and those coupling with the state  $\vert 101 \rangle$ of computational space. Since the key goal is to realize an iSWAP gate,  the transition between $ \vert 100 \rangle$ and $ \vert 001 \rangle $ is used to realize the target gate exactly. Therefore, the last four terms of Eq.~\eqref{Hamiltonian: computational basis}, describing the coupling processes between $\vert 101 \rangle$ and high-energy states  $\vert 200 \rangle$, $\vert 020 \rangle$, $\vert 002 \rangle$, are counted as parasitic  couplings. 
Specifically, when the driven pulses  are applied adiabatically, these couplings will lead to the additional phase, which results in a parasitic  control phase gate. Such interaction is often called ZZ crosstalk which becomes a performance-limiting factor for gate fidelity \cite{mckay2019three,barends2019diabatic,sheldon2016procedure,mckay2016universal,magesan2020effective} and quantum error correction \cite{takita2016demonstration}.
Unlike classical crosstalk, which can be removed through careful characterization and control optimization \cite{winick2020simulating}, ZZ parasitic crosstalk is hard to be mitigated \cite{xu2020zz}. 

Next,  reducing to the basis which consists of the lowest two energy levels of computational qubits, the effective Hamiltonian  is expressed as 
\begin{equation} \label{Hamiltonian:xy+zz}
\hat{H}''_{\rm eff} \approx  \frac{\tilde{\omega}_{q1}}{2} \hat{\sigma}_z^1 + \frac{\tilde{\omega}_{q2}}{2} \hat{\sigma}_z^2  +  \hat{H}^{\rm XY}_{\rm int}  +  \hat{H}^{\rm ZZ}_{\rm int},
\end{equation}
with  two different types of coupling reading
\begin{eqnarray}
 \hat{H}^{\rm XY}_{\rm int} &=& \frac{ g_{\rm eff}}{2} \left(\hat{\sigma}_{x}^{q1} \hat{\sigma}_{x}^{q2} + \hat{\sigma}_{y}^{q1} \hat{\sigma}_{y}^{q2}  \right), \label{Hamiltonian: XY} \\
 \hat{H}^{\rm ZZ}_{\rm int} &=& \zeta_{zz} \hat{\sigma}_{z}^{q1} \hat{\sigma}_{z}^{q2}, \label{Hamiltonian: ZZ}
\end{eqnarray}
where both XY and ZZ  coupling strengths, namely $ g_{\rm eff}$ and $ \zeta_{zz}$, can be derived analytically from  Hamiltonian $ {\hat{H}}'_{\rm eff} $ [Eq.~\eqref{Hamiltonian: computational basis}] within the regimes of interest. Moreover, the correctness of the analytical results can be further verified via numerically diagonalizing the system Hamiltonian $\hat{H}_{\rm Lab}$ [Eq.~\eqref{Hamiltonian:lab frame}]. In particular, $2 g_{\rm eff}$ is evaluated as the energy difference between $\omega_{100} $ and $\omega_{001}$,
 and $ \zeta_{zz}=\omega_{101} - \omega_{100}- \omega_{001}$. Here, $\omega_{q1,c,q2}$ denotes the eigenenergy of the system Hamiltonian, and $\omega_{000}$ is set to zero for simplification.  

As seen clearly from Eq.~\eqref{Hamiltonian: computational basis} that the parasitic ZZ coupling $ \zeta_{zz}$ originate from various  couplings between the states $\vert 101 \rangle$ and  $\vert 200 \rangle$,  $\vert 020 \rangle$, $\vert 002 \rangle$, $\vert 101 \rangle$.
In addition to the numerical results which can be solved trivially, previous investigations \cite{mundada2019suppression,zhao2020high,sung2020realization,zhao2020suppression} calculated $ \zeta_{zz}$ via diagonalizing the system Hamiltonian perturbatively. Apart from the extremely cumbersome calculations, one cannot obtain clear physical mechanisms. As a contrast, the effective Hamiltonian as well as the analytical results obtained in this paper have simple forms, and can be interpreted as the physical processes of parasitic couplings. 
For different parameter regime,  we find that different coupling term dominates. Here,  we  concentrate on three different regimes representing three typical physical processes. The first one is to consider the resonant process between $\vert 101 \rangle$ and $\vert 200 \rangle$ (or $\vert 002 \rangle$), which will be discussed in subsection \ref{Error from higher energy levels}. 
The second one is to consider the resonant process between $\vert 101 \rangle$ and $\vert 020 \rangle$, the high energy level of the coupler will play an important role; this will be discussed in subsection \ref{Error from the coupler}. The third one is to consider the dispersive regime, namely the effective coupling strengths are much smaller than the energy difference between $\vert 101 \rangle$ and $\vert 200 \rangle$ (or $\vert 002 \rangle$, $\vert 020 \rangle$),  which will be discussed in subsection \ref{dispersive regime}. 

\subsection{Parasitic couplings due to high energy levels of computational qubits  } 
\label{Error from higher energy levels}

In actual superconducting circuits experiments, in addition to computational space consisting of $\vert 000 \rangle$, $\vert 001 \rangle$, $\vert 100 \rangle$, $\vert 101 \rangle$, the effect of computational qubits' higher energy levels has to be considered as well. 
Even if without the coupler (namely two computational qubits couple directly \cite{barends2014superconducting, kelly2015state}), the usual Jaynes-Cummings interaction between two computational qubits, i.e., $(\hat{a}^{\dagger}_{q1} \hat{a}_{q2} + \hat{a}^{\dagger}_{q2} \hat{a}_{q1})$, will couple the states $\vert 11 \rangle$ and $\vert 20 \rangle$ (or $\vert 02 \rangle$).  Besides, the terms
$ (\hat{a}^{\dagger}_{q1} \hat{a}_{q1} \hat{a}_{q1} \hat{a}^{\dagger}_{q2} + 
  \hat{a}^{\dagger}_{q2} \hat{a}_{q2} \hat{a}_{q2} \hat{a}^{\dagger}_{q1} + H.c. )$ of effective Hamiltonian $\hat{H}^2_{\rm eff}$, i.e., Eq.~\eqref{Hamiltonian: new representation}, contribute to parasitic coupling as well. 
  In particular, it describes the transition between the states $\vert 101 \rangle$ and $\vert 200 \rangle$ (or $\vert 002 \rangle$) exactly, because the term has the relations: $\hat{a}^{\dagger}_{q1} \hat{a}_{q1} \hat{a}_{q1} \hat{a}^{\dagger}_{q2} \vert 200 \rangle = \sqrt{2} \vert 101 \rangle$ and $\hat{a}^{\dagger}_{q2} \hat{a}_{q2} \hat{a}_{q2} \hat{a}^{\dagger}_{q1} \vert 002 \rangle = \sqrt{2} \vert 101 \rangle$.
Specially, in the parameter regime with  $\tilde{\omega}_{q1} + \tilde{\omega}_{q2}  \approx 2 \tilde{\omega}_{q1} + \tilde{\alpha}_{q1}$ or  $\tilde{\omega}_{q1} + \tilde{\omega}_{q2}  \approx 2 \tilde{\omega}_{q2} + \tilde{\alpha}_{q2}$, namely the states $\vert 101 \rangle$ and $\vert 200 \rangle$ (or $\vert 002 \rangle$) are in resonant nearly, the resulting ZZ coupling $\zeta_{zz}$ will originate from this resonant process while the contributions from other dispersive couplings can be neglected. Using Eq.~\eqref{Hamiltonian: computational basis}, $\zeta_{zz}$ can be derived analytically. In particular, we obtain
\begin{eqnarray}\label{analytic:101 and 200}
 \vert \zeta_{zz} \vert &=& \frac{1}{2} \left(\sqrt{(\tilde{\Delta}_{12}+ \tilde{\alpha}_{q1})^2 +4 \tilde{g}^2_{200} } - \vert \tilde{\Delta}_{12} + \tilde{\alpha}_{q1}  \vert \right.  \\
&& + \left.  \sqrt{(\tilde{\Delta}_{12}- \tilde{\alpha}_{q2})^2 +4 \tilde{g}^2_{002} } - \vert \tilde{\Delta}_{12} - \tilde{\alpha}_{q2} \vert \right),  \nonumber
\end{eqnarray}
where $\tilde{\Delta}_{12}=\tilde{\omega}_{q1}-\tilde{\omega}_{q2}$ is the frequency detuning of two computational qubits in new representation, and  the  corresponding coupling strength $\tilde{g}_{200} $, $\tilde{g}_{002} $ was given in Eq.~\eqref{expressions: coupling strength}. The first (last) two terms in the bracket of Eq.~\eqref{analytic:101 and 200} corresponds to the resonant process between $\vert 101 \rangle$ and $\vert 200 \rangle$ ($\vert 002\rangle$). The  derivation of Eq.~\eqref{analytic:101 and 200}  is presented in Appendix \ref{ZZ coupling: Analytical expression}. 

\begin{figure}[htbp]
\centering
{\includegraphics[width=\columnwidth]{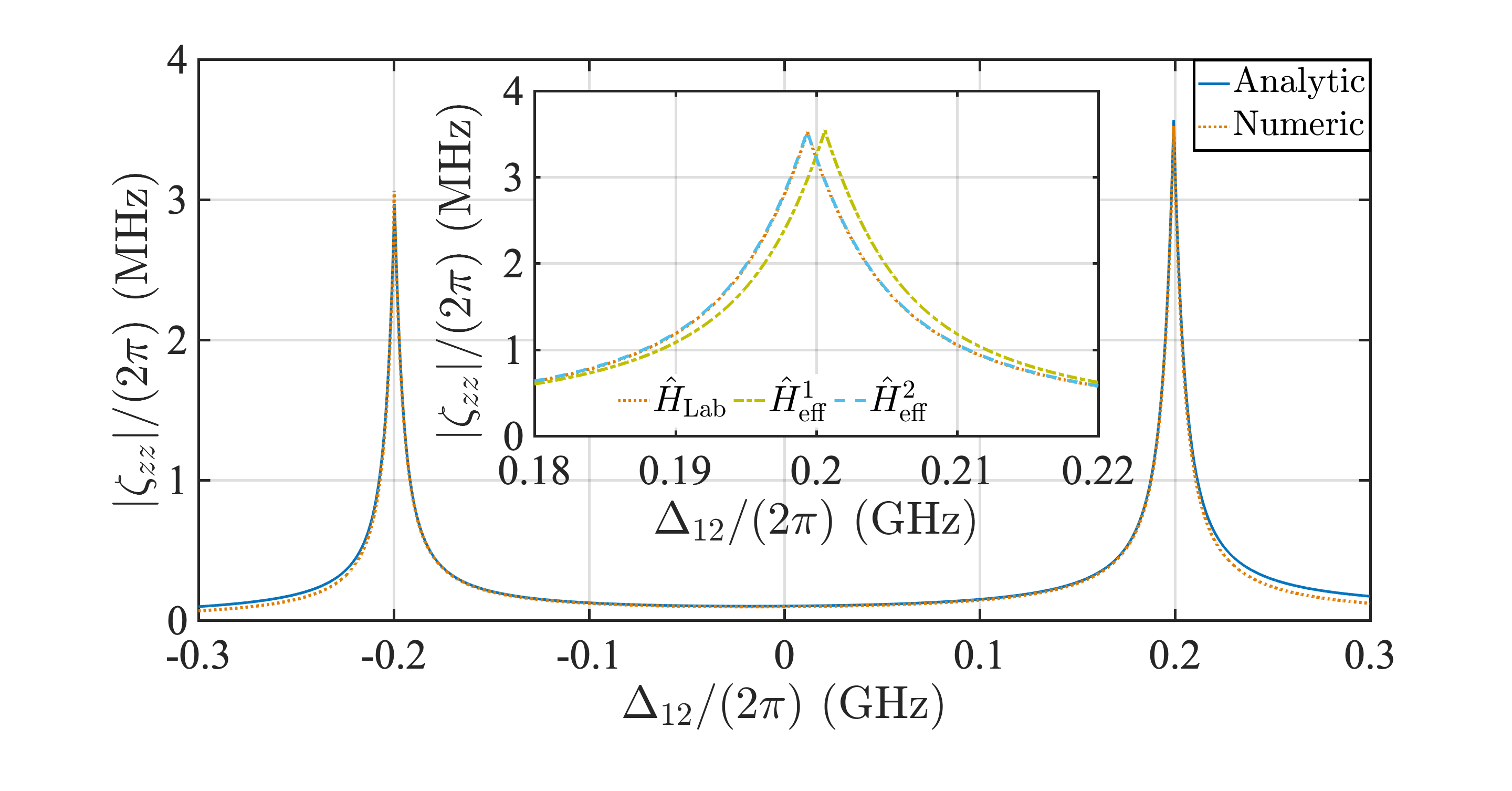}}
\caption{ZZ coupling strength $\vert \zeta_{zz} \vert$ characteristics in the regime that $\vert 101 \rangle$ is closely in resonance with $\vert 200\rangle$ or $\vert 002 \rangle$. The analytical result (blue solid) is computed using Eq.~\eqref{analytic:101 and 200}, while the numerical result (orange dotted) is evaluated through diagonalizing the system Hamiltonian $\hat{H}_{\rm Lab}$[Eq.~\eqref{Hamiltonian:lab frame}].
 (inset) ZZ coupling strength $\vert \zeta_{zz} \vert$ are evaluated numerically using three different Hamiltonian, including $\hat{H}_{\rm Lab}$ [Eq.~\eqref{Hamiltonian:lab frame}] in lab frame, $\hat{H}^1_{\rm eff}$ [Eq.~\eqref{Hamiltonian: 1st SW}] with 1 time SWT, $\hat{H}^2_{\rm eff}$ with 2 times SWT [Eq.~\eqref{Hamiltonian: new representation}]. Our results indicates the importance of  the 2nd SWT in effective Hamiltonian approach. The used parameters are: $\omega_{q2}/(2 \pi)=5~{\rm GHz}$,  $\omega_{c}/(2 \pi)=6~{\rm GHz}$,  $\alpha_{q1}/(2 \pi)=\alpha_{q2}/(2 \pi)=\alpha_{c}/(2 \pi)=-0.2~{\rm GHz}$,  $g1/(2 \pi)=g2/(2 \pi)=0.05~{\rm GHz}$,  $g_{12}=0$. } 
\label{fig:resonant 010 and 200}
\end{figure}

With the regime of interest,  we evaluate and plot  ZZ coupling strength $\vert \zeta_{zz} \vert$ with varying ${\Delta}_{12}$ (${\Delta}_{12}=\omega_{q1} - \omega_{q2}$) in Fig.~\ref{fig:resonant 010 and 200}.  As expected, ZZ parasitic coupling becomes pronounced when the state $\vert 101 \rangle$ is on resonance with the states  $\vert 200 \rangle$ or $\vert 002 \rangle$. Specially, when $\tilde{\Delta}_{12}= - \tilde{\alpha}_{q1}( \tilde{\alpha}_{q2})$  the ZZ coupling strength is evaluated as $\vert \zeta_{zz} \vert \approx  \vert \tilde{g}_{200} \vert (\vert \tilde{g}_{002} \vert)$ at the resonant point. Besides, ZZ coupling is largely suppressed once it is tuned away from the resonant processes. 
A similar result was also obtained in Ref.~\cite{zhao2020high} using numerical methods. 
The correctness of our analytical result (blue solid), plotted using Eq.~\eqref{analytic:101 and 200}, is verified through numerically diagonalizing the system Hamiltonian $\hat{H}_{\rm Lab}$ (orange dotted). It is obvious that the analytical result matches very well with the numerical one.
Another thing we want to point out is that the second SWT is very important in the regime considered. As shown in the inset of Fig.~\ref{fig:resonant 010 and 200}, the ZZ coupling strengths $\vert \zeta_{zz} \vert$  are evaluated using three different Hamiltonian, namely $\hat{{H}}_{\rm Lab}$[Eq.~\eqref{Hamiltonian:lab frame}] in lab frame, $\hat{{H}}^1_{\rm eff}$ [Eq.~\eqref{Hamiltonian: 1st SW}] with 1 time SWT, and $\hat{{H}}^2_{\rm eff}$ with 2 times SWT [Eq.~\eqref{Hamiltonian: new representation}].
In absence of 2nd SWT, the result solved from  $\hat{{H}}^1_{\rm eff}$  does not match very well with that of the original Hamiltonian $\hat{{H}}_{\rm Lab}$, which implies the effective Hamiltonian method with two time SWT gives an accurate result. 

\subsection{Parasitic couplings due to high energy levels of coupler }
\label{Error from the coupler}


As an auxiliary qubit, the coupler's main purpose is to generate tunable coupling between two computational qubits and serve as a switch if necessary. Another advantage is that the external driven noise induced through the coupler can be suppressed largely in dispersive regimes, comparing with directly driving computational qubits.  However, due to the existence of the coupler's high energy levels, we may have to pay the price of additional parasitic coupling. Using the effective Hamiltonian $\hat{H}^2_{\rm eff}$, the generation of parasitic coupling can be explained. In particular, the term $(\hat{a}^{\dagger}_{c} \hat{a}^{\dagger}_{c} \hat{a}_{q1}\hat{a}_{q2} +H.c. )$ of $\hat{H}^2_{\rm eff}$, i.e., Eq.~\eqref{Hamiltonian: new representation}, exactly describes this process. 
 It reflects the transition between the states ${\vert 101 \rangle}$ and ${\vert 020 \rangle}$, because the term has the relations:  $\hat{a}^{\dagger}_{c} \hat{a}^{\dagger}_{c} \hat{a}_{q1}\hat{a}_{q2} \vert 101 \rangle = \sqrt{2} \vert 020 \rangle $
 and $\hat{a}_{c} \hat{a}_{c} \hat{a}^{\dagger}_{q1}\hat{a}^{\dagger}_{q2} \vert 020 \rangle = \sqrt{2} \vert 101 \rangle $. Furthermore, when we consider the parameter regime with  $\tilde{\omega}_{q1} + \tilde{\omega}_{q2}  \approx 2 \tilde{\omega}_{c} + \tilde{\alpha}_{c}$, the  states $\vert 101 \rangle$ and $\vert 020 \rangle$ are in resonant nearly. Under this regime, the ZZ parasitic coupling $\zeta_{zz}$ mainly originate from this resonant process and other non-resonant processes can be neglected. As a further step, the analytical results of $\zeta_{zz}$ can be derived. We obtain
\begin{equation} \label{analytic:101 and 020}
\vert \zeta_{zz} \vert =  \left[\sqrt{\left(\tilde{\omega}_{c}-\tilde{\omega}^{*}_{c} \right)^2 + \tilde{g}^2_{020} } - \vert \tilde{\omega}_{c}-\tilde{\omega}^{*}_{c}\vert \right],
\end{equation}
where the central frequency $\tilde{\omega}^{*}_{c}$ is given as
$\tilde{\omega}^{*}_{c} = \left(\tilde{\omega}_{q1} + \tilde{\omega}_{q2} - \tilde{\alpha}_{c} \right)/2$, and the  corresponding coupling strength $\vert \tilde{g}_{020} \vert$ was given in Eq.~\eqref{expressions: coupling strength}.
The derivation of Eq.~\eqref{analytic:101 and 020}  is presented in Appendix \ref{ZZ coupling: Analytical expression}.

\begin{figure}[htbp]
\centering
\begin{minipage}[t]{\columnwidth}
\centering
\includegraphics[width= \columnwidth]{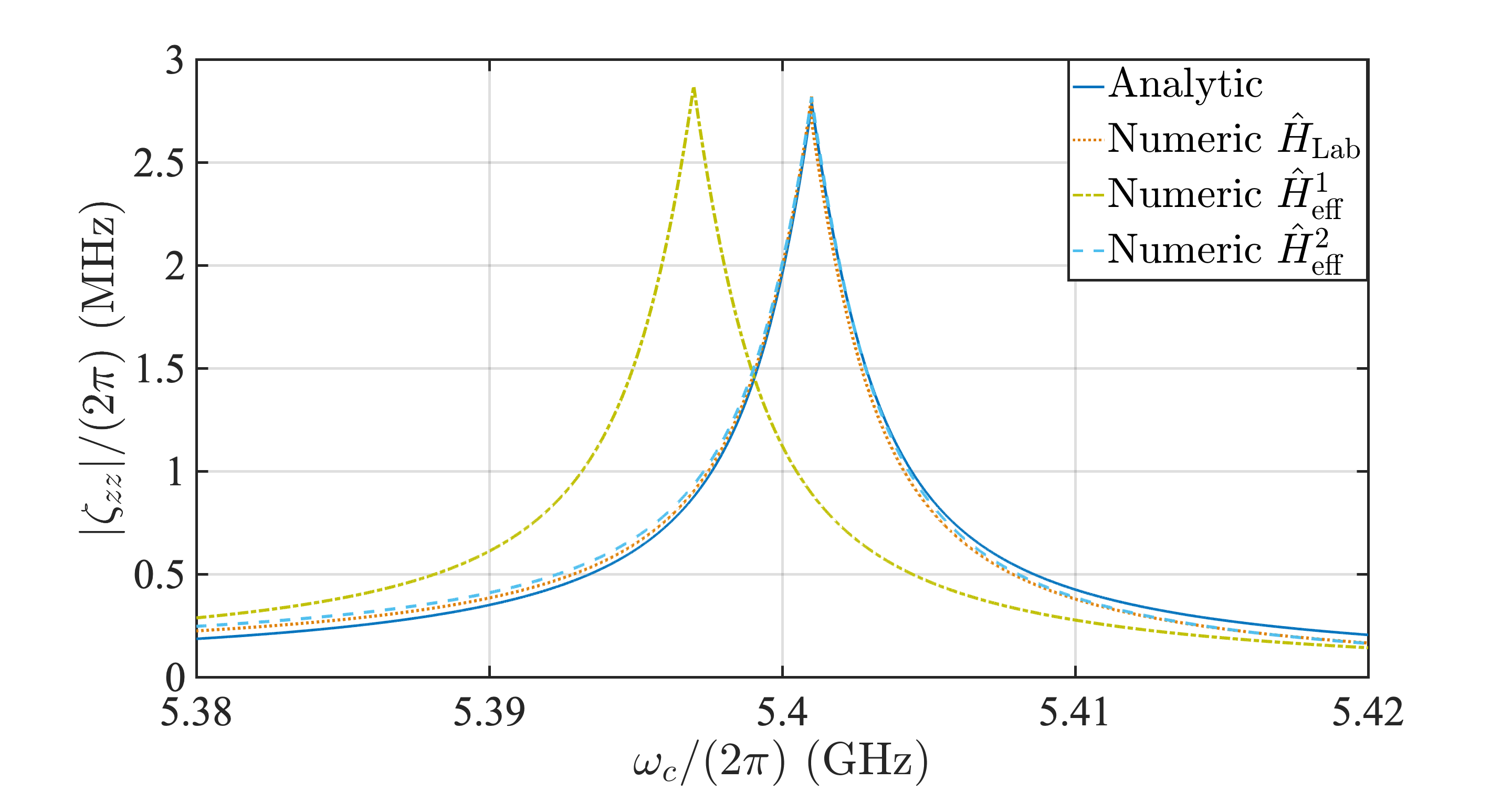}
\end{minipage}
\begin{minipage}[t]{\columnwidth}
\centering
\includegraphics[width=0.9 \columnwidth]{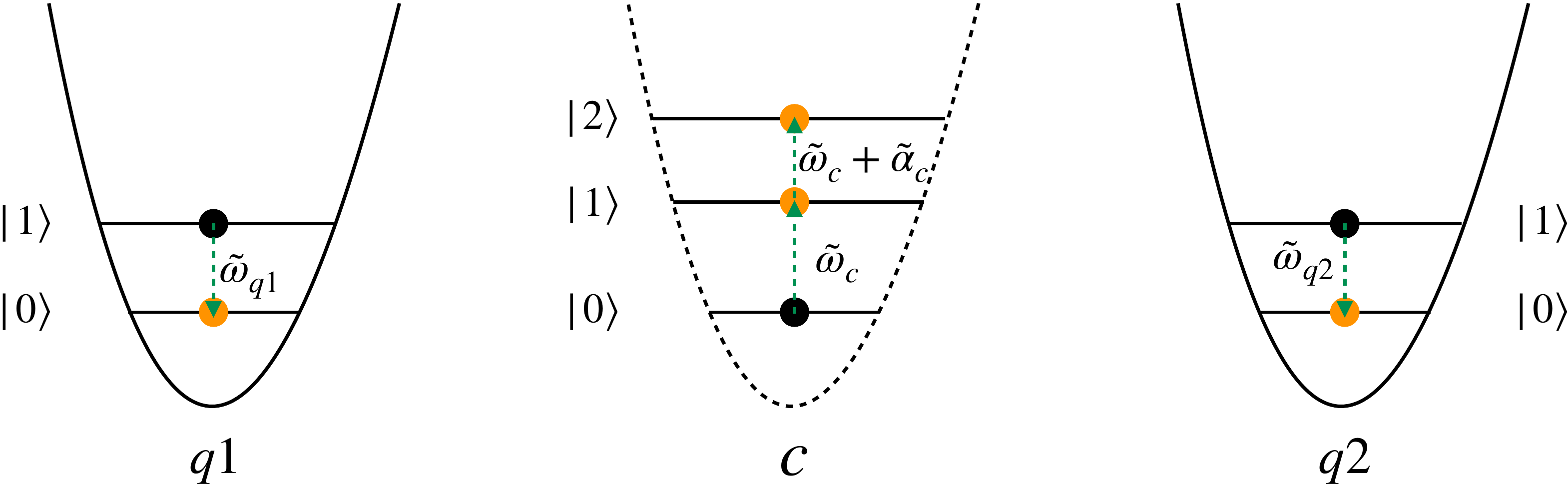}
\end{minipage}
\caption{(upper) ZZ coupling strength $\vert \zeta_{zz} \vert$ characteristics in the regime that the state $\vert 101 \rangle$ is closely in resonance with $\vert 020 \rangle$. 
The analytical result (blue solid) is plotted using Eq.~\eqref{analytic:101 and 020}. To verify the correctness of the analytical result, we compute numerically using the Hamiltonian $\hat{H}_{\rm Lab}$. Besides, we also show the numerical results with the Hamiltonian $\hat{H}^1_{\rm eff}$ and $\hat{H}^2_{\rm eff}$. The used parameters are: $\omega_{q1}/(2 \pi)=\omega_{q2}/(2 \pi)=5~{\rm GHz}$,   $\alpha_{q1}/(2 \pi)=\alpha_{q2}/(2 \pi)=-0.2~{\rm GHz}$, $\alpha_{c}/(2 \pi)=-0.8~{\rm GHz}$, $g1/(2 \pi)=g2/(2 \pi)=0.02~{\rm GHz}$, $g_{12}=0$.
 (lower) Energy level diagrams corresponds to the resonant process between the states $\vert 101 \rangle$ and $\vert 020 \rangle$. It can be used to explain the physical mechanism of parasitic coupling due to the coupler's second excite state. The initial state of the system is prepared with $\vert 101 \rangle$, one is able to create  the resonant  process between $\vert 101 \rangle$ and $\vert 020 \rangle$ (governed by the term  $\hat{a}^{\dagger}_{c} \hat{a}^{\dagger}_{c} \hat{a}_{q1}\hat{a}_{q2} +H.c.  $ of $\hat{H}^2_{\rm eff}$) when the system parameters satisfy the condition $\tilde{\omega}_{q1} + \tilde{\omega}_{q2} \approx 2\tilde{\omega}_c + \tilde{\alpha}_c $.   }
\label{fig:resonant 010 and 020}
\end{figure}

With the parameter regime of interest, we evaluate and plot  ZZ coupling strengths $\vert \zeta_{zz} \vert$ as a function of coupler frequency $\omega_c$ in Fig.~\ref{fig:resonant 010 and 020}. The analytical result (blue solid) is plotted using Eq.~\eqref{analytic:101 and 020}. To verify the correctness of analytical result, we also compute numerically $\zeta_{zz}$ using three different system Hamiltonian, including $\hat{{H}}_{\rm Lab}$ [Eq.~\eqref{Hamiltonian:lab frame}] in lab frame, $\hat{H}^1_{\rm eff}$ [Eq.~\eqref{Hamiltonian: 1st SW}] with 1 time SWT, and $\hat{H}^2_{\rm eff}$ with 2 time SWT [Eq.~\eqref{Hamiltonian: new representation}]. It is shown that the analytical result matches very well with the numerical result. Apart from that, 
we verify again that the necessity of 2nd SWT  in the resonant regimes.  It is noticeable that the central frequency for larger ZZ coupling is $\tilde{\omega}^{*}_{c}$, which corresponds to the resonant process between the states $\vert 101 \rangle$ and $\vert 020 \rangle$. The maximum ZZ coupling strength is evaluated as $\zeta_{zz}(\tilde{\omega}_c=\tilde{\omega}^{*}_{c})=\tilde{g}_{020}$. Once the coupler frequency is tuned away from $\tilde{\omega}^{*}_{c}$,  ZZ couplings are suppressed gradually.

As the effects of computational qubits' higher energy levels  were discussed before, it is quite natural to think about the consequence of possible resonant process between the states $\vert 101 \rangle$  and $\vert 020\rangle$. If it happens, a larger ZZ parasitic coupling $\zeta_{zz}$ may appear.  To explore the physical mechanism of resonant parasitic coupling induced by the coupler, we study it from the perspective of energy level diagrams. Initially, the coupler is prepared in ground state  $\vert 0 \rangle$.  Firstly, the question we want to ask is, is it possible to excite the coupler from the ground state to the first excited state, i.e.,  $\vert 0 \rangle \rightarrow \vert 1 \rangle$? To make it happen, an external energy $\omega_c$ is required. However, we realize this is almost impossible under the dispersive conditions $g_k \ll \vert \Delta_k \vert$, $k=1,2$. 
Secondly, is it possible to excite the coupler from the ground state to the second excited state, i.e., $\vert 0 \rangle \rightarrow \vert 2 \rangle$? To make it happen, it requires an energy $2 \tilde{\omega}_c + \tilde{\alpha}_c$ ($ \tilde{\alpha}_c <0$ for transmon qubit). This becomes possible if both qubit 1 and qubit 2 drop from $\vert 1 \rangle$ to $\vert 0 \rangle$. As seen from the  effective  Hamiltonian  $\hat{H}_{\rm eff}^{2}$ [Eq.~\eqref{Hamiltonian: new representation}], the term with $(\hat{a}^{\dagger}_{c} \hat{a}^{\dagger}_{c} \hat{a}_{q1}\hat{a}_{q2} +H.c. ) $ describes exactly this process. 
The physical picture for this resonant process is explained using energy levels diagrams in the lower one of Fig.~\ref{fig:resonant 010 and 020}. To focus on the effect of the coupler, we restrict the computational qubits as two-level systems (for simplicity and no loss of generality) and regard the coupler as a qutrit.  The transition between the states $\vert 101 \rangle$ and $\vert 020 \rangle$ may occur when we consider the parameter regime  $\tilde{\omega}_{q1} + \tilde{\omega}_{q2}  \approx 2\tilde{\omega}_c + \tilde{\alpha}_c$. 
In particular, the initial state of the system is prepared with $\vert 101 \rangle$ (black dots), after applying the operation $\hat{a}^{\dagger}_{c} \hat{a}^{\dagger}_{c} \hat{a}_{q1}\hat{a}_{q2} $, it transforms to the state $\vert 020 \rangle$ (orange dots).
As seen from the upper plot, the more close to this resonant condition, the larger $\vert \zeta_{zz} \vert$ obtained. In current experiments with coupler architecture, a larger detuning between qubits and coupler frequency and a relatively small negative  $\alpha_c$ are frequently used \cite{arute2019quantum,li2020tunable}, hence the resonant condition does not hold. Consequently, the parasitic coupling raised due to the coupler's higher energy levels is largely suppressed and thus can be ignored.  Very recently, novel parameter regimes (beyond dispersive approximation) of coupler architecture were proposed and experimentally realized \cite{xu2020high,PhysRevLett.125.240502}. In this new regime, the resonant process between $\vert 101 \rangle$ and $\vert 020 \rangle$ may occur, and therefore the coupler's higher energy plays an essential role in the resulting ZZ parasitic coupling. 



\subsection{Parasitic couplings in dispersive regime}
\label{dispersive regime}

In additional to the two special resonant regimes discussed above, we turn to explore a different parameter regions: dispersive regime, i.e.,  $\tilde{g}_{200} \ll \vert (\tilde{\omega}_{q1} + \tilde{\omega}_{q2})- (2\tilde{\omega}_{q1} + \tilde{\alpha}_{q1})  \vert$, $\tilde{g}_{020} \ll \vert (\tilde{\omega}_{q1} + \tilde{\omega}_{q2})- (2\tilde{\omega}_{c} + \tilde{\alpha}_{c})  \vert$, and $\tilde{g}_{002} \ll \vert (\tilde{\omega}_{q1} + \tilde{\omega}_{q2})- (2\tilde{\omega}_{q2} + \tilde{\alpha}_{q2})  \vert$.
Comparing with the two resonant regimes in which only one specific term dominates, all the terms of the effective Hamiltonian $\hat{H}^2_{\rm eff}$ [Eq.~\eqref{Hamiltonian: new representation}] contribute in dispersive regime.  Summing up all different kinds of contributions, we arrive as a quite concise  and meaningful  analytical result for $\vert \zeta_{zz} \vert $, expressing as
\begin{equation} \label{analytic:dispersive}
\vert \zeta_{zz} \vert \approx 2 \tilde{g}_{12}^2 \left\vert \frac{1}{\alpha_{q1}} + \frac{1}{\alpha_{q2}} + \frac{4}{\alpha_{c}-2\Delta}\right\vert,
\end{equation}
where for simplification we set $g_{12}=0$ and $\Delta_{12}=0$ ($\Delta_{12}=\omega_{q1}-\omega_{q2}$). As for the general cases with finite coupling $g_{12}$ and fine detunings $\Delta_{12}$, we also derive the corresponding analytical expression for $\zeta_{zz}$ [see Eq.~\eqref{analytic:general regime}].
Besides, $\tilde{g}_{12}$ is the effective coupling between computational qubits given in Eq.~\eqref{coupling: q1 q2}, and the three terms correspond to the contribution from the coupling between the states $\vert 101 \rangle$  and $\vert 200 \rangle$,  $\vert 002 \rangle$,  $\vert 020 \rangle$, respectively. The derivation of Eq.~\eqref{analytic:dispersive}  is presented in Appendix \ref{ZZ coupling: Analytical expression}. 

\begin{figure}[htbp]
\centering
\includegraphics[width=\columnwidth]{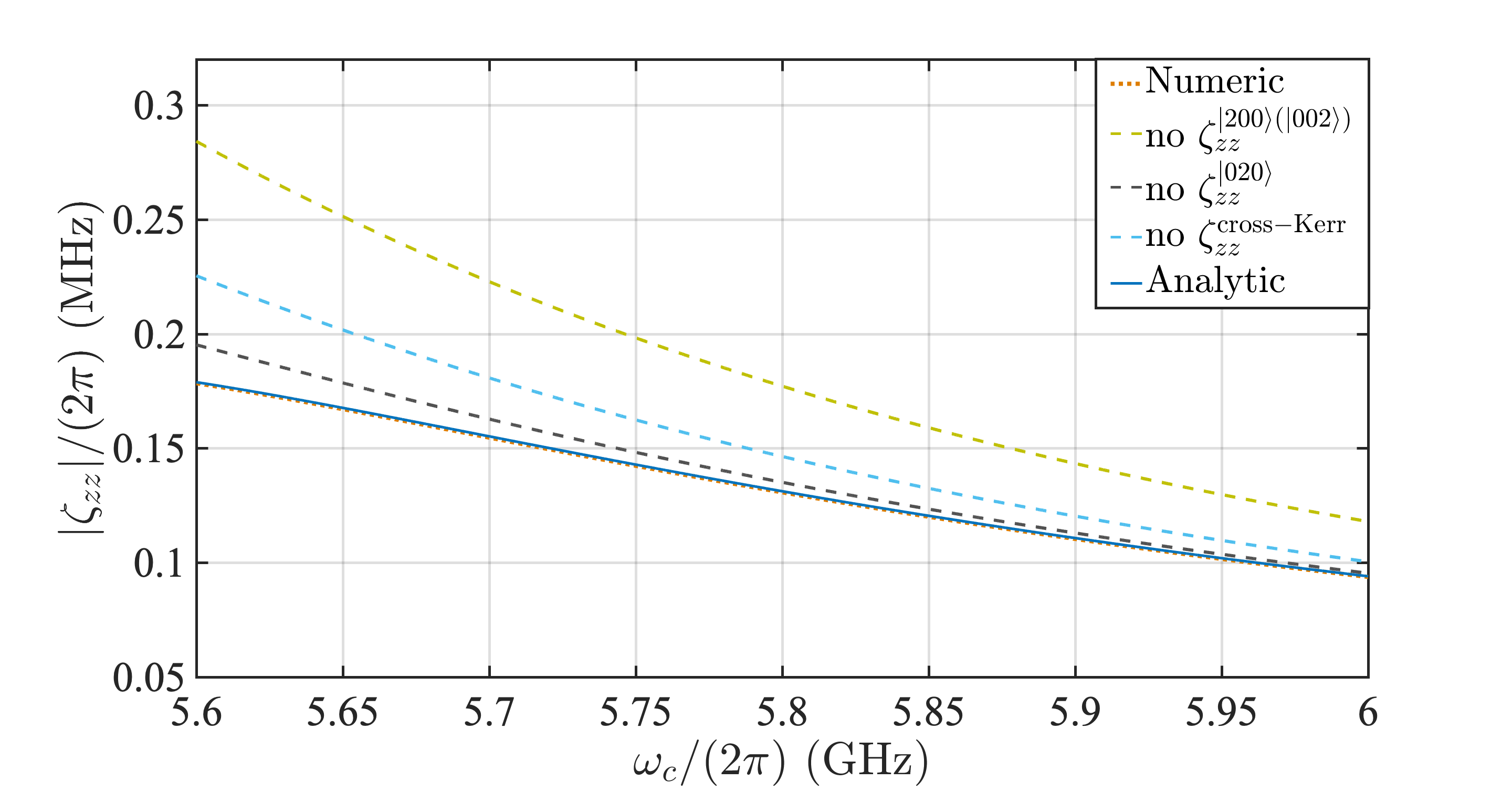}
\caption{ZZ coupling strength $\zeta_{zz}$ characteristics in  dispersive coupling regime. The analytical result (blue solid) is plotted using Eq.~\eqref{analytic:dispersive}.  The numerical result (orange dotted), obtained via diagonalizing the system Hamiltonian $\hat{H}_{\rm Lab}$, match very well with the analytical result. To verify the importance for each coupling term, we plot the results with different Hamiltonian in which the couplings with the states $\vert 200(002) \rangle$, $\vert 020 \rangle$, as well as the cross-Kerr term  are not involved, respectively.  The used parameters are: $\omega_{q1}/(2 \pi)=\omega_{q2}/(2 \pi)=5~{\rm GHz}$,   $\alpha_{q1}/(2 \pi)=\alpha_{q2}/(2 \pi)=-0.2~{\rm GHz}$, $\alpha_{c}/(2 \pi)=-0.4~{\rm GHz}$, $g1/(2 \pi)=g2/(2 \pi)=0.05~{\rm GHz}$, $g_{12}=0$.   }
\label{fig:dispersive regime}
\end{figure}

In Fig.~\ref{fig:dispersive regime}, the ZZ parasitic coupling dependent of coupler frequency is evaluated and plotted. First and foremost, our analytical result (blue solid), namely Eq.~\eqref{analytic:dispersive}, is verified via numerically diagonalizing the Hamiltonian $\hat{H}_{\rm Lab}$ (orange dotted). It is shown clearly that the analytical result matches very well with the numerical one. As a further step, it is also interesting to study the contribution for each of the coupling processes,  named as $\zeta_{zz}^{\vert 200 \rangle}$, $\zeta_{zz}^{\vert 002 \rangle}$,  $\zeta_{zz}^{\vert 020 \rangle}$, and $\zeta_{zz}^{\rm cross-Kerr}$. To do so, we evaluate and plot the resulting $\vert \zeta_{zz} \vert $ in absence of one of them.  As seen from the plots, the result cannot match well with the exact one without involving any one of them. This implies that all the coupling terms matter to the resulting ZZ parasitic coupling. In our specific case with chosen parameters specified in Fig.~\ref{fig:dispersive regime}, the contribution of $\zeta_{zz}^{\vert 200 \rangle (\vert 002 \rangle)}$ is the largest one while that of $\zeta_{zz}^{\vert 020 \rangle}$ is relatively smaller. 
Another characteristic is that the strength for $\zeta_{zz}$ is suppressed apparently in dispersive regimes comparing with the resonant regimes discussed before.  In the next section, we will mainly focus on the dispersive regime and further explore the elimination of ZZ parasitic couplings. 

\section{Novel parameter regions for eliminating ZZ parasitic couplings}
\label{sec:New parameters regions}


In this section, we further explore ZZ parasitic coupling characteristics.  Especially, we pay attention to those parameter regions in which ZZ parasitic couplings can be suppressed or even eliminated. 
 We first examine ZZ  coupling characteristics with current existing experimental parameters in subsection \ref{ZZ coupling strength}, and further figure out the optimized parameter regions for minimizing ZZ couplings. 
In addition to the general numerical results, we provide a clear physical understanding benefiting from the analytical result obtained. More importantly, inspired by the analytical results, we also propose four novel parameter regions and the related experimental realizations in which ZZ couplings are expected to be eliminated; this will be discussed in subsection \ref{New parameter regions for eliminating ZZ couplings}. 

\subsection{ZZ coupling characteristics in existing experimental parameter regions}
\label{ZZ coupling strength}

\begin{figure}[htbp]
\centering
\begin{minipage}[t]{0.5\textwidth}
\centering
\includegraphics[width=\columnwidth]{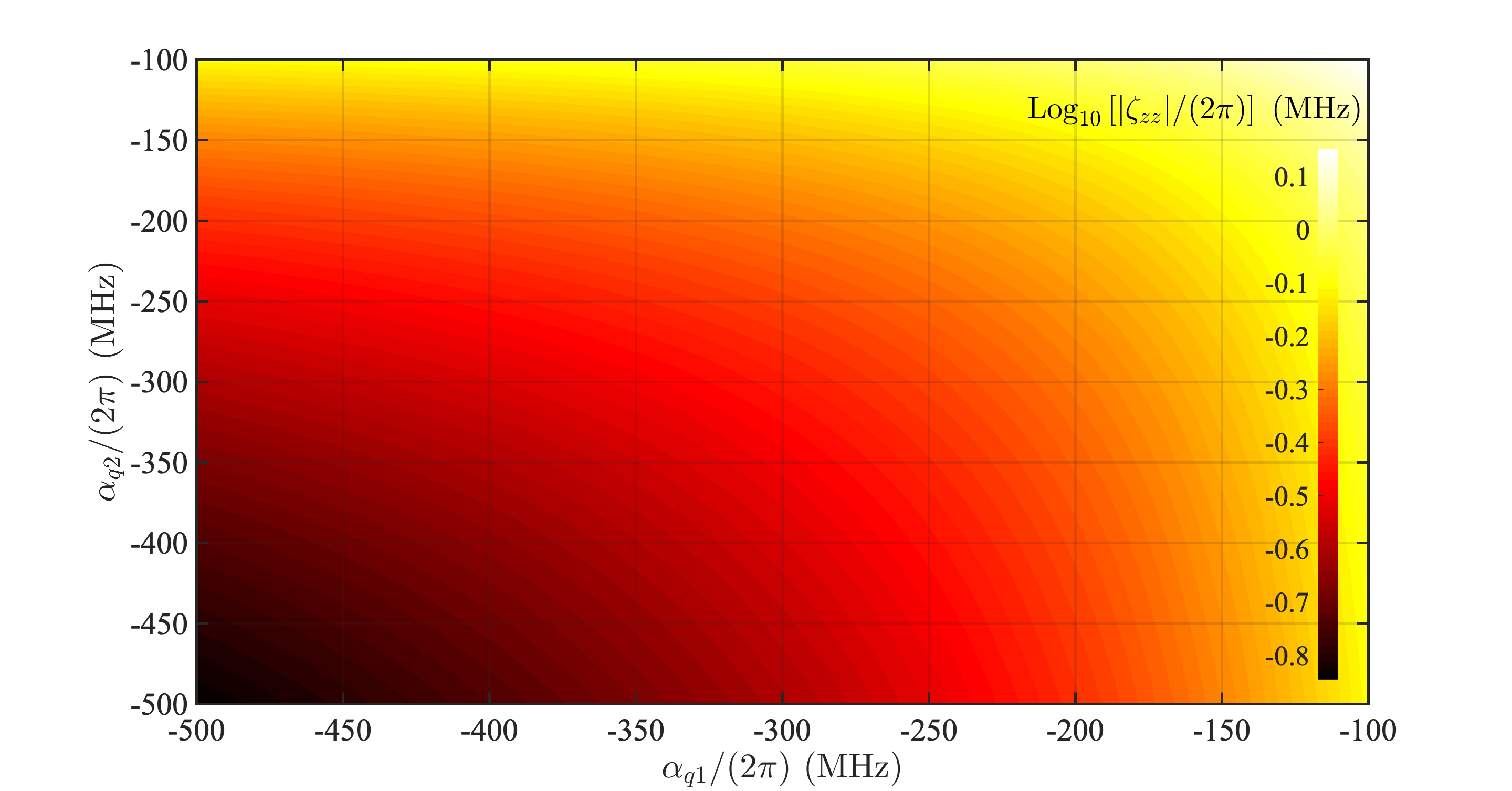}
\end{minipage}
\begin{minipage}[t]{0.5\textwidth}
\centering
\includegraphics[width=\columnwidth]{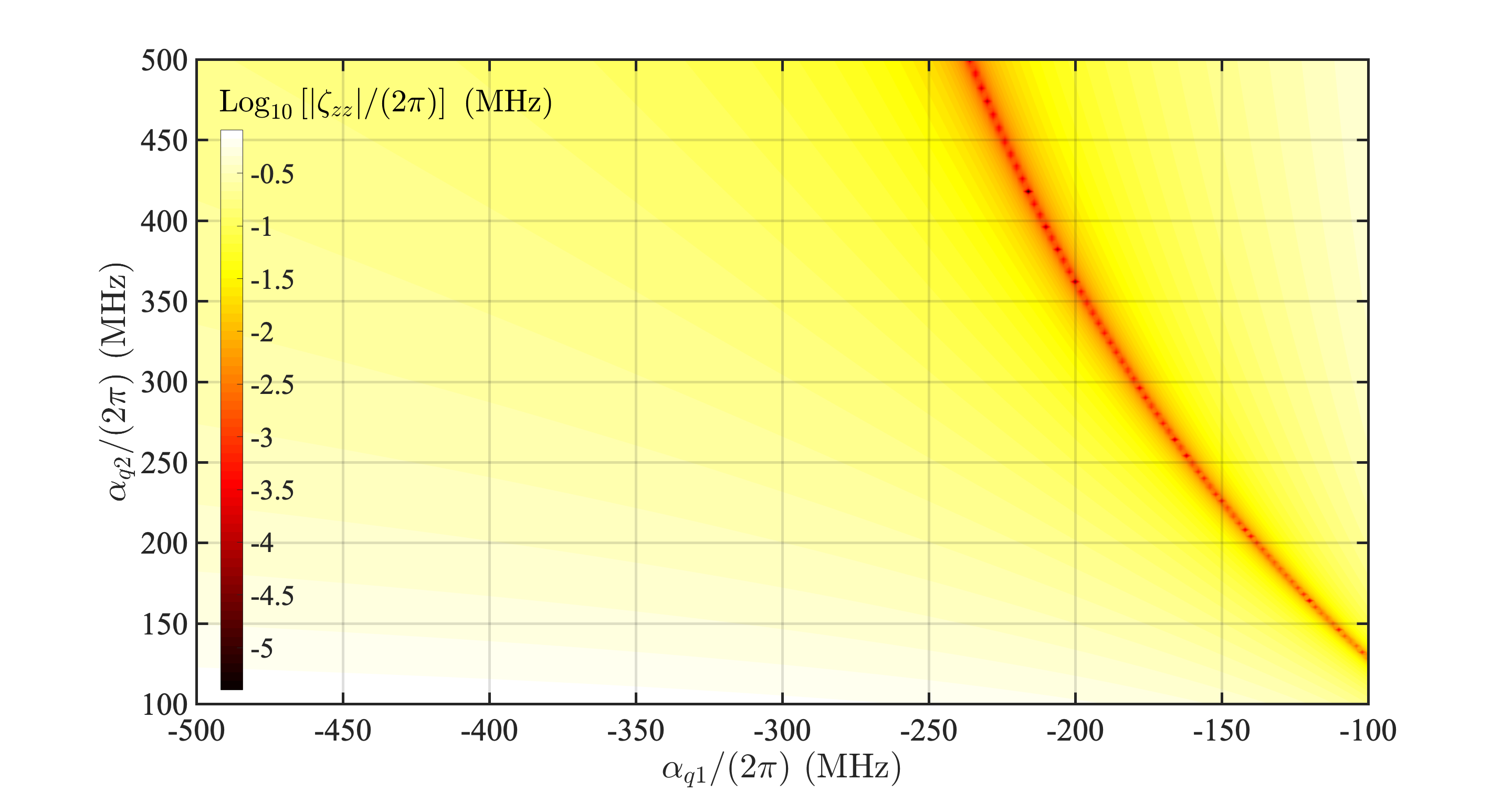}
\end{minipage}
\caption{ZZ coupling $\zeta_{zz}$ is computed numerically via diagonalizing the system Hamiltonian $\hat{H}_{\rm Lab}$ in lab frame. (upper) For qubit anharmonicities with the same sign, e.g., $\alpha_{q1,q2}<0$, $\zeta_{zz}$ decreases with increasing qubit anharmonicities $\vert \alpha_{q1} \vert$, $\vert \alpha_{q2} \vert$; (lower) For qubit anharmonicities with different signs, $\zeta_{zz}$ could be eliminated at certain region (red band). The used parameters are:  $\omega_{q1}/(2 \pi)=\omega_{q2}/(2 \pi)=5~{\rm GHz}$,  $\omega_{c}/(2 \pi)=6~{\rm GHz}$,  $\alpha_{c}/(2 \pi)=-0.25~{\rm GHz}$,  $g1/(2 \pi)=g2/(2 \pi)=0.08~{\rm GHz}$, $g_{12}=0$. } 
 \label{fig-Jzz_varyingAnharmonicty}
\end{figure}

Concentrating on dispersive regime and 
using current experimental parameters, e.g.,  $\omega_{q1}/(2 \pi)=\omega_{q2}/(2 \pi)=5~{\rm GHz}$,  $\omega_{c}/(2 \pi)=6~{\rm GHz}$,  $\alpha_{c}/(2 \pi)=-0.25~{\rm GHz}$,  $g1/(2 \pi)=g2/(2 \pi)=0.08~{\rm GHz}$, $g_{12}=0$, we evaluate and plot the ZZ coupling strength in Fig.~\ref{fig-Jzz_varyingAnharmonicty}. In particular, we study  ZZ coupling characteristics with varying qubits anharmonicities $\alpha_{q1}$ and $\alpha_{q2}$. 

Let us first consider the general case that the computational qubits' anharmonicities has the same sign, for instance both of qubits are traditional transmon \cite{koch2007charge} with $\alpha_{q1},\alpha_{q2}<0$,  ZZ couplings $\zeta_{zz}$ are estimated with the order of $\sim  1~ {\rm MHz}$, which will lead to gate error definitely. 
As seen from the upper one of Fig.~\ref{fig-Jzz_varyingAnharmonicty}, 
 we further find that
ZZ couplings are suppressed through increasing the strength of qubit anharmonicities, i.e., $\vert \alpha_{q1} \vert$ and $\vert \alpha_{q2} \vert$.  
The larger anharmonicities chosen, the weaker ZZ parasitic coupling we will obtain. 
Unfortunately, larger anharmonicities are restricted by current technology and qubit anharmonicities are usually small (around $-100 ~ {\rm to}  -300 ~{\rm MHz}$) for the frequently used transmon qubits.

Interestingly, $\vert \zeta_{zz} \vert$ can be suppressed when the two qubits’ anharmonicities have different signs; for instance, one is transmon qubit \cite{koch2007charge} with $\alpha_{q1}<0$ and the other one is capacitively shunted flux qubit (CSFQ) \cite{steffen2010high,chow2011simple,yan2016flux} with $\alpha_{q2}>0$.  In this regime,  
it is seen from the lower figure that  ZZ coupling characteristics are quite different from the traditional one (the upper figure).
In particular,  we find that the resulting $\vert \zeta_{zz} \vert$ are suppressed evidently for arbitrary negative  $\alpha_{q1}$ and positive $\alpha_{q2}$; more interestingly, ZZ couplings could be eliminated at specific regions. 
 Actually, such parameters regime was studied and discussed in Ref.\cite{zhao2020high, zhao2020switchable} and the ZZ coupling suppressing effects were also verified in a very recent experiment \cite{ku2020suppression}. 
 Although high-fidelity two-qubit gates were realized in such parameter regime, a couple of essential questions were not explored. For instance,  are we able to figure out the specific parameter region (i.e., red band in the lower figure) for $\zeta_{zz} \rightarrow 0$? Furthermore, what is the physical mechanism for ZZ coupling elimination? To solve these critical issues, we drive the explicit expression for $\zeta_{zz}$ with the effective Hamiltonian approach introduced before.

Considering the regime of our interest, i.e., $\Delta_1 =\Delta_2=\Delta$ (i.e., $\Delta_{12}=0$),  we obtain the explicit form of  $\zeta_{zz}$ (see detailed derivation in Appendix \ref{ZZ coupling: Analytical expression}):
\begin{eqnarray} \label{zz coupling: analytic}
\zeta_{zz} &=& - 2 g_{\rm eff} \left[ g_{\rm eff}   \left(\frac{1}{\alpha_{q1}} + \frac{1}{\alpha_{q2}} \right) \right. \nonumber\\
&& +\left. \frac{4 \tilde{g}_{12}}{\alpha_c -2 \Delta} \right] + 4 \frac{g_{12}  \tilde{g}_{12} \alpha_c }{\Delta (\alpha_c -2 \Delta)}. 
\end{eqnarray}
As  for the more general case,  i.e., $\Delta_1 \neq \Delta_2$, the result  is given  in Eq.~\eqref{analytic:general regime} of Appendix \ref{ZZ coupling: Analytical expression}, which can be used to explore the  regime with finite detuning between qubits' frequencies. When the effective coupling between qubits $q1$ and $q2$ is tuned off (i.e., $g_{\rm eff}=0$),
 the above equation \eqref{zz coupling: analytic} reduces to a simple form, i.e., $4 {g_{12}  \tilde{g}_{12} \alpha_c }/{[\Delta (\alpha_c -2 \Delta)]}$. This is normally a small value which implies that the parasitic ZZ coupling is extremely weak. This conclusion is indeed verified by 
 the very recent experiment \cite{li2020tunable}. 
 Beyond the commonly accepted view that  ZZ parasitic coupling is suppressed by tuning off $g_{\rm eff}$, we find an alternative means to mitigate ZZ couplings: adjusting system parameters to make $ [g_{\rm eff}   \left({1}/{\alpha_{q1}} + {1}/{\alpha_{q2}} \right)  + {4 \tilde{g}_{12}}/{(\alpha_c -2 \Delta)}] \rightarrow 0$. One advantage of this new means is that XY coupling $g_{\rm eff}$ can be maintained while mitigating ZZ crosstalk. The physical mechanism behind is that the coupler's high-energy state plays a vital role, it can neutralize the energy shift induced by high-energy states of computational qubits.
 Actually, our findings can be used to explain the key results of Ref.~\cite{zhao2020suppression}: two separated branches for mitigating ZZ coupling are obtained.  More importantly, we can even figure out the explicit condition for the two branch using Eq.~\eqref{zz coupling: analytic}. Choosing $\alpha_c=0$ and $\alpha_{q1}=\alpha_{q2}=\alpha_{q}$, the upper branch corresponds to $g_{12}=-g_1 g_2/\Delta $ [i.e., Eq.~\eqref{omega C:switch condition}], and the lower branch can also be solved easily as $g_{12}=-(g_1 g_2/\Delta) (1- \alpha_{q}/\Delta)$. It is obvious the lower branch will  get close to the upper branch with larger detuning $\vert \Delta \vert$ (which is normally the  case for current experiments); however, they split into two branches once the ratio $\alpha_{q}/\Delta$ play a role. 
 
In the remainder of this paper, we concentrate on the regime $ g_{12}  \ll \vert \tilde{g}_{12} \vert $ (the regime beyond this will be studied in future work), then Eq.~\eqref{zz coupling: analytic} reduces approximately to a simple form:
$\vert \zeta_{zz} \vert \approx 2 \tilde{g}_{12}^2 \vert {1}/{\alpha_{q1}} + {1}/{\alpha_{q2}} + {4}/{(\alpha_{c}-2\Delta)} \vert$ [namely Eq.~\eqref{analytic:dispersive}]. Next, we use 
this analytical result to explain the above numerical results shown in Fig.~\ref{fig-Jzz_varyingAnharmonicty}.  
 If all of the three elements are transmon qubits, moreover $\vert \Delta \vert  \gg \alpha_{\lambda}$, $\lambda=q1,q2,c$, $\vert \zeta_{zz} \vert$ is estimated as $\tilde{g}_{12}^2/ \vert \alpha_{\lambda} \vert $. It is obvious that ZZ coupling strength is proportional to the effective XY coupling $\tilde{g}_{12}$, implying stronger XY coupling has to pay the price of larger ZZ parasitic couplings.
Moreover,  ZZ coupling strength is inversely proportional to qubit anharmonicities. As the anharmonicities for transmon qubits are usually small, ZZ coupling becomes one of the leading gate error sources for coupler architecture with traditional parameter region.
Then, we turn to the case that qubit anharmonicities have different signs.  Apparently, $\zeta_{zz}$ becomes weaker compared with the general case with using transmon qubits. To suppress largely ZZ coupling,  the choice of $\alpha_{q1}$ and $\alpha_{q2}$ should satisfy some specific condition. 
If one simplify choose $\alpha_{q1}=-\alpha_{q2}$, the ZZ coupling can not be eliminated completely. To further eliminate ZZ coupling, we need to let ${1}/{\alpha_{q1}} + {1}/{\alpha_{q2}} + {4}/{(\alpha_{c}-2\Delta)}=0$, from which we solve the explicit analytical condition for zero ZZ coupling, namely
\begin{equation} \label{zero ZZ coupling: alpha_q1}
\alpha_{q1} =  \left[ \frac{4}{2\Delta - \alpha_c} - \frac{1}{\alpha_{q2}}   \right]^{-1},
\end{equation}
which corresponds exactly to the red band in the lower one of Fig.~\ref{fig-Jzz_varyingAnharmonicty}. This tells us that one has to design proper superconducting circuit parameters to realize lower ZZ coupling. In this paper, we provide an explicit condition which could be applied to real experiments.



\begin{figure}[htbp]
\centering
\begin{minipage}[t]{0.5\textwidth}
\centering
\includegraphics[width= \columnwidth]{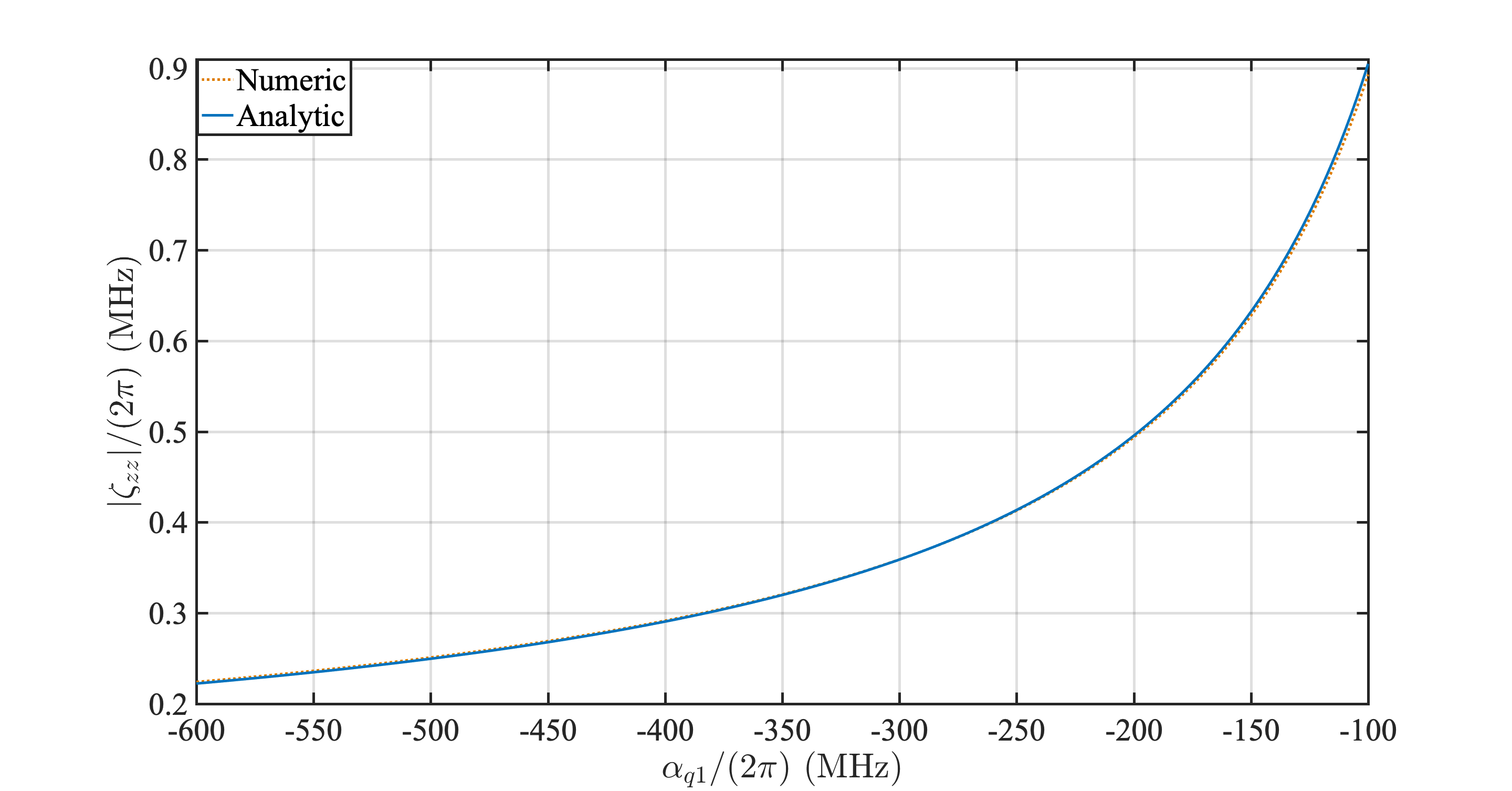}
\end{minipage}
\begin{minipage}[t]{0.5 \textwidth}
\centering
\includegraphics[width= \columnwidth]{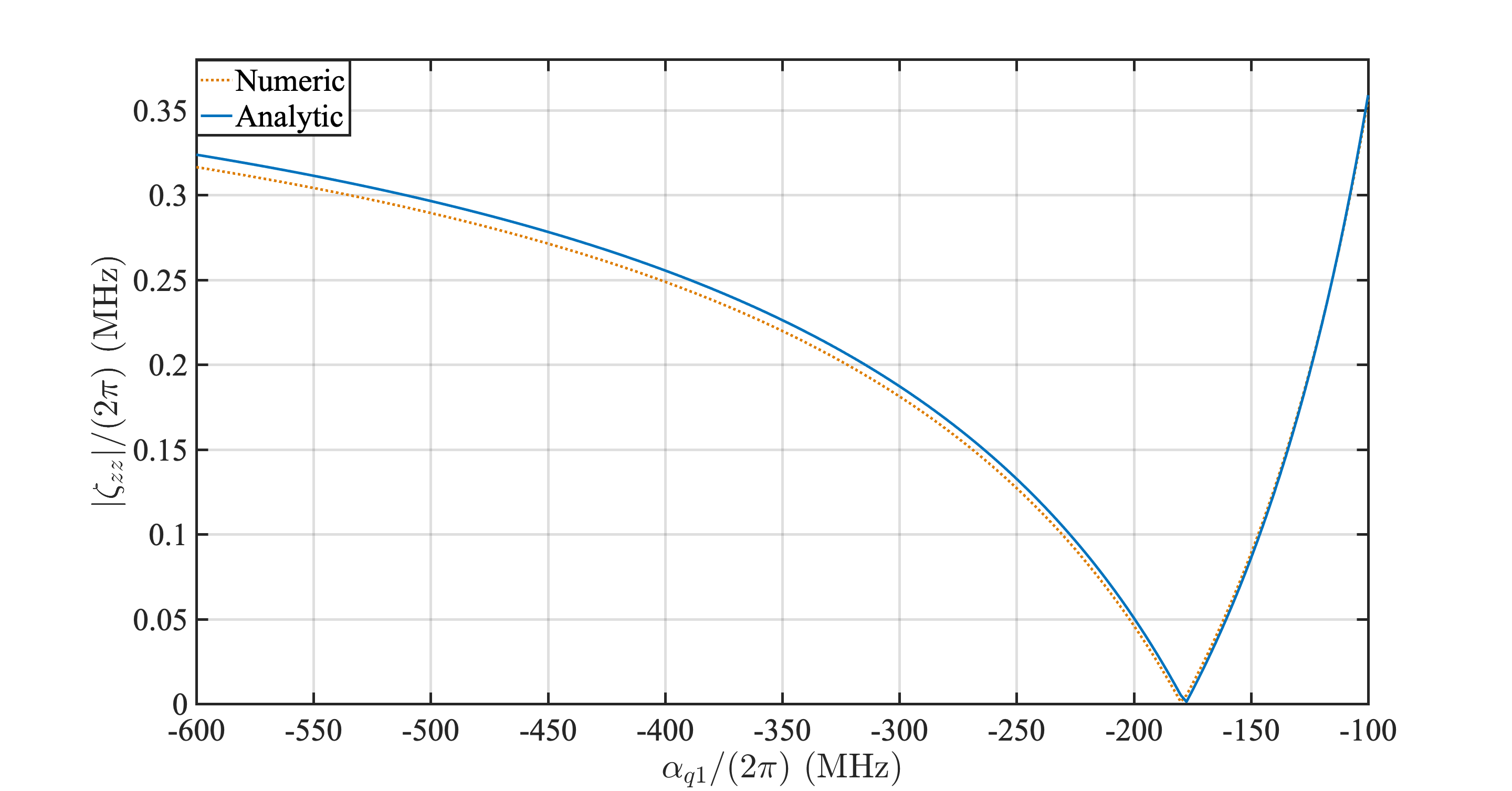}
\end{minipage}
\caption{ZZ coupling strength $\vert \zeta_{zz} \vert$ dependent of qubit anharmonicity $\alpha_{q1}$ with fixed negative anharmonicity (upper) and positive anharmonicity  (lower), respectively. The numerical results (orange dotted) are computed via numerically diagonalizing  the Hamiltonian in lab frame $\hat{H}_{\rm Lab}$ [Eq.~\eqref{Hamiltonian:lab frame}], while the analytical results (blue solid) are plotted using Eq.~\eqref{analytic:dispersive}.  The used parameters are identical to Figure \ref{fig-Jzz_varyingAnharmonicty}, and the anharmonicity for the upper and lower figure are $\alpha_{q2}/(2\pi)=-0.3~{\rm GHz}$ and $\alpha_{q2}/(2\pi)=0.3~{\rm GHz}$, respectively.  
}  \label{fig: Jzz_q1Anhar_Nag q2}
\end{figure}

With choosing fixed $\alpha_{q2}$ (either negative or positive),  we plot $\vert \zeta_{zz} \vert$ dependent of  $\alpha_{q1}$  in Fig.~\ref{fig: Jzz_q1Anhar_Nag q2}. 
 Here, $\zeta_{zz}$ is computed numerically and analytically, respectively.  In particular, the numerical results (orange dotted) are computed via numerically diagonalizing the Hamiltonian $\hat{H}_{\rm Lab}$ [Eq.~\eqref{Hamiltonian:lab frame}], while the analytical results (blue solid) are plotted using Eq.~\eqref{analytic:dispersive}.
As expected, the ZZ coupling strength decreases with larger $\vert \alpha_{q1} \vert$  for negative $\alpha_{q2}$ (the upper one). As for positive $\alpha_{q2}$ (the lower one), the ZZ coupling strength can be eliminated at certain $\alpha_{q1}$. Using Eq.~\eqref{zero ZZ coupling: alpha_q1} it is estimated as $\alpha_{q1}/(2\pi) \approx - 178 ~{\rm MHz}$ with the parameters used.
Moreover, we verify that our derived analytical expression, i.e., Eq.~\eqref{analytic:dispersive}, matches well with the numerical results. 

\subsection{Novel parameter regions for eliminating ZZ couplings}
\label{New parameter regions for eliminating ZZ couplings}

Although the regime for qubit anharmonicity with different signs could reduce ZZ parasitic couplings, 
in practice the qubits with positive anharmonicity are usually unstable and own relatively short coherence time.
As a consequence, the resulting gate error would suffer from the decoherence of computational qubits. 
Inspired by the analytical expression [i.e., Eq.~\eqref{analytic:dispersive}] obtained, we may ask a question: are there other parameter regions existing for mitigating ZZ coupling?
Actually, beyond the regimes discussed above, we find four novel parameter regions (unexplored yet), in which ZZ coupling elimination may be expected. The four types of parameter regions are introduced as follows. Some typical system parameters for different type are listed in Table~\ref{table}.

\begin{table*}[htbp]
\caption{\label{table}Some typical parameters for the four novel parameter regions I, II, III, IV, and the resulting ZZ coupling characteristics as well as the corresponding experimental realization using superconducting circuits (SC). All the units of parameters are $\rm GHz$ and comparable with realistic experimental parameters.}
\begin{ruledtabular}
\begin{tabular}{ccccccccc}
& $\omega_{q1}/{2\pi}$ & $\omega_{q2}/2\pi$ & $\omega_{c}/2\pi$ & $\alpha_{c}/2\pi$ & $g_1/2\pi$ & $g_2/2\pi$ & $\zeta_{zz}$ characteristics & SC realization \\
\midrule  
I & 5 & 5& 5.4& -0.3& 0.04& 0.04 & Fig.~\ref{fig-Jzz} (a) & Fig.~\ref{fig:circuitRealization} (a) \\
II & 5 & 5& 5.6& -0.8& 0.06& 0.06  & Fig.~\ref{fig-Jzz} (b) & Fig.~\ref{fig:circuitRealization} (a) \\
III & 5.8 & 5.8& 5.4& 1.2& 0.04& 0.04  & Fig.~\ref{fig-Jzz} (c) & Fig.~\ref{fig:circuitRealization} (c) \\
IV & 5.8 & 5.8& 5& 0.6& 0.06& 0.06  & Fig.~\ref{fig-Jzz} (d) & Fig.~\ref{fig:circuitRealization} (d) \\
\end{tabular}
\end{ruledtabular}
\end{table*}

\emph{Type I}: the first parameter regions we suggest are, both the computational qubits and coupler have negative anharmonicity and could be typical transmon qubits, but the frequency detunings between qubits and coupler  as well as the coupling strengths $g_{1}$, $g_{2}$ are relatively small comparing with the general case. It is noted that the dispersive couplings $g_{k}/\vert \Delta_k \vert \ll 1$, $k=1,2$ always hold. 

\emph{Type II}: the second parameter regions we suggest are, both computational qubits are transmon with negative anharmonicity; moreover the frequency detunings between qubits and coupler are similar to the general case, but the coupler is realized with a  strong negative anharmonicity \cite{mundada2019suppression}.

\emph{Type III}: the third parameter regions we suggested are, both computational qubits are transmon with negative anharmonicity, while the coupler is chosen with a positive anharmonicity (namely $\alpha_{c} >0$), which could be realized with CSFQ.

\emph{Type IV}: the fourth parameter regions we suggested are, both computational qubits and coupler have positive anharmonicities, i.e., $\alpha_{q1}, \alpha_{q2}, \alpha_{c}>0$, all of them may be realized with CSFQ.

\begin{figure*}[htbp]
\centering
\begin{minipage}[t]{0.49\textwidth}
\centering
\includegraphics[width= \columnwidth]{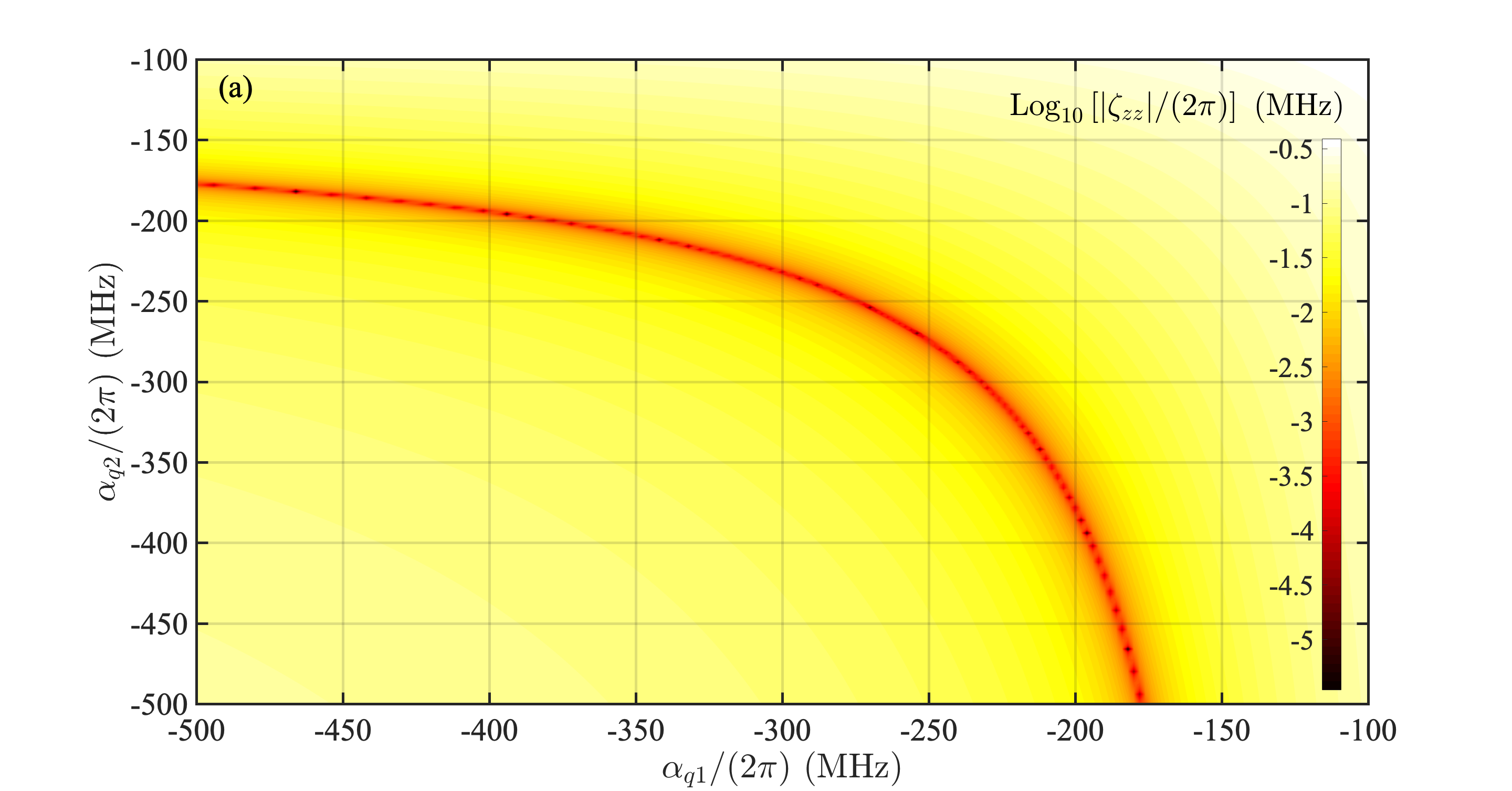}
\end{minipage}
\begin{minipage}[t]{0.49\textwidth}
\centering
\includegraphics[width= \columnwidth]{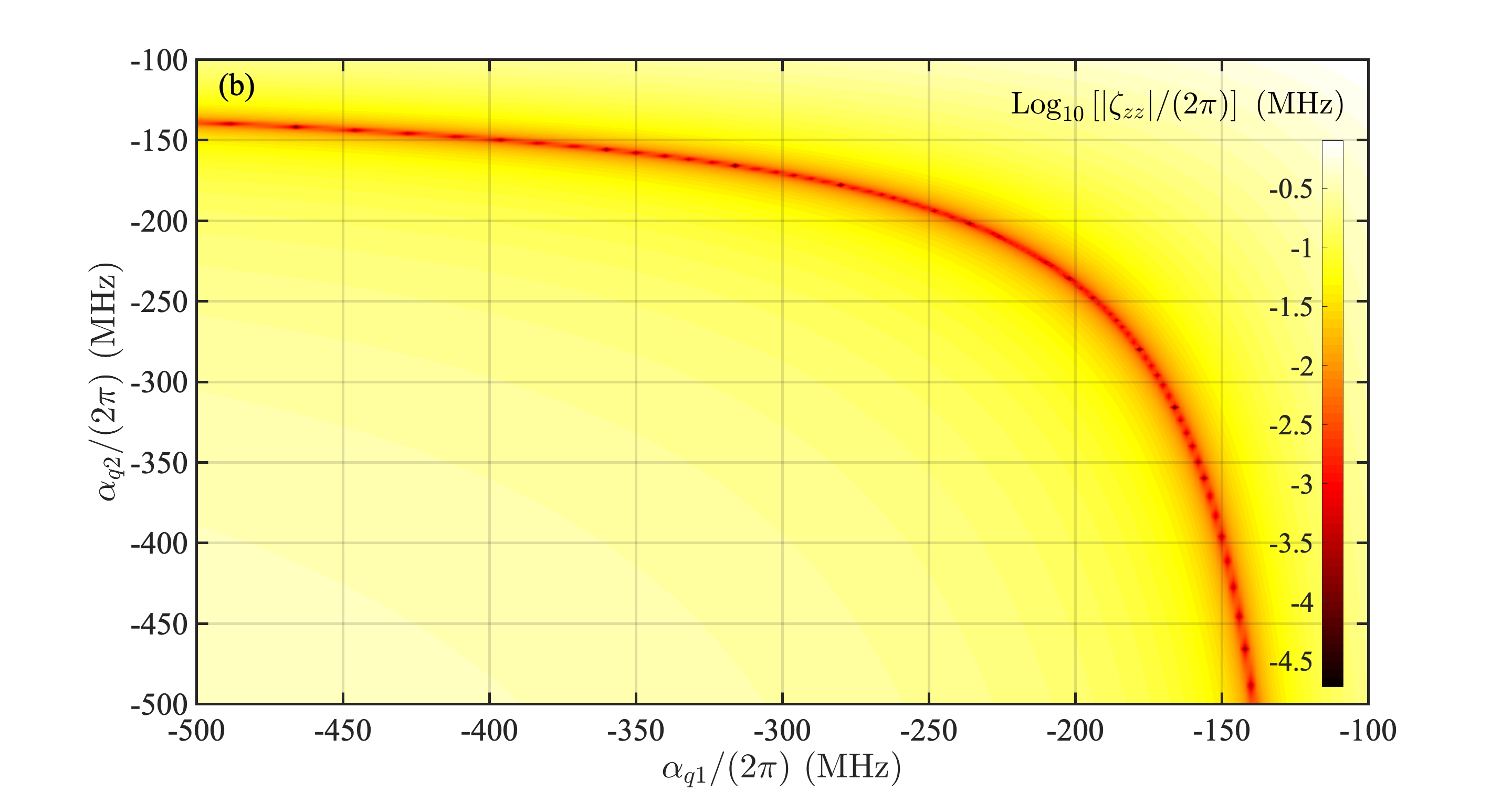}
\end{minipage}
\begin{minipage}[t]{0.49\textwidth}
\centering
\includegraphics[width= \columnwidth]{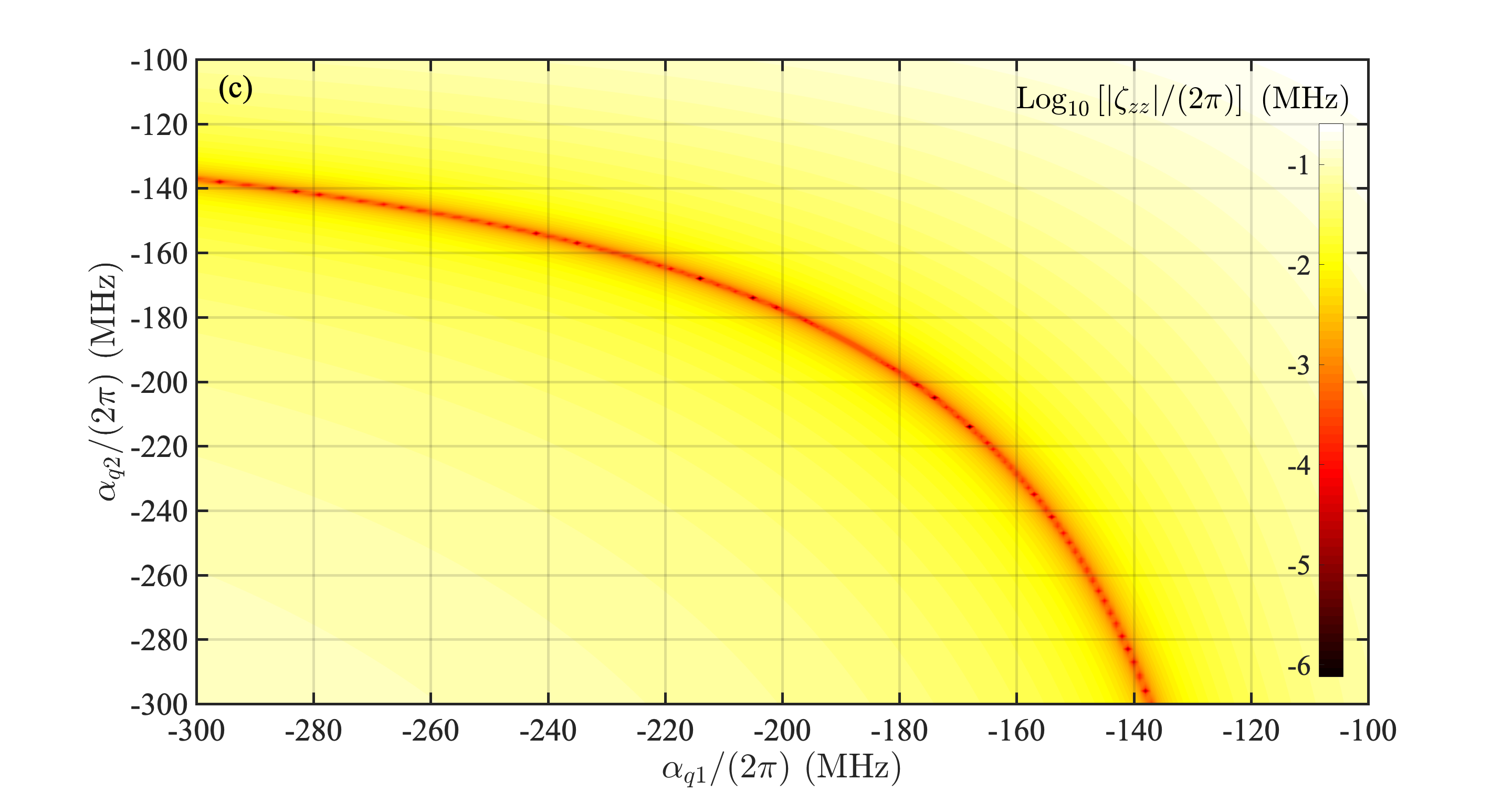}
\end{minipage}
\begin{minipage}[t]{0.49\textwidth}
\centering
\includegraphics[width= \columnwidth]{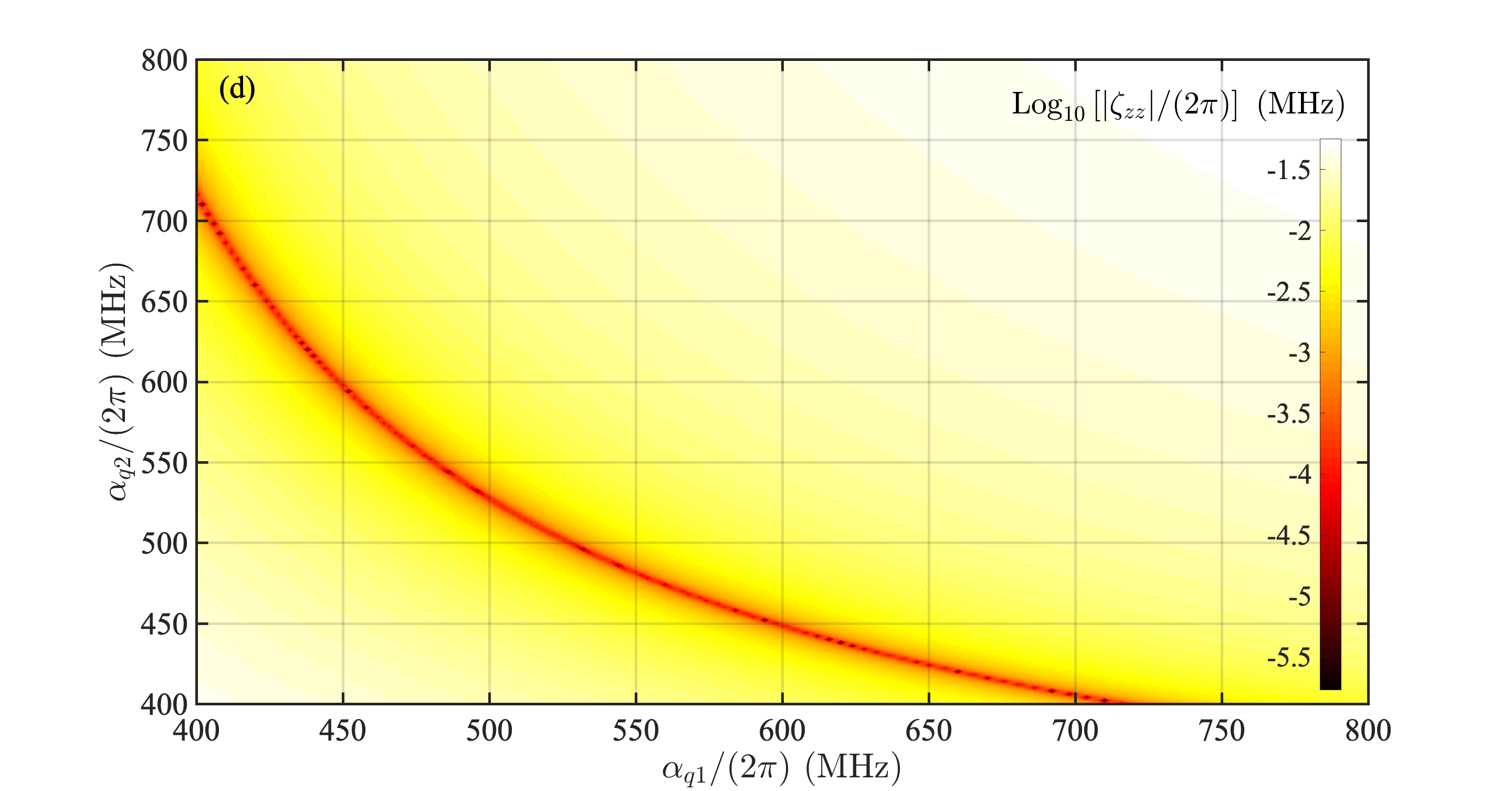}
\end{minipage}
\caption{ZZ coupling $\zeta_{zz}$ characteristics with our suggested parameter regions I, II, III, IV. $\zeta_{zz}$ are computed numerically via diagonalizing the system Hamiltonian $\hat{H}_{\rm Lab}$. The superconducting circuit parameters used for the four different parameter regions are listed in Table~\ref{table}, respectively. Our results indicate that ZZ couplings can be eliminated at certain region (red bands).  } 
 \label{fig-Jzz}
\end{figure*}

Using the superconducting circuit parameters given in Table \ref{table}, we evaluate and plot  ZZ coupling strengths $\vert \zeta_{zz} \vert$ dependent of qubit anharmonicities $\alpha_{q1}$ and $\alpha_{q2}$ with different type of parameter regions in Fig.~\ref{fig-Jzz} (a)-(d).  $\zeta_{zz}$ are computed through numerically diagonalizing the lab frame Hamiltonian $\hat{H}_{\rm Lab}$. Comparing with the result with traditional parameter regions (i.e., the top one of Fig.~\ref{fig-Jzz_varyingAnharmonicty}), ZZ  couplings are largely suppressed with the novel system parameters suggested. 
As seen from Fig.~\ref{fig-Jzz} (a)-(d), ZZ coupling characteristics for different parameter regions are slightly different. However, they share a similar property: to achieve lower ZZ couplings, qubit anharmonicities have to be chosen properly to satisfy specific conditions (the red band), which can be figured out using the analytical expression, i.e., Eq.~\eqref{zero ZZ coupling: alpha_q1}. Once the superconducting circuit parameters are tuned away from the red band, the resulting gate fidelity will be affected unavoidably by ZZ parasitic couplings. 

\begin{figure}[htbp]
\centering
\includegraphics[width=0.9   \columnwidth]{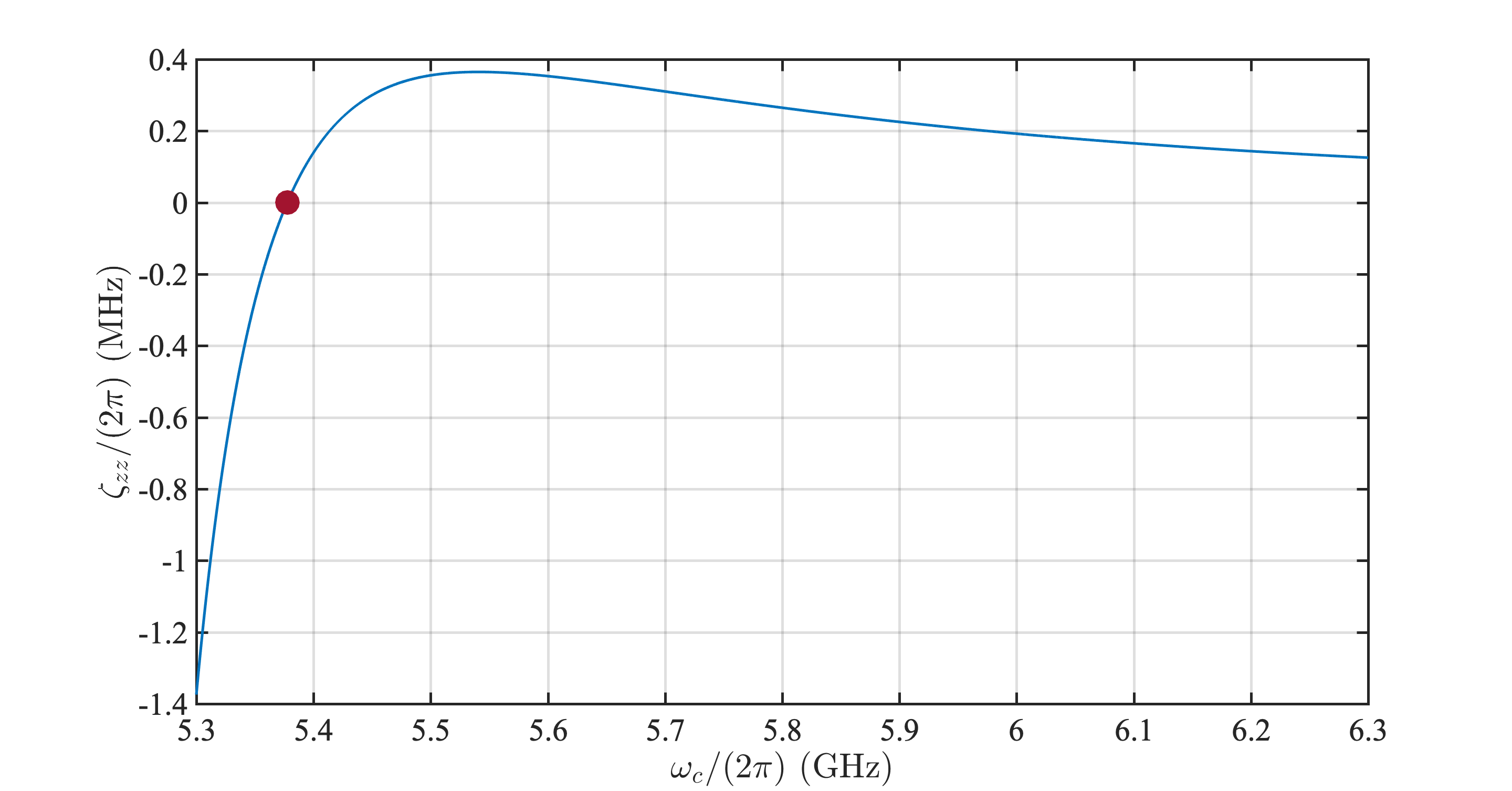}
\caption{ZZ coupling  $\zeta_{zz}$ dependent of coupler frequency $\omega_c$. With varying $\omega_c$,  $\zeta_{zz}$ changes continually from negative to positive, which implies that zero $\zeta_{zz}$ (red dot)  can be  achieved via adjusting the coupler frequency. The parameter used are: $\omega_{q1}/(2 \pi)=\omega_{q2}/(2 \pi)=5~{\rm GHz}$,  $\alpha_{q1}/(2 \pi)=-0.2~{\rm GHz}$,   $\alpha_{q2}/(2 \pi)=\alpha_{c}/(2 \pi)=-0.3~{\rm GHz}$,  $g1/(2 \pi)=g2/(2 \pi)=0.04~{\rm GHz}$. 
}  \label{fig: zz_couplerFre}
\end{figure}

According to ZZ coupling characteristics in our suggested parameter regions, it seems that the superconducting circuit parameters have to be designed carefully for suppressing ZZ couplings. Moreover, the parameter regions for $\zeta_{zz} \rightarrow 0$ are relatively narrow, which implies that it might not be easy to reach these specific parameter regions. Fortunately, we can still tune off ZZ coupling even if the parameters are not optimized perfectly. This is true because ZZ coupling strength $\zeta_{zz}$ can also be controlled by tuning the coupler frequency $\omega_c$ \cite{mundada2019suppression}. In Fig.~\ref{fig: zz_couplerFre}, using the same parameters as in parameter region I, and choosing qubit anharmonicities randomly with $\alpha_{q1}/(2\pi)=-0.2~{\rm GHz}$ and $\alpha_{q2}/(2\pi)=-0.3~{\rm GHz}$, we get a finite $\zeta_{zz}$. However, through further tuning the coupler frequency $\omega_c$, ZZ coupling $\zeta_{zz}$ can be tuned continuously from negative to positive. This means that one can always eliminate ZZ coupling by further adjusting coupler frequency.

To further understand the resulting ZZ coupling characteristics and the mechanisms of ZZ coupling elimination, we use the language of energy level diagrams to explain.  As discussed before, the origin of ZZ  coupling $\zeta_{zz}$ comes from the parasitic coupling between the state $\vert 101 \rangle$ (in computational space) and other states $\vert 200 \rangle$, $\vert 020 \rangle$, $\vert 002 \rangle$ (out of computational space). In particular, with considering adiabatically external drive, these couplings will result in a shift of energy level $\vert 101 \rangle$. Depending on the specific energy level interacted with $\vert 101 \rangle$, the energy shift to  $\vert 101 \rangle$ could be either positive or negative. Under dispersive regime, ZZ parasitic coupling contains the contribution from different energy levels, incl.,  $\vert 200 \rangle$, $\vert 020 \rangle$, $\vert 002 \rangle$. Once the positive energy shift equals exactly with the negative energy shift, the consequence of the overall effect will keep the energy level of $\vert 101 \rangle$ remain unchanged. This is indeed the physical mechanism of eliminating ZZ coupling.
Our four suggested parameter regions I-IV and the case with considering two computational qubits' anharmonicity with different signs share a similar mechanism. According to the derivations and analysis, we realize that the effect of $\vert 020 \rangle$ is vital for eliminating ZZ couplings, which was usually ignored in previous work. 
For general parameter regions, namely the traditional transmon qubits are used for both computational qubits and coupler, the usual large energy difference between $\vert 101 \rangle$ and $\vert 020 \rangle$ results in a very small energy shift, therefore cannot neutralize the energy level shift induced by $\vert 200 \rangle$ and $\vert 002 \rangle$. By contrast, the situation is quite different in the four parameter regions we proposed in this paper. Through choosing proper system parameters, the energy shifts induced by $\vert 200 \rangle$ and $\vert 002 \rangle$ are always able to be neutralized by that induced by $\vert 020 \rangle$.
 
\begin{figure}[h]
\centering
\includegraphics[width= \columnwidth]{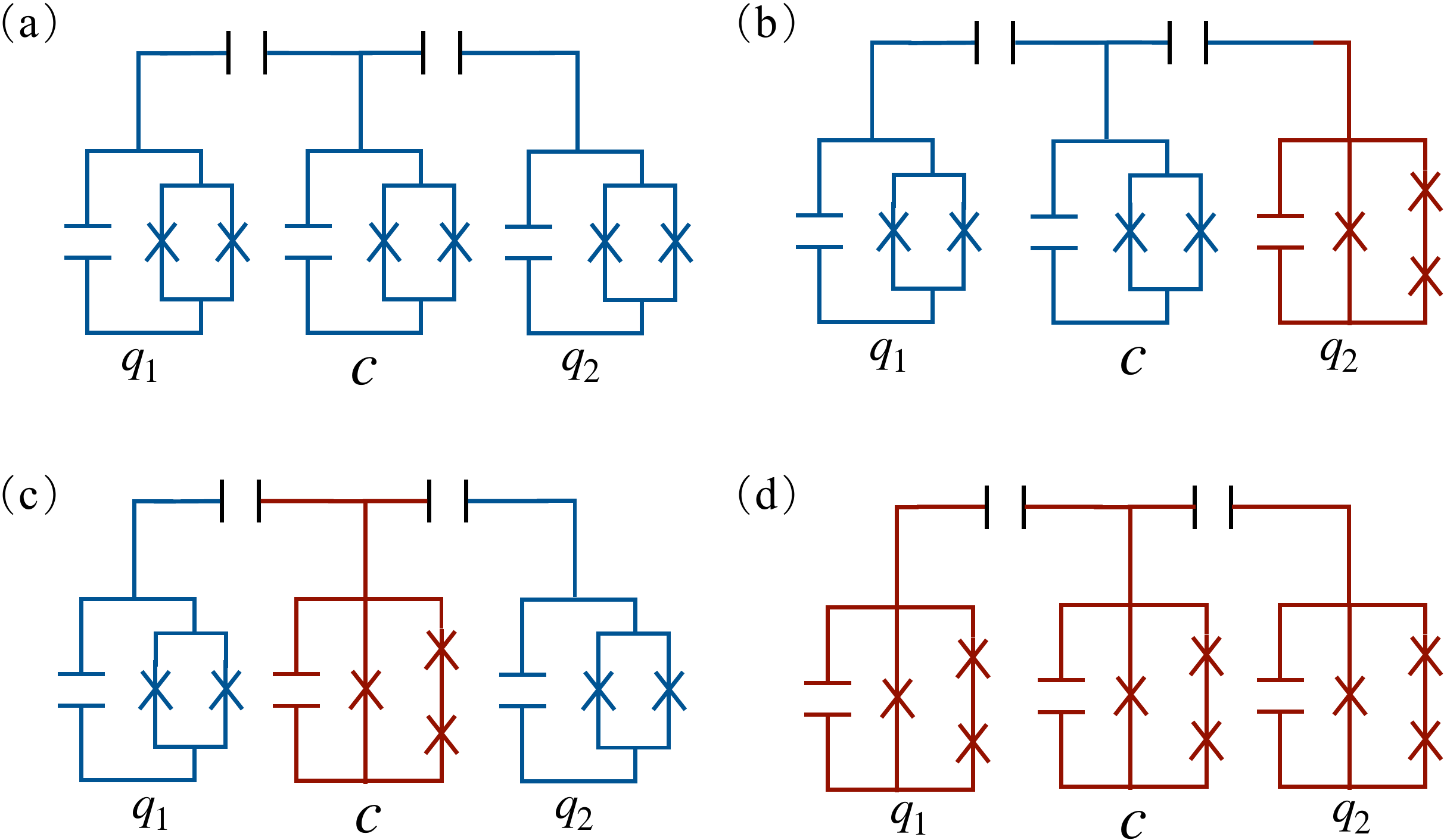}
\caption{Superconducting circuit realization with different parameter regions (Type I-IV discussed before). (a) Type I, II: both computational qubits and coupler are realized with transmon qubits (blue); (b) The two computational qubits are realized with transmon qubit (blue) and CSFQ (red), respectively, while the coupler is realized with transmon qubit (blue); (c) Type III: both computational qubits are realized with transmon qubits (blue), while the coupler is realized with CSFQ (red); (d) Type IV: both computational qubits and coupler are realized with CSFQs (red). All of these four different superconducting circuits are realizable with current experimental technology.}
\label{fig:circuitRealization}
\end{figure}

 Finally, the natural thing is to think about experimental realizations for these four novel parameter regions. 
 The good thing is that every suggested parameter region can be realized within current experimental technology. As shown in Fig.~\ref{fig:circuitRealization}, for different parameter regions I-IV, the superconducting circuit architecture is the same and the main difference is the qubit type for computational qubits and coupler.  Currently, for most of the existing experiments with coupler architecture \cite{arute2019quantum,mundada2019suppression,li2020tunable,han2020error,sung2020realization,collodo2020implementation,xu2020high}, both computational qubits and coupler are transmon qubits, as shown in Fig.~\ref{fig:circuitRealization} (a).  Our suggested parameter regions I, II are realized with such superconducting circuit as well, but the specific parameter regimes are different from those of the general ones. For Fig.~\ref{fig:circuitRealization} (b), two computational qubits are realized with different types of qubits which correspond to the anharmonicities with different signs. 
As discussed before, lower ZZ couplings and high-fidelity two-qubit gate  were investigated \cite{zhao2020high} and realized in such hybrid systems \cite{ku2020suppression}. Our suggested parameter region III is realized with the superconducting circuit shown in Fig.~\ref{fig:circuitRealization} (c), i.e., two transmon qubits are sandwiched by a CSFQ. Comparing with the superconducting circuit of Fig.~\ref{fig:circuitRealization} (b), CSFQ is changed from the computational qubit to the coupler. Benefited from this change, higher two-qubit gate fidelity may be expected and realized. 
As for the parameter region IV, it can be realized with the same type of qubit, as shown in  Fig.~\ref{fig:circuitRealization} (d). Comparing with the general superconducting circuit in Fig.~\ref{fig:circuitRealization} (a), all the transmon qubits are replaced by positive-anharmonicity qubits, i.e., CSFQ.
We expect these superconducting circuits with novel parameter regions could be fabricated and studied in future experiments. 

\section{Implementing low-error two-qubit gates with suggested parameter regions}
\label{sec:Gate error analysis}

In this section, we further study two-qubit gate errors in coupler architecture, especially focus on the novel parameter regions proposed in this paper. 
There exist several different kinds of noises which may affect the desired two-qubit gate fidelity. In particular, the first one could be some noises induced by external driven pulses on computational qubits, e.g., pulse amplitude fluctuations \cite{krantz2019quantum}, classical crosstalk \cite{winick2020simulating}, etc.. The second one could be quasistatic flux noise induced by flux fluctuations of the coupler. Benefiting from the dispersive couplings between computational qubits and coupler, such kind of noise would be largely suppressed. The third one is energy relaxation of computational qubits and coupler. The fourth one could be the parasitic couplings (discussed in Sec.~\ref{sec:Physical mechanisms of parasitic couplings} and \ref{sec:New parameters regions}) raised due to higher energy levels of both computational qubits and coupler.  This work mainly concentrates on two dominant noises: qubits' energy relaxation effects and ZZ parasitic couplings.

Our goal is to realize a high-fidelity iSWAP gate \cite{krantz2019quantum}.
Considering two computational qubits with identical frequency and involving only XY type interaction, an ideal iSWAP gate is expected naturally.  Therefore, the residual ZZ parasitic couplings as well as qubits' energy relaxation are counted as noise sources and thus bring gate errors. 
In the following, several interesting problems will be studied and discussed. How do these two different kinds of noises affect the resulting gate errors? What are the gate error characteristics with the novel parameter regions? Could we estimate conveniently the two-qubit gate error in coupler architecture? 
In additional to the numerical results  (shown in subsection \ref{Sec: Numerical results-gate error characteristics}) solved from Lindblad equations, we  also derive an analytical result (shown in subsection \ref{Sec: Analytical results and Discussion})  from which not only the physical mechanisms of gate error characteristics can be understood deeply, but also one is able to estimate straightforward the average two-qubit gate error in coupler architecture.

\subsection{Numerical results: gate error characteristics}
\label{Sec: Numerical results-gate error characteristics}
 
Involving different kinds of noises, the dynamics of the density matrix $\rho(t)$ is govern by the well-known Lindblad equation \cite{nielsen2002quantum}:  
$\partial_t {\rho}(t) = - i [\hat{H}_{\rm Lab},  {\rho}(t)] 
+  \sum_{i=q1,c,q2} \gamma_i [{\hat{a}}_i {\rho}(t) {\hat{a}}_i^{\dagger} - \{ {\hat{a}}_i^{\dagger}  {\hat{a}}_i,  {\rho}(t)  \} /2 ]$. Here, $\hat{H}_{\rm Lab}$ is the system Hamiltonian given in Eq.~\eqref{Hamiltonian:lab frame}, the Lindblad operators $\hat{a}_{i}$ ($\hat{a}^{\dagger}_{i}$) is annihilation (creation) operator for computational qubits and coupler, and $\gamma_i$ represents the energy relaxation rate of computational qubits or coupler (it often relates to the qubit energy relaxation time $T^i_1=1/\gamma_i$). Besides, $\{A,B \}=AB+BA$ denotes the anti-commutator of two elements A and B.
Using the new representation introduced in this paper,  we transform the Lindblad equation to a new form. 
To distinguish the new representation from the origin lab frame, we add a symbol ``tilde" to every quantity in the Lindblad equation: $\partial_t \tilde{\rho}(t) = - i [\tilde{\hat{H}},  \tilde{\rho}(t)] 
+  \sum_{i=q1,c,q2} {\gamma}_i [\tilde{\hat{a}}_i \tilde{\rho}(t) \tilde{\hat{a}}_i^{\dagger} - \{ \tilde{\hat{a}}_i^{\dagger}  \tilde{\hat{a}}_i,  \tilde{\rho}(t)  \} /2 ]$. In particular, with the help of SW transformations specified in Sec.~\ref{sec:Coupler architecture and Hamiltonian},  $\hat{H}_{\rm Lab}$ is transformed to $\tilde{\hat{H}} \simeq \hat{H}_{\rm eff}^2$, and the Lindblad operators are transformed to
$\tilde{\hat{a}}_{qk} \simeq {\hat{a}}_{qk} - ({g_k}/{\Delta_k}) \hat{a}_c, 
\tilde{\hat{a}}_{c} \simeq {\hat{a}}_{c} +\sum_{k=1,2} ({g_k}/{\Delta_k}) \hat{a}_{qk}$, ~ $k=1,2$.
Considering all of these and reducing to computational space,
 we ultimately obtain the dynamical equation for the density matrix $\tilde{\rho}(t)$ in Eq.~\eqref{eq:Lindblad XY AND ZZ}. The validation of Eq.~\eqref{eq:Lindblad XY AND ZZ} is verified numerically through comparing with the corresponding results solved from the lab frame Hamiltonian.
\begin{eqnarray}\label{eq:Lindblad XY AND ZZ}
&&\frac{\partial}{\partial t} \tilde{\rho}(t) \simeq  - i \left[\hat{H}^{\rm XY}_{\rm int},  \tilde{\rho}(t)\right]  - i \left[ \hat{H}^{\rm ZZ}_{\rm int},  \tilde{\rho}(t)\right] \\
&& +\sum_{i,j=q1,q2}  \tilde{\gamma}_{i,j}  \left( {\hat{\sigma}}^{i}_{-} \tilde{\rho}(t) {\hat{\sigma}}^{j}_{+} - \frac 1 2 \left\{{\hat{\sigma}}^{i}_{+}  {\hat{\sigma}}^{j}_{-}, \tilde{\rho}(t) \right \}  \right), \nonumber
\end{eqnarray} 
where $\hat{H}^{\rm XY}_{\rm int}$ and $\hat{H}^{\rm ZZ}_{\rm int}$ were given in Eqs.~\eqref{Hamiltonian: XY} and \eqref{Hamiltonian: ZZ}, the operators $\hat{\sigma}^{i}_{-}$ ($\hat{\sigma}^{i}_{+}$) are annihilation (creation) operators for computational qubits, and the effective energy relaxation rates of computational  qubits are affected by  the coupler, obtaining as
\begin{eqnarray}\label{effective decay rates}
\tilde{\gamma}_{qk,qk} &=& \gamma_{qk} + \left(\frac{g_k}{\Delta_k}\right)^2 \gamma_{c},~
k=1,2, \nonumber\\
\tilde{\gamma}_{q1,q2} &=& \tilde{\gamma}_{q2,q1} = \frac{g_1 g_2}{\Delta_1 \Delta_2} {\gamma}_{c} .
\end{eqnarray}

In the equation above, $\gamma_{q1}$, $\gamma_{q2}$, $\gamma_{c}$ are the energy relaxation  rates of qubit 1, qubit 2, and the coupler, respectively.  It is interesting to see that the effective energy relaxation  rates for  computational qubits are still dominant by their own energy relaxation  rate, while the influence induced by the coupler is suppressed by a prefactor $(g_k/\Delta_k)^2$ (which is a smaller value in dispersive regime). 
Benefiting from the robustness to the noise induced by the coupler, high-fidelity two-qubit gates are still realizable even with a noised coupler. 
In addition to realize tunable coupling, this can be seen as another advantage for coupler architecture. 
Beyond qualitative analysis, we further obtain a quantitative result to clarify how strong does the coupler noise affects the resulting gate error. Using the above analytical result [i.e., Eq.~\eqref{effective decay rates}], we can obtain approximately the critical point at which the noise from the coupler matters. In particular, the critical point is estimated roughly as
${T}_1^{c}  \approx \frac{g_1 g_2}{\Delta_1 \Delta_2 } T_1^{i} ~ (i =q1, q2)$. Using current experimental parameters (e.g., taking from Ref.\cite{li2020tunable}), we find that the resulting two-qubit gate errors are almost independent of the coupler's energy relaxation once $T_1^{c} \gg 1 ~\mu s$. Normally this is the case with current superconducting circuits technology. 
Besides, we notice that an additional effective bath induced by the coupler appears in Eq.~\eqref{eq:Lindblad XY AND ZZ}. Apart from the independent local bathes for each computational qubit, the non-negligible term with $i \neq j$ in Eq.~\eqref{eq:Lindblad XY AND ZZ} can be interpreted as a global bath for the composed system qubit 1 and qubit 2.  

Next, we use the idea of process tomography \cite{nielsen2002quantum} to compute gate fidelity. In particular, setting randomly $N$ initial input  states $\tilde{\rho}_0^k(0)  $ in computational space  with $k=1,2,3, \cdots, N$ indicating the $k$-th  initial state, and then let the state evolve in noise and noise-free cases, respectively. After a certain time (e.g., gate time $t_g$), the final states $\tilde{\rho}^k(t_g)$ (noise case) and $\tilde{\rho}_0^k(t_g)$ (noise-free case) are solved from the above Lindblad equation. In this particular case, it is reasonable to define the average  gate fidelity as \cite{jozsa1994fidelity,liang2019quantum}
\begin{equation} \label{defination: fidelity}
{\cal F} =\frac 1 N \sum^N_{k=1} {\rm tr} \left( \tilde{\rho}^k(t_g)  \tilde{\rho}^k_0(t_g) \right). 
\end{equation}
Choosing proper system parameters and numerically solving the Lindblad equations \eqref{eq:Lindblad XY AND ZZ},  we can ultimately obtain the average gate error $\varepsilon=1- {\cal F}$. 

\begin{figure}[htbp]
\centering
\includegraphics[width=\columnwidth]{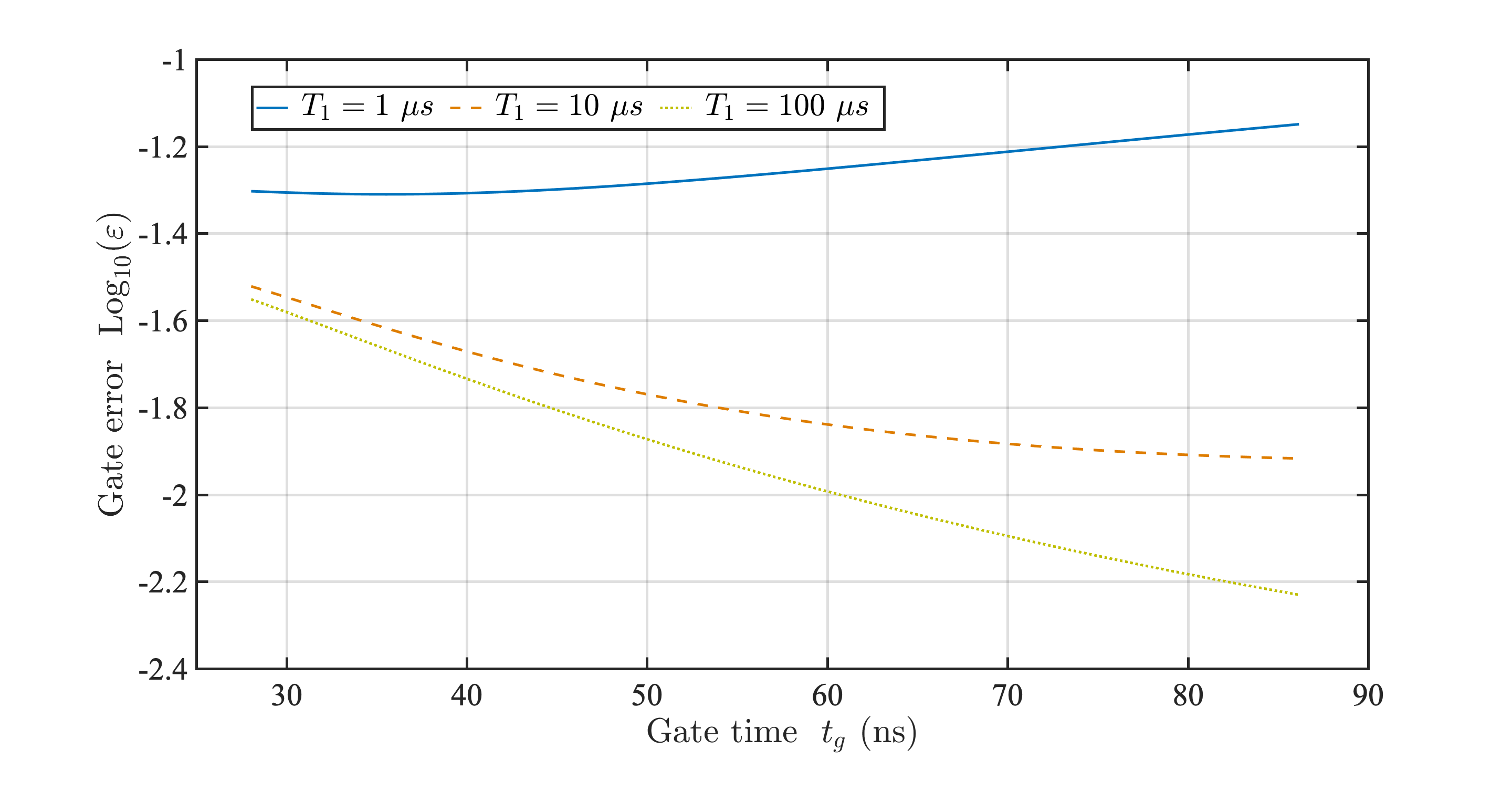}
\caption{Average gate error $\varepsilon$  dependent of gate time $t_g$ with different energy relaxation time $T_1$ for the general case (corresponds to the parameter regimes that are frequently used in experiments). Both computational qubit and coupler are realized with transmon qubits. The relaxation time for computational qubits and coupler is identical for simplicity. The used parameters are:  $\omega_{q1}/(2 \pi)=\omega_{q2}/(2 \pi)=5~{\rm GHz}$,  $\alpha_{q1}/(2 \pi)=\alpha_{q2}/(2 \pi)=-0.2~{\rm GHz}$,  $\alpha_{c}/(2 \pi)=-0.25~{\rm GHz}$,  $g1/(2 \pi)=g2/(2 \pi)=0.08~{\rm GHz}$; the number of random initial states $N=10^5$. }  
\label{fig:gateError_general}
\end{figure}

Let us first consider the general case with frequently used parameters of coupler architecture \cite{arute2019quantum}, namely both computational qubits and coupler are realized with transmon qubits. In this regime, the qubits' anharmonicities are negative and the strengths are usually designed to be around $0.1 - 0.3 ~ {\rm GHz}$. Moreover, the frequency detunings between computational qubits and coupler are usually large, e.g., $\sim 1~{\rm GHz}$.
Using the experimental parameter regimes, 
we evaluate and plot the gate error $\varepsilon$ dependent of gate time $t_g$ with different energy relaxation time $T_1$ in Fig.~\ref{fig:gateError_general}.  Without loss of generality, the relaxation time for computational qubits and coupler is identical for simplicity. 
It is shown that the gate errors are dominant by the energy relaxation for shorter qubit energy relaxation time, e.g., $T_1 \sim 1~{\rm \mu s}$. With increasing gate time, the gate error grows approximately with a linear behavior as expected \cite{yan2018tunable}. We will see later this linear characteristics can be explained theoretically under the approximate condition $t_g \ll  T_1$. As a contrast, the characteristics become different for larger $T_1$. The resulting gate error decreases with increasing gate time. 
As the noise induced by qubit energy relaxation does not  dominate for longer $T_1$, ZZ parasitic coupling starts to play a vital role. Longer gate time corresponds to a weaker effective coupling $g_{\rm eff}$ between computational qubits, hence results in weaker $\zeta_{zz}$ and lower gate errors $\varepsilon$.   The gate error induced by ZZ coupling is estimated as $\sim 10^{-2}$ with the parameters used. As discussed before, the gate error in this regime can be further reduced with larger qubits' anharmonicities. 

\begin{figure}[htbp]
\centering
\begin{minipage}[t]{0.49\textwidth}
\centering
\includegraphics[width=  \columnwidth]{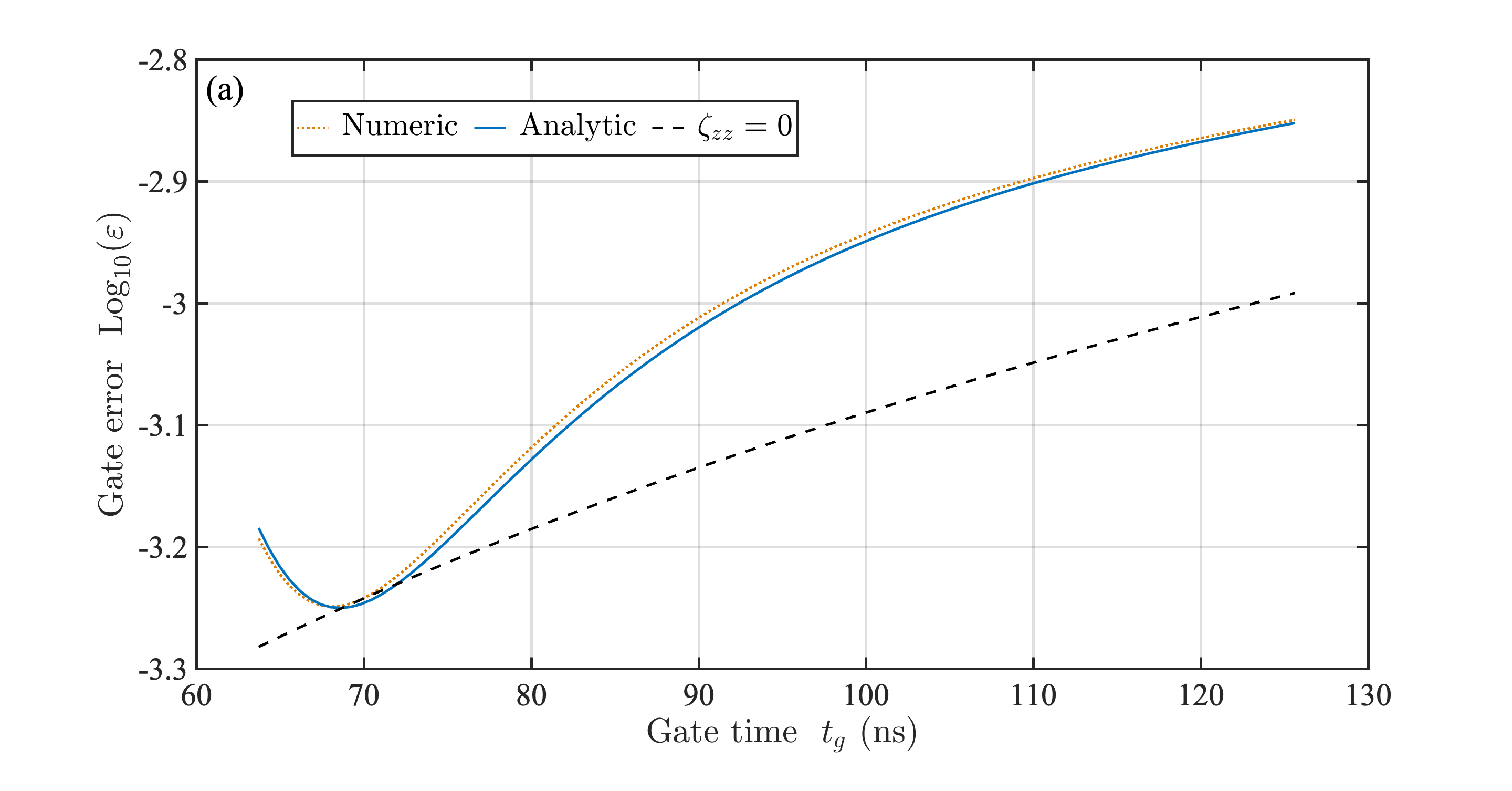}
\end{minipage}
\begin{minipage}[t]{0.49\textwidth}
\centering
\includegraphics[width=  \columnwidth]{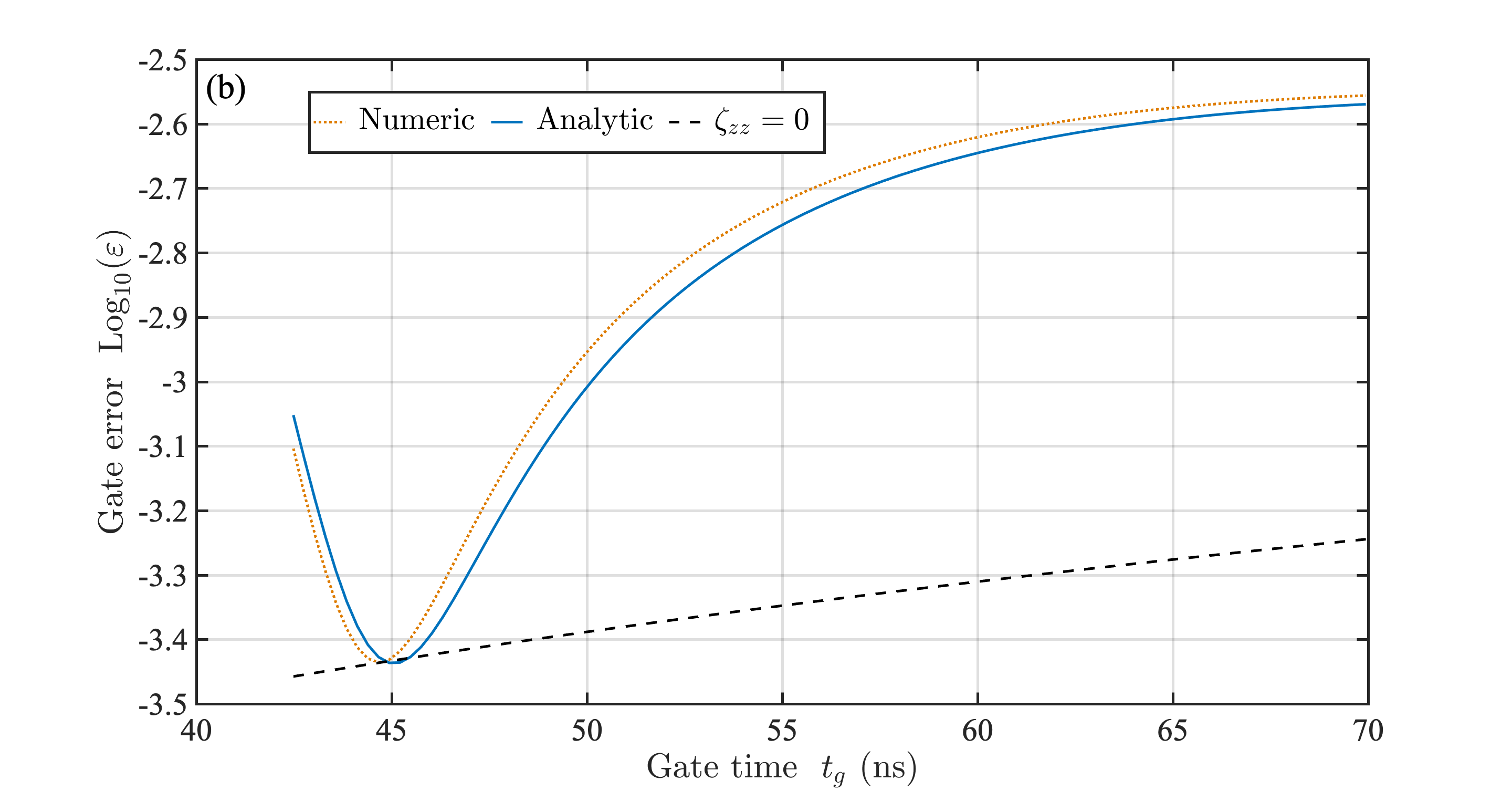}
\end{minipage}
\begin{minipage}[t]{0.49\textwidth}
\centering
\includegraphics[width=   \columnwidth]{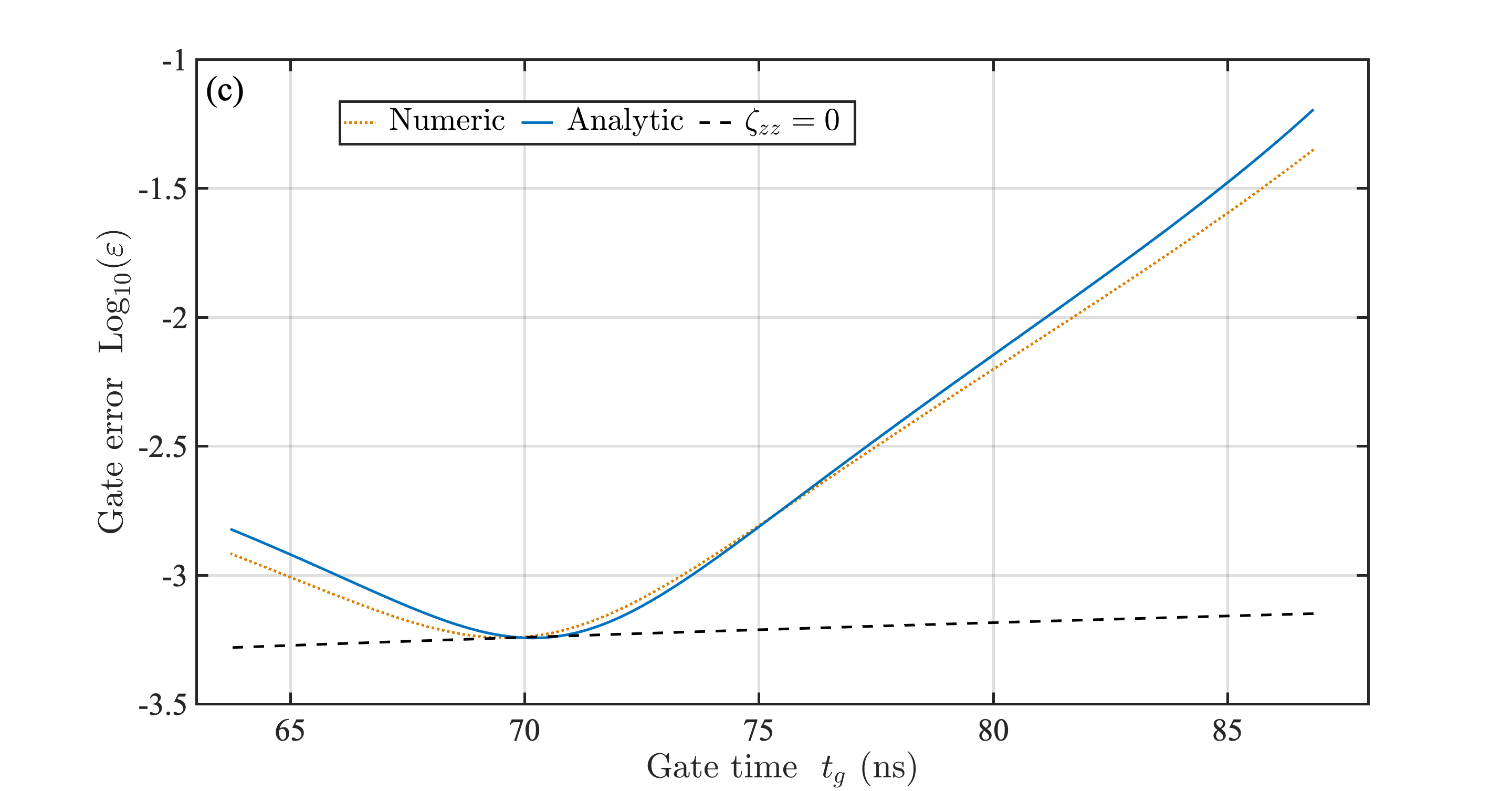}
\end{minipage}
\begin{minipage}[t]{0.49\textwidth}
\centering
\includegraphics[width=  \columnwidth]{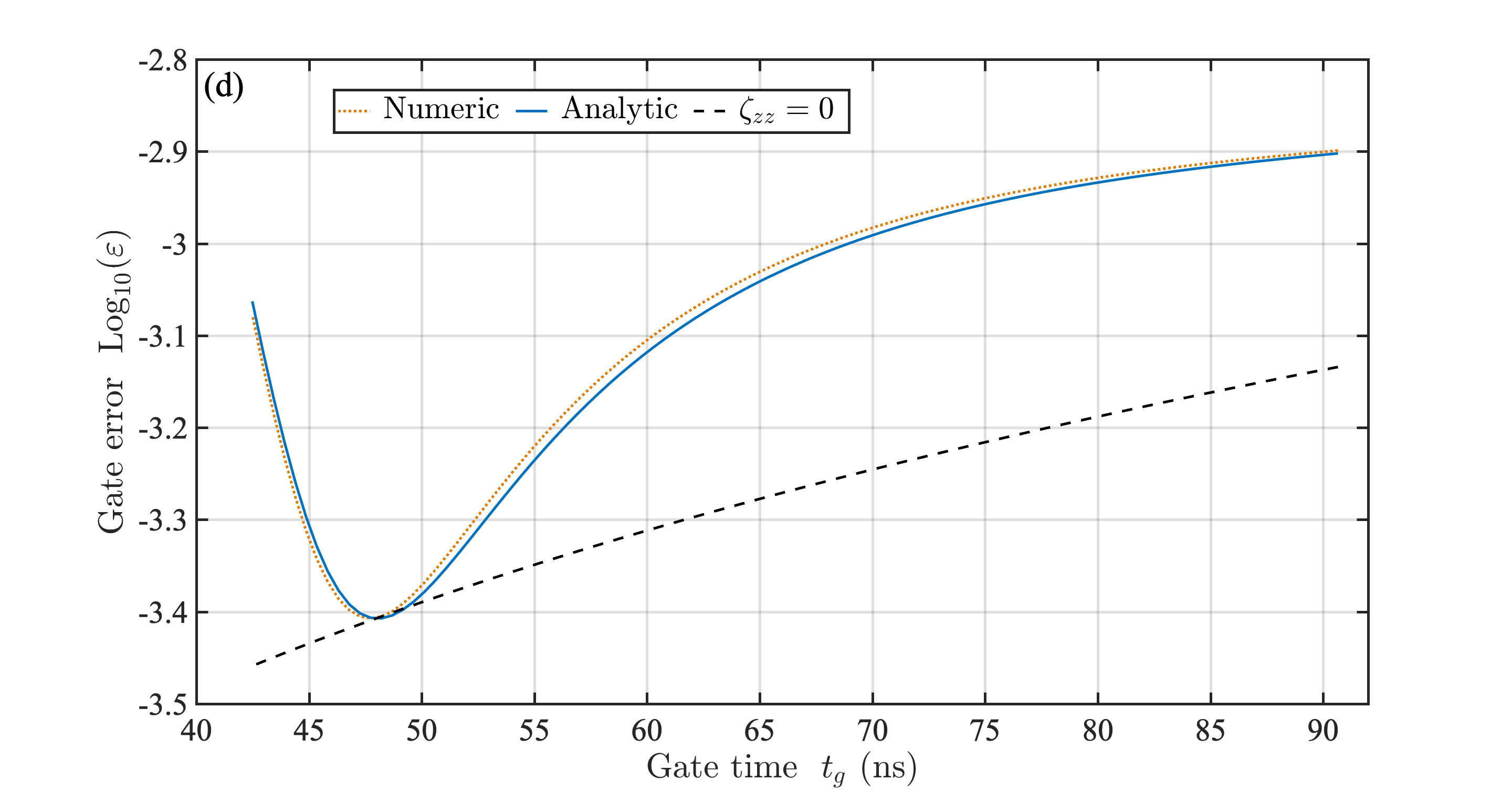}
\end{minipage}
\caption{Average gate error $\varepsilon$ characteristics with our suggested parameter regions I-IV.
The numerical results (orange dotted) are solved via  diagonalizing the system Hamiltonian $\hat{H}_{\rm Lab}$ while the analytical results (blue solid) are plotted using  Eq.~\eqref{error-analytic}. The black dashed lines corresponds to the cases in absence of ZZ couplings.
The used parameters are identical to Table \ref{table} and the anharmonicity of two computational qubits are chosen as: (a) regions I:
$\alpha_{q1}/(2 \pi)=\alpha_{q2}/(2 \pi)=-0.3~{\rm GHz}$;
  (b) regions II: $\alpha_{q1}/(2 \pi)=\alpha_{q2}/(2 \pi)=-0.25~{\rm GHz}$; 
  (c) regions III: $\alpha_{q1}/(2 \pi)=\alpha_{q2}/(2 \pi)=-0.15~{\rm GHz}$;
  (d) regions IV: $\alpha_{q1}/(2 \pi)=\alpha_{q2}/(2 \pi)=0.4~{\rm GHz}$; the number of random initial states $N=10^5$, and the energy relaxation time $T_1=100~\mu s$ for both computational qubits and coupler.} 
 \label{fig:gateError_region}
\end{figure}

Next, we turn to study the gate error characteristics with our suggested parameter regions. As referred before, an iSWAP gate is expected in coupler architecture. 
We evaluate and plot the resulting gate errors $\varepsilon$ dependent of gate time $t_g$ in Fig.~\ref{fig:gateError_region} (a)-(d) with four novel parameter regions: Type I (a),  Type II (b), Type III (c),  and Type IV (d), respectively.
In real experiments, varying gate time $t_g$ is equivalent to tuning $\omega_c$, because the gate time is directly related to the effective qubit-qubit coupling which is tuned by varying the coupler frequency $\omega_c$. 
The specific definitions for these novel parameter regions were explained in Sec.~\ref{sec:New parameters regions} and the typical system parameters can be found  in Table \ref{table}.  
Although these four suggested regions correspond to very different parameter regimes, they share a common physical mechanism and exhibit similar gate error characteristics. 
To concentrate on gate error characteristics induced by ZZ parasitic effects, we choose a longer energy relaxation time, e.g., $T_1=100~\mu s$.  Obviously, the gate errors obtained in these novel parameter regions are much lower than those in the general case (e.g., Fig.~\ref{fig:gateError_general}).  Moreover, the gate error reaches a minimum value at certain gate time. 
It is clearly seen that these novel realizable parameter regions provide a new way to reach lower error two-qubit gates without changing circuit architecture. 
Here, the gate errors  $\varepsilon$  are evaluated with various methods that correspond to different lines of each figure. The dotted orange lines (labelled as ``Numeric") are accurate numerical results obtained through solving Lindblad equation \eqref{eq:Lindblad XY AND ZZ} and using Eq.~\eqref{defination: fidelity}. In particular, both the coupling strengths $g_{\rm eff}$ and $\zeta_{zz}$ are solved via diagonalizing numerically the system Hamiltonian $\hat{H}_{\rm Lab}$.   Besides, the  solid blue lines (labelled as ``Analytic") are plotted using Eq.~\eqref{error-analytic} which will be derived in next subsection. The good agreement between the analytical and numerical results indicates that Eq.~\eqref{error-analytic} would be a good approximated expression to estimate the gate errors as well as explore the gate error's physical mechanism in coupler architecture. 
To further analyze the gate error components, we also evaluate and plot the gate error in absence of ZZ couplings, i.e., taking $\vert \zeta_{zz} \vert = 0$. The results (black dashed lines, labelled as ``$\zeta_{zz}=0$"), containing only qubits' energy relaxation contributions, behave linear approximately as expected.  Specially, we find that gate error reaches a minimum value at a specific gate time $t_g^{*}$. This critical point $t_g^{*}$ corresponds to a minimum $\vert \zeta_{zz} \vert$.  At this point, the gate error is limited mainly by the energy relaxation of computational qubits and coupler. Therefore, the coupler architecture with our suggested parameter regions is viable in the long term as superconducting qubits' coherence time continues to improve \cite{place2021new}. 

Finally, we add some remarks to gate error characteristics with the four suggested parameter regions discussed above.
For parameter regions I, II, III, the computational qubits $q1, q2$ are realized with transmon-type qubits that are stable and own longer coherent time. The main difference among these three regions is the specific character of the coupler.  For region I, the coupler is realized with transmon qubit as well. Besides, it requires a relatively small detuning between computational qubits and coupler. Restricting by dispersive relations, one also has to choose weaker qubit-coupler couplings $g_{1}$, $g_{2}$. As a consequence, it results in longer gate time $t_g$. This might be a disadvantage, especially when qubits' energy relaxation time is shorter.  
By contrast, the couplers in regions II and III are realized by either large-anharmonicity transmon qubit or CSFQ, whose energy relaxation time is usually shorter than mature transon-type qubits. Thus, one may worry that this will lead to larger gate errors. However, this is not the case. In coupler architecture, we find that the noises (of course, including the qubits' energy relaxation) induced by the coupler are suppressed largely by the dispersive couplings between the computational qubits and coupler. 
This is very different from the case of choosing two qubits' anharmonicity with different signs \cite{zhao2020high, zhao2020switchable, ku2020suppression}. Therefore, we expect that our suggested parameter regions may exhibit higher gate fidelity.  
As for region IV, the situation is different from the other three regions. Since all the elements are positive-anharmonicity qubits realized with CSFQ, one may worry about the ultimate performance (e.g., gate fidelity) because the stability and coherence of CSFQ cannot be comparable with mature transmon qubits. However, the situation may be improved rapidly in the future with mature CSFQ technology. 
Indeed, some promising progress has already been made very recently that the coherent time of CSFQ achieved in the range of 50 ${\rm \mu s}$ to 100 ${\rm \mu s}$ \cite{yan2016flux,abdurakhimov2019long}. Therefore, high-fidelity two-qubit gates may be realized in the parameter region IV as well.


\subsection{Analytical results and Discussion}
\label{Sec: Analytical results and Discussion}

In addition to numerical results, we are also interested in deriving the corresponding analytical results using the method ``vectorization of the Lindblad equation" \cite{jakob2003variational,mascarenhas2015matrix}.
 In particular, the density matrix $\tilde{\rho}(t)$ is vectorized as a column vector $\vert \tilde{\rho}(t) \rangle \rangle$.
It is noted that the double bracket notation serves to remind us that this is not the standard Hilbert space of state vectors.
 Using the  vectorization property $\vert A \tilde{\rho}(t) B \rangle \rangle \rightarrow B^{T} \otimes A \vert \tilde{\rho}(t) \rangle \rangle$ ($A,B$ are matrices, $B^T$ denotes the transpose of matrix $B$), the Lindblad equation \eqref{eq:Lindblad XY AND ZZ} is rewritten as
\begin{eqnarray}\label{eq:vectorization}
\frac{\partial}{\partial t} \vert \tilde{\rho}(t) \rangle \rangle &=&\hat{{\cal L}} \vert \tilde{\rho}(t) \rangle \rangle,
\end{eqnarray}
 where $\hat{{\cal L}}=\hat{{\cal L}}_0+\hat{{\cal L}}^{\rm ZZ}_{\rm noise}+\hat{{\cal L}}^{\rm decay}_{\rm noise}$ consists of three parts: $\hat{{\cal L}}_0$ corresponds to the noise-free case while $\hat{{\cal L}}^{\rm ZZ}_{\rm noise}$, $\hat{{\cal L}}^{\rm decay}_{\rm noise}$ represents the noise contribution induced by ZZ parasitic coupling and qubits decay effect, respectively. The specific expressions are obtained as
\begin{eqnarray}\label{L0}
\hat{{\cal L}}_0 &=& - i \hat{I} \otimes  \hat{H}^{\rm XY}_{\rm Int}   + i \left( \hat{H}^{\rm XY}_{\rm Int} \right)^{T} \otimes \hat{I}, \\
\hat{{\cal L}}_{\rm noise}^{\rm ZZ} &=& - i \hat{I} \otimes  \hat{H}^{\rm ZZ}_{\rm Int}    + i ( \hat{H}^{\rm ZZ}_{\rm Int} )^{T} \otimes \hat{I}, \\
\hat{{\cal L}}_{\rm noise}^{\rm decay} &=&  \sum_{i,j=q1,q2}  \tilde{\gamma}_{i,j}  \left[({\hat{\sigma}}^{i}_{+})^T \otimes  {\hat{\sigma}}^{j}_{-} - \frac 1 2 \hat{I} \otimes {\hat{\sigma}}^{i}_{+} {\hat{\sigma}}^{j}_{-} \right. \nonumber\\
&& - \left. \frac 1 2  ({\hat{\sigma}}^{i}_{+} {\hat{\sigma}}^{j}_{-} )^T  \otimes \hat{I}
 \right], 
\end{eqnarray}
where the expressions for Hamiltonian $ \hat{H}^{\rm XY}_{\rm Int}$ and $\hat{H}^{\rm ZZ}_{\rm Int}$ were given in Eqs.~\eqref{Hamiltonian: XY} and \eqref{Hamiltonian: ZZ} respectively, $\hat{I}$ is identity operator, and the effective decay rates $\tilde{\gamma}_{i,j} $ were given in Eq.~\eqref{effective decay rates}.

With the new notations, it is very natural to make a connection using  the  relation
$
{\rm tr} ( \tilde{\rho}^k_0(t_g) \tilde{\rho}^k(t_g) ) = \langle \langle \tilde{\rho}^k_0(t_g) \vert \tilde{\rho}^k(t_g)  \rangle \rangle
$,
where the trace of two density matrix can be evaluated by computing the inner product of the two corresponding ``vectorization states" $\vert \tilde{\rho}_0^k(t_g)  \rangle \rangle$ and $\vert \tilde{\rho}^k(t_g)  \rangle \rangle$. 
With considering these and using Eqs.~\eqref{defination: fidelity} and \eqref{eq:vectorization}, we obtain
\begin{equation} \label{Fidelity:vectorization}
{\cal F} = \frac 1 N \sum^N_{k=1} \langle \langle \tilde{\rho}^k_0(0) \vert  e^{- \hat{{\cal L}}_0 t_g} e^{\hat{{\cal L}} t_g} \vert  \tilde{\rho}^k_0(0)  \rangle \rangle,
\end{equation}
where we used $ \vert \tilde{\rho}^k(0)  \rangle \rangle =  \vert \tilde{\rho}^k_0(0)  \rangle \rangle$ because the initial states for the noise and noise-free cases are identical. 
Next, the task becomes to compute the quantity $e^{-\hat{{\cal L}}_0 t_g} e^{\hat{{\cal L}} t_g} $. In particular, this calculation can be simplified through making  representation transformation upon $\hat{{\cal L}}$ with choosing $V(t)=e^{\hat{{\cal L}}_0 t/2}$,  obtaining 
$\hat{\bar{{\cal L}}} =  V^{-1}(t)  \hat{{\cal L}_0} V(t) + V^{-1}(t)  (\hat{{\cal L}}_{\rm noise}^{ZZ} + \hat{{\cal L}}_{\rm noise}^{\rm decay}) V(t)  - 2 V^{-1}(t) \partial V(t)$. Here, the prefactor of the last term is ``2" (instead of ``1") because we perform representation transformation on two independent operators. Under the new representation, we obtain
\begin{equation}
\hat{\bar{{\cal L}}} =  \hat{{\cal L}}^{\rm ZZ}_{\rm noise} +  e^{-\hat{{\cal L}}_0 t/2}  \hat{{\cal L}}^{\rm decay}_{\rm noise}  e^{\hat{{\cal L}}_0 t/2},
\end{equation}
where the commute relation $[\hat{\sigma}_x^{q1} \hat{\sigma}_x^{q2} + \hat{\sigma}_y^{q1} \hat{\sigma}_y^{q2}, \hat{\sigma}_z^{q1} \hat{\sigma}_z^{q2}  ]=0$  was used.
Consequently, the key quality in new representation is computed as
\begin{eqnarray}\label{LL0-new representation}
e^{-\hat{{\cal L}}_0 t_g} e^{\hat{{\cal L}} t_g}  & \rightarrow & \exp \left( \hat{{\cal L}}^{\rm ZZ}_{\rm noise} t_g  + e^{-\hat{{\cal L}}_0 t_g/2}  \hat{{\cal L}}^{\rm decay}_{\rm noise} t_g e^{\hat{{\cal L}}_0 t_g/2}  \right) \nonumber\\
& \approx & \hat{I}   + e^{-\hat{{\cal L}}_0 t_g/2}  \hat{{\cal L}}^{\rm decay}_{\rm noise} t_g e^{\hat{{\cal L}}_0 t_g/2}+\frac{ \left( \hat{{\cal L}}^{\rm ZZ}_{\rm noise}  \right)^2 t_g^2}{ 2} , \nonumber\\
\end{eqnarray}
where in the last step we used the approximations that  gate time is much smaller than the decay time, i.e.,  $t_g \ll T_1$, and the ZZ coupling strength is much smaller than the XY coupling strength, i.e., $ \vert \zeta_{zz} \vert \ll \vert g_{\rm eff} \vert $. 
Ultimately, substituting Eq.~\eqref{LL0-new representation} back into Eq.~\eqref{Fidelity:vectorization}, the average  gate error arrives as  a simple form. 
\begin{eqnarray} \label{error-analytic}
\varepsilon (t_g) & \approx &  \lambda_{\rm decay} \frac{t_g}{T_1} \\
&&+  \lambda_{\rm zz}   \left(\frac{1}{\alpha_{q1}} + \frac{1}{\alpha_{q2}} + \frac{1}{\frac{\alpha_{c}}{4}- {\rm sgn}(\Delta) \frac{ g_1 g_2 t_g} {\pi}} \right)^2 \frac{1}{t_g^2},\nonumber
\end{eqnarray}
where $\rm sgn$ is signum function; $\lambda_{\rm zz} $  and $ \lambda_{\rm decay} $ can be evaluated with randomized initial states, namely
$\lambda_{\rm decay} = -  \frac 1 N \sum^N_{k=1} \langle \langle \tilde{\rho}^k_0(0) \vert 
e^{-\hat{{\cal L}}_0 t_g/2} \hat{{\cal L}}^{\rm decay}_{\rm noise} T_1 e^{\hat{{\cal L}}_0 t_g/2} 
   \vert  \tilde{\rho}^k_0(0)  \rangle \rangle \approx 0.81$  and
$ \lambda_{zz} = - \frac{\pi^4}{8}  \frac 1 N \sum^N_{k=1} \langle \langle \tilde{\rho}^k_0(0) \vert 
  (\hat{{\cal L}}^{\rm ZZ}_{\rm noise}/\zeta_{zz})^2
   \vert  \tilde{\rho}^k_0(0)  \rangle \rangle \approx 18.55$. The correctness of the analytical result is verified using numerical results with different parameter regimes (see Fig.~\ref{fig:gateError_region}). 
We find it gives a good result whose behaviors are very close to the accurate numerical results. Therefore,  with the help of Eq.~\eqref{error-analytic}, one is able to estimate the average two-qubit gate errors of coupler architecture conveniently with given superconducting circuit parameters.    This would be very helpful to experimental scientists when they design superconducting circuits.    
   
Furthermore, more interesting physics can be reflected from this analytical result. As seen clearly from Eq.~\eqref{error-analytic}, the first term represents the gate error induced by qubits' energy relaxation while the second term corresponds to the error induced by ZZ parasitic couplings. The characteristics are different for different parameter regime.
 When the energy relaxation effect dominates, the average gate errors are proportional to gate time $t_g$ and inversely proportional to qubits' energy relaxation time $T_1$. 
This is why short pulses  are usually favorable. When ZZ parasitic coupling dominates,  the average gate errors are inversely proportional to the gate time's square; hence lower gate errors favor longer gate time. This is understandable because longer gate time corresponds to weaker effective qubit-qubit coupling, resulting in weaker ZZ coupling.  Consequently, to realize fast quantum gates, the price we have to pay is larger parasitic couplings. Obviously, there exists a trade-off effect between these two different kinds of noises.  
As proper gate time is chosen to eliminate ZZ parasitic coupling, the resulting gate error becomes $\varepsilon  \approx   \lambda_{\rm dacay} {t_g}/{T_1}$. In absence of the contributions of coupler, the optimized gate time $t_g^{*}$ can be obtained from our analytical results, arriving as $t_g^{*} =[2  \lambda_{\rm zz} T_1 ({1}/{\alpha_{q1}} + {1}/{\alpha_{q2}} )^2/\lambda_{\rm dacay}]^{1/3} $ and the corresponding minimum gate error is estimated as $\varepsilon (t_g^{*})$. 


\section{Summary and Perspectives}
\label{Sec:Summary and Perspectives}

In summary, we studied systematically the physical mechanisms of ZZ parasitic coupling and the resulting two-qubit gate error characteristics in coupler architecture using effective Hamiltonian approach. Through applying two times Schrieffer-Wolf transformation, we obtained an effective Hamiltonian containing some intriguing terms (unexplored yet) which can be interpreted as parasitic couplings between the state $\vert 101 \rangle$ in computational space and high-energy states $\vert 200 \rangle$, $\vert 002 \rangle$, and $\vert 020 \rangle$ out of computational space. These parasitic couplings are counted as noise source for realizing iSWAP gates.
Benefiting from the effective Hamiltonian, we successfully derived the analytical result for ZZ coupling strength $\zeta_{zz}$ for different regime; using it, some previous impressive research \cite{zhao2020high,zhao2020suppression,ku2020suppression} can be thoroughly explained and understood. Beyond it, we also provided an explicit quantitative condition for eliminating ZZ parasitic couplings. As applications, this can be used to fix the proper parameters in designing superconducting quantum processor. Last but not least, we proposed four novel parameter regions in which minimum ZZ couplings and high-fidelity two-qubit gates are expected. Using the novel parameter regions, we found that the coupler's high energy levels play a vital role (ignored in previous research), which can neutralize the energy shift induced by computational qubits' high energy levels. 
We hope the coupler architecture with these suggested parameter regions and  the predicted characteristics could be realized and verified in future experiments. 

Numerically solving the Lindblad equation containing both energy relaxation effects and ZZ parasitic couplings, we verified that high-fidelity two-qubit gates are realizable with our suggested parameter regions. In particular, ZZ parasitic coupling can be eliminated with proper system parameters and the resulting gate errors are limited mainly by qubits' energy relaxation. Apart from the numerical results, we also successfully derived an analytical expression [Eq.~\eqref{error-analytic}] of the average gate error via vectorizing the Lindblad equation. We found the trade-off effect between the error induced by qubit energy relaxation and ZZ parasitic couplings with different gate time. Moreover, this can be applied to estimate the average gate error in coupler architecture with given system parameters conveniently.  

Beyond the architecture and dispersive regime discussed in this paper, new coupler architecture connecting fixed-frequency floating qubits and new regime were also studied and high-fidelity gates were realized very recently \cite{sung2020realization,stehlik2021tunable,sete2021floating}. Therefore, it would be also interesting to study the interesting physical mechanism behind and explore more possibility. 
 
\vspace{0.5 cm} 

\begin{acknowledgments}
We would like to thank Runyao Duan for helpful discussions. 
\end{acknowledgments}

\appendix

\begin{widetext}

\section{Derivation of the Effective Hamiltonian}
\label{appendix: new representation}

In this appendix, we show the detailed derivation of Eq.~\eqref{Hamiltonian: new representation} (which is the cornerstone of this paper) using Schrieffer–Wolff transformation (SWT).
In particular, we apply two times SWT and consider the fourth-order contribution. Comparing with previous work \cite{yan2018tunable} in which only 1 time SWT was applied and only the first-order contribution was involved, our results are more accurate and contain richer physics.  Benefiting from the substantial derivation, the irrelevant terms will be got rid from the lab Hamiltonian eventually. More importantly, the physical mechanisms can be understood clearly using the resulting effective Hamiltonian.

\subsection{1st SWT}

In order to obtain the indirect coupling between computational qubits, the first and foremost task is to decouple the coupler using SWT. In particular, we need to find a suitable $\hat{s}_1$ and compute  $e^{\hat{s}_1} \hat{H}_{\rm Lab} e^{-\hat{s}_1}$. The explicit form for $\hat{s}_1$ is given by
\begin{equation}
\hat{s}_1 = \sum_{k=1,2} \frac{g_k}{\Delta_k} \left(\hat{a}^{\dagger}_{qk}  \hat{a}_{c}  - \hat{a}_{qk}       \hat{a}^{\dagger}_{c} \right) - \frac{g_k}{\sum_{k}} \left(\hat{a}^{\dagger}_{qk}  \hat{a}^{\dagger}_{c} - \hat{a}_{qk}  \hat{a}_{c}  \right),
\end{equation}
where $\Delta_k = \omega_{qk} - \omega_c$ and  $\sum_k = \omega_{qk} + \omega_c$.
For easy  reference, we write down again the lab frame Hamiltonian. 
\begin{equation} \label{circuit Hamiltonian}
 \hat{H}_{\rm Lab} =  \sum_{\lambda=q1,q2,c}  \omega_{\lambda} \hat{a}^{\dagger}_{\lambda} \hat{a}_{\lambda} + \frac{\alpha_{\lambda}}{2} \hat{a}^{\dagger}_{\lambda} \hat{a}^{\dagger}_{\lambda} \hat{a}_{\lambda} \hat{a}_{\lambda} + g_{12} \left( \hat{a}^{\dagger}_{q1}  \hat{a}_{q2}  - \hat{a}^{\dagger}_{q1}  \hat{a}^{\dagger}_{q2} + H.c. \right) + \sum_{k=1,2}  g_k \left( \hat{a}^{\dagger}_{qk}  \hat{a}_{c}   - \hat{a}^{\dagger}_{qk}  \hat{a}^{\dagger}_{c} + H.c.   \right).
\end{equation}
Next, the task becomes to compute $e^{\hat{s}_1} \hat{H}_{\rm Lab} e^{-\hat{s}_1} = \hat{H}_{\rm Lab} + [\hat{s}_1, \hat{H}_{\rm Lab}] +[\hat{s}_1, [\hat{s}_1, \hat{H}_{\rm Lab}]]/2! +  \cdots $. First of all, the zero-order contribution is $\hat{H}_{\rm Lab}$ itself. Since the calculation for the higher order contributions always rely on the results of lower orders, we compute it order by order respectively. 

\subsection*{The first-order contribution}
\label{appendix:1st order}

As a starting point, let us compute the first-order contribution $[\hat{s}_1, \hat{H}_{\rm Lab}]$. Since $\hat{s}_1$ contains four terms, they can be computed separately. Here, 
 we show only  the first term calculation, i.e., $[\hat{a}^{\dagger}_{q1}  \hat{a}_{c}  - \hat{a}_{q1}       \hat{a}^{\dagger}_{c}, \hat{H}_{\rm Lab} ] $, as an example. Besides, it is noticeable there are nine terms in Hamiltonian $ \hat{H}_{\rm Lab}$, we compute them independently and then combine.  At the end, we  obtain 
\begin{eqnarray} \label{SW:first-order:1st term}
\left[\hat{a}^{\dagger}_{q1}  \hat{a}_{c}  - \hat{a}_{q1}       \hat{a}^{\dagger}_{c}, \hat{H}_{\rm Lab} \right] &=& - \Delta_1 \left(\hat{a}^{\dagger}_{q1}  \hat{a}_{c}  + \hat{a}_{q1}       \hat{a}^{\dagger}_{c}\right) + 2 g_1 \left(\hat{a}^{\dagger}_{q1}  \hat{a}_{q1} - \hat{a}^{\dagger}_{c}  \hat{a}_{c} \right)  + g_2  \left( \hat{a}^{\dagger}_{q1}  \hat{a}_{q2}  + \hat{a}_{q1}       \hat{a}^{\dagger}_{q2} - \hat{a}^{\dagger}_{q1}  \hat{a}^{\dagger}_{q2} - \hat{a}_{q1}  \hat{a}_{q2} \right) \nonumber\\
&& - g_1 \left( \hat{a}^{\dagger}_{q1} \hat{a}^{\dagger}_{q1} +  \hat{a}_{q1} \hat{a}_{q1} \right) + g_1 \left(  \hat{a}^{\dagger}_{c}  \hat{a}^{\dagger}_{c}+  \hat{a}_{c} \hat{a}_{c}  \right) - g_{12} \left( \hat{a}^{\dagger}_{q2}  \hat{a}_{c}  + \hat{a}_{q2}       \hat{a}^{\dagger}_{c} - \hat{a}^{\dagger}_{q2}  \hat{a}^{\dagger}_{c} - \hat{a}_{q2}  \hat{a}_{c}   \right) \nonumber\\
&& - \alpha_{q1} \left( \hat{a}^{\dagger}_{q1}  \hat{a}^{\dagger}_{q1}\hat{a}_{q1} \hat{a}_{c} + \hat{a}^{\dagger}_{q1} \hat{a}_{q1} \hat{a}_{q1} \hat{a}^{\dagger}_{c}\right) + \alpha_c \left( \hat{a}^{\dagger}_{q1}   \hat{a}^{\dagger}_{c}   \hat{a}_{c}     \hat{a}_{c}    + \hat{a}_{q1}  \hat{a}^{\dagger}_{c}  \hat{a}^{\dagger}_{c}     \hat{a}_{c}   \right) .
\end{eqnarray}
Using the same procedure, we obtain 
\begin{eqnarray} \label{SW:first-order:2nd term}
\left[\hat{a}^{\dagger}_{q1}  \hat{a}^{\dagger}_{c}  - \hat{a}_{q1}       \hat{a}_{c}, \hat{H}_{\rm Lab} \right] &=& - \Sigma_1 \left(\hat{a}^{\dagger}_{q1}  \hat{a}^{\dagger}_{c}  + \hat{a}_{q1}       \hat{a}_{c}\right) + 2 g_1 \left(\hat{a}^{\dagger}_{q1}  \hat{a}_{q1} + \hat{a}^{\dagger}_{c}  \hat{a}_{c} \right) + g_2  \left( \hat{a}^{\dagger}_{q1}  \hat{a}_{q2}  + \hat{a}_{q1}       \hat{a}^{\dagger}_{q2} - \hat{a}^{\dagger}_{q1}  \hat{a}^{\dagger}_{q2} - \hat{a}_{q1}  \hat{a}_{q2} \right) \nonumber\\
&& - g_1 \left( \hat{a}^{\dagger}_{q1} \hat{a}^{\dagger}_{q1} +  \hat{a}_{q1} \hat{a}_{q1} \right) - g_1 \left(  \hat{a}^{\dagger}_{c}  \hat{a}^{\dagger}_{c}+  \hat{a}_{c} \hat{a}_{c}  \right) + g_{12} \left( \hat{a}^{\dagger}_{q2}  \hat{a}_{c}  + \hat{a}_{q2}       \hat{a}_{c} - \hat{a}^{\dagger}_{q2}  \hat{a}^{\dagger}_{c} - \hat{a}_{q2}  \hat{a}^{\dagger}_{c}   \right) \nonumber\\
&& - \alpha_{q1} \left( \hat{a}^{\dagger}_{q1}  \hat{a}^{\dagger}_{q1}\hat{a}_{q1} \hat{a}^{\dagger}_{c} + \hat{a}^{\dagger}_{q1} \hat{a}_{q1} \hat{a}_{q1} \hat{a}_{c}    \right) - \alpha_c \left( \hat{a}^{\dagger}_{q1}   \hat{a}^{\dagger}_{c}   \hat{a}^{\dagger}_{c}     \hat{a}_{c}    + \hat{a}_{q1}  \hat{a}^{\dagger}_{c}  \hat{a}_{c}     \hat{a}_{c}   \right) .
\end{eqnarray}
Similarly, through simply exchanging the subscripts ``q1" and ``q2", the other two terms $[\hat{a}^{\dagger}_{q2}  \hat{a}_{c}  - \hat{a}_{q2} \hat{a}^{\dagger}_{c}, \hat{H}_{\rm Lab}]$ and $[\hat{a}^{\dagger}_{q2}  \hat{a}^{\dagger}_{c}  - \hat{a}_{q2}\hat{a}_{c}, \hat{H}_{\rm Lab}]$ are computed as well. 
Finally, using the equations above and considering the related prefactors, we are able to obtain the first-order contribution. It reads 
\begin{eqnarray} \label{SW:first-order}
[\hat{s}_1, \hat{H}_{\rm Lab}] &=&  - \sum_{k=1,2} g_k \left(\hat{a}^{\dagger}_{qk}  \hat{a}_{c}  - \hat{a}^{\dagger}_{qk}  \hat{a}^{\dagger}_{c}  + H.c. \right)  + \sum_{k=1,2}\left[ \left( \frac{2 g^2_k}{\Delta_k} - \frac{2 g^2_k}{\Sigma_k} \right)  \hat{a}^{\dagger}_{qk}  \hat{a}_{qk}  - \left( \frac{2 g^2_k}{\Delta_k} + \frac{2 g^2_k}{\Sigma_k}  \right)   \hat{a}^{\dagger}_{c}  \hat{a}_{c} \right] \nonumber\\
&& + g_1 g_2 \left(\frac{1}{\Delta_1}+\frac{1}{\Delta_2}-\frac{1}{\Sigma_1}-\frac{1}{\Sigma_2} \right) \left( \hat{a}^{\dagger}_{q1}  \hat{a}_{q2}   - \hat{a}^{\dagger}_{q1}  \hat{a}^{\dagger}_{q2}+ H.c. \right) \nonumber\\
&& - \sum_{k=1,2} \left( \frac{g_k \alpha_{qk}}{\Delta_k}  \hat{a}^{\dagger}_{qk} \hat{a}^{\dagger}_{qk} \hat{a}_{qk} \hat{a}_{c}  + \frac{g_k \alpha_{qk}}{\Sigma_k}  \hat{a}^{\dagger}_{qk} \hat{a}^{\dagger}_{qk} \hat{a}_{qk} \hat{a}^{\dagger}_{c}  + \frac{g_k \alpha_c}{\Delta_k}  \hat{a}^{\dagger}_{qk} \hat{a}^{\dagger}_{c} \hat{a}_{c} \hat{a}_{c}  + \frac{g_k \alpha_c}{\Sigma_k}  \hat{a}^{\dagger}_{qk} \hat{a}^{\dagger}_{c}  \hat{a}^{\dagger}_{c} \hat{a}_{c} +H.c. \right) \nonumber\\
&& + \sum_{k=1,2} \left[ g^2_k \left(\frac{1}{\Sigma_k} - \frac{1}{\Delta_k} \right)
 \hat{a}^{\dagger}_{qk}  \hat{a}^{\dagger}_{qk}  + g^2_k \left(\frac{1}{\Sigma_k} + \frac{1}{\Delta_k} \right) \hat{a}^{\dagger}_{c}  \hat{a}^{\dagger}_{c}  + H.c. \right]
\nonumber\\
&& - \sum_{k=1,2} \left[ \frac{ g_k g_{12}}{\Delta_k} \left( \hat{a}^{\dagger}_{q\bar{k}}  \hat{a}_{c}  + \hat{a}_{q\bar{k}}       \hat{a}^{\dagger}_{c} - \hat{a}^{\dagger}_{q\bar{k}}  \hat{a}^{\dagger}_{c} - \hat{a}_{q\bar{k}}  \hat{a}_{c}   \right)  + \frac{g_k g_{12}}{\Sigma_k} \left( \hat{a}^{\dagger}_{q\bar{k}}  \hat{a}_{c}  + \hat{a}_{q\bar{k}}       \hat{a}_{c} - \hat{a}^{\dagger}_{q\bar{k}}  \hat{a}^{\dagger}_{c} - \hat{a}_{q\bar{k}}  \hat{a}^{\dagger}_{c}   \right)\right] ,
\end{eqnarray}
where for simplification we defined $\bar{k}$: $\bar{1}=2$ and $\bar{2}=1$ .

It is necessary to point out that the above equation is the exact result without any approximations. In particular, 
the first term of the equation's right side is used to cancel $\hat{H}_{qc}$, i.e.,  Eq.~\eqref{Hamiltonian:qc}. The second term implies the frequency shift of the qubits induced by the couplings. The third term represents the effective indirect coupling between two computational qubits due to the coupler. It is exactly this term that made the coupling between computational qubits tunable. For the fourth term, it originates from the three nonlinear Kerr interaction terms of $\hat{H}_0$, i.e., Eq.~\eqref{Hamiltonian:0}. Since they describe higher energy levels of superconducting qubits,  we will see very fruitful physics originated from these nonlinear terms. 
The contribution of the fourth term was usually ignored in previous research \cite{yan2018tunable}, but we find they are indeed very important especially when the condition $\alpha_{\lambda} \ll \Delta_{1,2}$, $\lambda=q1,c,q2$  violates (for instance when the qubits are not transmon type). In the following, we keep only the first four terms and neglect the last two terms. The reason is as follows:   firstly, the fifth term is high-frequency rotating and thus can be ignored; meanwhile, even if the second-order contribution of the fifth term produces some terms describing the coupling between computational qubits and coupler but with a small prefactor $(g_k/\Delta_k)^2$ or $(g_k/\Sigma_k)^2$, $k=1,2$, thus the strength goes to zero under dispersive regime; secondly, the last term can also be ignored with further considering $g_{12} \ll \Delta_{k}, \Sigma_{k}$, namely assuming the direct coupling between the computational qubits are rather weak. Comparing the last term and the first term, it is obvious the last term can be ignored. Meanwhile, for the corresponding second-order contribution, all of the terms are along with small prefactors $(g_k/\Delta_k)^2$ or $(g_k/\Sigma_k)^2$; therefore, we can safely neglect them as well. 

\subsection*{The second-order contribution}

Using the same procedure, we continue to compute the second-order contribution. 
  Although there are many terms in $[\hat{s}_1, \hat{H}_{\rm Lab}] $, i.e., Eq.~\eqref{SW:first-order}, we will see most of them can be neglected under dispersive regime. To present the calculation in a simple way, the contribution from each term of $[\hat{s}_1, \hat{H}_{\rm Lab}] $ is labelled as $[\hat{s}_1, [\hat{s}_1, \hat{H}_{\rm Lab}] ]^{(k)} $ ($k$ indicates the $k$-th term of Eq.~\eqref{SW:first-order}). In the following, they are  computed  one by one. First of all, we consider
\begin{equation}
 \left[\hat{s}_1, [\hat{s}_1, \hat{H}_{\rm Lab}] \right]^{(1)} = \sum_{k=1,2} \left[   \left(  \frac{2 g^2_k}{\Sigma_k}-\frac{2 g^2_k}{\Delta_k}  \right)  \hat{a}^{\dagger}_{qk}  \hat{a}_{qk}  +  \left( \frac{ 2g^2_k}{\Delta_k} + \frac{ 2g^2_k}{\Sigma_k}  \right)   \hat{a}^{\dagger}_{c}  \hat{a}_{c} 
 - g_1 g_2 \left(\frac{1}{\Delta_k}-\frac{1}{\Sigma_k} \right) \left( \hat{a}^{\dagger}_{q1}  \hat{a}_{q2}   - \hat{a}^{\dagger}_{q1}  \hat{a}^{\dagger}_{q2} + H.c. \right) \right] . 
\end{equation}
For the second and third terms of $[\hat{s}_1, \hat{H}_{\rm Lab}]$, namely $ [\hat{s}_1, [\hat{s}_1, \hat{H}_{\rm Lab}] ]^{(2)}$ and $ [\hat{s}_1, [\hat{s}_1, \hat{H}_{\rm Lab}] ]^{(3)}$, we  can straightforward compute them using the previous results.  However, they do not generate new terms (comparing with the first-order result, i.e., Eq.~\eqref{SW:first-order}) but with small prefactors $(g_k/\Delta_k)^2$, $(g_k/\Sigma_k)^2$, or $(g_k/\Delta_k) (g_k/\Sigma_k)$, $k=1,2$, hence can be neglected under the regime of interest, namely $g_k \ll \vert \Delta_k \vert, \Sigma_k$. 

Next, we turn to look into the nonlinear terms, namely $[\hat{s}, [\hat{s}, \hat{H}_{\rm Lab}] ]^{(4)} $.  We obtain 
\begin{eqnarray}
 \left[\hat{s}_1 ,  [\hat{s}_1 ,  \hat{H}_{\rm Lab} ] \right]^{(4)} &\simeq & \sum_{k=1,2} \left[  - \frac{2 g_k^2 \alpha_{qk}}{\Delta_k^2}  \hat{a}^{\dagger}_{qk}   \hat{a}^{\dagger}_{qk}   \hat{a}_{qk}  \hat{a}_{qk}   +  \frac{4 g_k^2 \alpha_{qk}}{\Delta_k^2}  \hat{a}^{\dagger}_{qk}   \hat{a}_{qk}   \hat{a}^{\dagger}_{c} \hat{a}_{c}  -  \frac{g_1 g_2 \alpha_{qk}}{\Delta_1 \Delta_2}
 \left( \hat{a}^{\dagger}_{qk} \hat{a}_{qk} \hat{a}_{qk} \hat{a}^{\dagger}_{q\bar{k}} +  H.c.  \right) \right. \nonumber\\
 && \left. - \frac{2 g_k^2 \alpha_{c}}{\Delta_k^2}  \hat{a}^{\dagger}_{c}   \hat{a}^{\dagger}_{c}   \hat{a}_{c}  \hat{a}_{c} +  \frac{4 g_k^2 \alpha_{c}}{\Delta_k^2} \hat{a}^{\dagger}_{qk}   \hat{a}_{qk}   \hat{a}^{\dagger}_{c} \hat{a}_{c}   \right] + \frac{2 g_1 g_2 \alpha_{c}}{\Delta_1 \Delta_2} \left(\hat{a}^{\dagger}_{q1} \hat{a}^{\dagger}_{q2} \hat{a}_{c} \hat{a}_{c}   + 2 \hat{a}^{\dagger}_{q1}  \hat{a}_{q2} \hat{a}^{\dagger}_{c} \hat{a}_{c}  + H.c.  \right), 
\end{eqnarray}
for simplification we defined $\bar{k}$: $\bar{1}=2$ and $\bar{2}=1$.  Moreover, we neglected those high rotating terms, e.g., $\hat{a}_{qk}^{\dagger} \hat{a}_{qk}^{\dagger} \hat{a}_{c} \hat{a}_{c} + H.c.$ in the equation above. The same rule will be applied in the following calculations. 
 
As discussed before, the last two terms of Eq,~\eqref{SW:first-order} does not generate new term and along with some small prefactors, and hence can be ignored in first and second contributions.

Combining all of these  terms,  the second order contribution arrives as 
\begin{eqnarray}\label{SW:second-order}
 \left[\hat{s}_1,  [\hat{s}_1,  \hat{H}_{\rm Lab}] \right] &=&
  - \sum_{k=1,2} \left[ \left( \frac{2 g^2_k}{\Delta_k} - \frac{2 g^2_k}{\Sigma_k} \right)  \hat{a}^{\dagger}_{qk}  \hat{a}_{qk}   +  \left( \frac{ 2g^2_k}{\Delta_k} + \frac{ 2g^2_k}{\Sigma_k}  \right)   \hat{a}^{\dagger}_{c}  \hat{a}_{c}    -   \frac{2 g_k^2 \alpha_{qk}}{\Delta_k^2}  \hat{a}^{\dagger}_{qk}   \hat{a}^{\dagger}_{qk}   \hat{a}_{qk}  \hat{a}_{qk}   +    \frac{2 g_k^2 \alpha_{c}}{\Delta_k^2}   \hat{a}^{\dagger}_{c}   \hat{a}^{\dagger}_{c}   \hat{a}_{c}  \hat{a}_{c} \right] \nonumber\\
&& - g_1 g_2 \sum_{k=1,2} \left(\frac{1}{\Delta_k}-\frac{1}{\Sigma_k} \right) \left( \hat{a}^{\dagger}_{q1}  \hat{a}_{q2}   - \hat{a}^{\dagger}_{q1}  \hat{a}^{\dagger}_{q2} + H.c.\right)  + \sum_{k=1,2} \frac{4 g_k^2 (\alpha_{qk}+\alpha_{c})}{\Delta_k^2}  \hat{a}^{\dagger}_{qk}   \hat{a}_{qk}   \hat{a}^{\dagger}_{c} \hat{a}_{c}   \nonumber\\
 &&  - \sum_{k=1,2} \frac{g_1 g_2 \alpha_{qk}}{\Delta_1 \Delta_2}
 \left( \hat{a}^{\dagger}_{qk} \hat{a}_{qk} \hat{a}_{qk} \hat{a}^{\dagger}_{q\bar{k}} +  H.c.  \right)   + \frac{2 g_1 g_2 \alpha_{c}}{\Delta_1 \Delta_2} \left(\hat{a}^{\dagger}_{q1} \hat{a}^{\dagger}_{q2} \hat{a}_{c} \hat{a}_{c}   + 2 \hat{a}^{\dagger}_{q1}  \hat{a}_{q2} \hat{a}^{\dagger}_{c} \hat{a}_{c}  + H.c. \right). 
\end{eqnarray}

As seen from the equation above, both the qubits' frequency and anharmonicity are shifted; this would be vital for some specific parameters regime. Besides, the resulting effective couplings between the computational qubits are generated as expected. More importantly, some interesting interacting terms (e.g., the last two terms) arise. We will see they describe very fruitful and clear physics when we study ZZ parasitic couplings and  two-qubit gate error sources. 


\subsection*{The third-order contribution}

Due to the existence of Kerr terms in lab frame Hamiltonian, we expect the cross-Kerr interaction (which is related to ZZ crosstalk) term, i.e.,  $\hat{a}^{\dagger}_{q1} \hat{a}_{q1} \hat{a}^{\dagger}_{q2} \hat{a}_{q2}$, will appear when the fourth-order contribution is involved.
 
 The same procedure is used to compute the third order contribution $  [\hat{s}_1 , [\hat{s}_1 ,  [\hat{s}_1 ,  \hat{H}_{\rm Lab} ] ]] $.  As we keep in mind that the goal is to derive the cross-Kerr interaction  $\hat{a}^{\dagger}_{q1} \hat{a}_{q1} \hat{a}^{\dagger}_{q2} \hat{a}_{q2}$,  only the terms related to $\hat{a}^{\dagger}_{q1} \hat{a}_{q1} \hat{a}^{\dagger}_{q2} \hat{a}_{q2}$ are kept, while other irrelevant terms will be neglected. It is not hard to recognize that only the last three terms of Eq.~\eqref{SW:second-order} contribute effectively. Hence, we compute $ [\hat{s}_1, [\hat{s}_1,  [\hat{s}_1,  \hat{H}_{\rm Lab} ] ]]^{(k)}$ ($k$ represents the $k$-th term of Eq.~\eqref{SW:second-order} )   one by one independently.
 The 3rd, 4th, and 5th terms are obtained as follows, respectively.
 \begin{eqnarray}
  \left[\hat{s}_1 , [\hat{s}_1 ,  [\hat{s}_1 ,  \hat{H}_{\rm Lab} ] ]\right]^{(3)} &=& \frac{4 g_1 g_2 }{\Delta_1 \Delta_2} \sum_{k=1,2} \frac{ g_k  (\alpha_{qk}+\alpha_{c})}{\Delta_k } \left(\hat{a}^{\dagger}_{qk}   \hat{a}_{qk}  \hat{a}^{\dagger}_{q\bar{k}} \hat{a}_{c}  + H.c.  \right), \\
  \left[\hat{s}_1 , [\hat{s}_1 ,  [\hat{s}_1 ,  \hat{H}_{\rm Lab} ] ]\right]^{(4)} &=& \frac{2 g_1 g_2 }{\Delta_1 \Delta_2}  \sum_{k=1,2}  \frac{g_k \alpha_{qk}}{\Delta_k}  \left(\hat{a}^{\dagger}_{qk}   \hat{a}_{qk}  \hat{a}^{\dagger}_{q\bar{k}} \hat{a}_{c} +H.c.
  \right) , \\
  \left[\hat{s}_1 , [\hat{s}_1 ,  [\hat{s}_1 ,  \hat{H}_{\rm Lab} ] ]\right]^{(5)} &=& 
   8 \frac{g_1 g_2 \alpha_{c}}{\Delta_1 \Delta_2}  \sum_{k=1,2}  \frac{g_k}{\Delta_k} \left( \hat{a}^{\dagger}_{qk}  \hat{a}_{qk} \hat{a}^{\dagger}_{q\bar{k}} \hat{a}_{c} +H.c. \right).
 \end{eqnarray}
 During the derivation, we used the approximated condition $g_k/\Sigma_k \ll g_k/\vert \Delta_k \vert$, $k=1,2$, so those terms containing $(g_k/\Sigma_k)^2$ were neglected for simplification. 
 Collecting these contributed terms, the third order contribution arrives as
  \begin{equation}\label{SW:third-order}
   \left[\hat{s}_1 , [\hat{s}_1 ,  [\hat{s}_1 ,  \hat{H}_{\rm Lab} ] ]\right] = \frac{6 g_1 g_2}{\Delta_1 \Delta_2} \sum_{k=1,2} \left[\frac{g_k (\alpha_{qk}+2\alpha_{c})}{\Delta_k } \left(\hat{a}^{\dagger}_{qk}   \hat{a}_{qk}  \hat{a}^{\dagger}_{q\bar{k}} \hat{a}_{c}  + H.c. \right) \right].
 \end{equation}

\subsection*{The forth-order contribution}

Using the result of third-order contribution, we continue compute the fourth-order contribution. There are two terms in Eq.~\eqref{SW:third-order}. We compute  $ [\hat{s}_1 ,  [\hat{s}_1 , [\hat{s}_1 ,  [\hat{s}_1 ,  \hat{H}_{\rm Lab} ] ]]]^{(k)}$ ($k$ represents the $k$-th term)   independently. 
\begin{equation}
[\hat{s}_1 ,  [\hat{s}_1 , [\hat{s}_1 ,  [\hat{s}_1 ,  \hat{H}_{\rm Lab} ] ]]]^{(k)}  =12 \left(\frac{ g_1 g_2 }{\Delta_1 \Delta_2}\right)^2  \left(\alpha_{qk}+2\alpha_{c}\right)  \hat{a}^{\dagger}_{q1}   \hat{a}_{q1}  \hat{a}^{\dagger}_{q2} \hat{a}_{q2},~~ k=1,2 .
\end{equation}
Summing up the two contributed terms, we finally obtain
 \begin{equation}\label{SW:forth-order}
[\hat{s}_1 ,  [\hat{s}_1 , [\hat{s}_1 ,  [\hat{s}_1 ,  \hat{H}_{\rm Lab} ] ]]]
=12 \left(\frac{ g_1 g_2 }{\Delta_1 \Delta_2}\right)^2  \left(\alpha_{q1}+\alpha_{q2}+4\alpha_{c}\right)  \hat{a}^{\dagger}_{q1}   \hat{a}_{q1}  \hat{a}^{\dagger}_{q2} \hat{a}_{q2} .
\end{equation}
As expected, the strength for the cross-Kerr interaction is expressed in the fourth-order. Besides,  the amplitude is related to the anharmonicity of both computational qubits and coupler. 

\subsection*{The effective Hamiltonian after 1st SWT}
 
 Using the results we obtained above, including Eqs.~\eqref{circuit Hamiltonian}, \eqref{SW:first-order}, \eqref{SW:second-order}, \eqref{SW:third-order}, \eqref{SW:forth-order}, and considering the prefactors, the Hamiltonian after 1st SWT is obtained as 
\begin{eqnarray} \label{Hamiltonian: 1st SW}
  \hat{H}_{\rm eff}^{1} & \approx & \sum_{\lambda=q1,q2,c} \omega'_{\lambda} \hat{a}^{\dagger}_{\lambda} \hat{a}_{\lambda} + \frac{\alpha'_{\lambda}}{2} \hat{a}^{\dagger}_{\lambda} \hat{a}^{\dagger}_{\lambda} \hat{a}_{\lambda} \hat{a}_{\lambda} + g_{\rm eff} \left( \hat{a}^{\dagger}_{q1}  \hat{a}_{q2}  + \hat{a}_{q1}       \hat{a}^{\dagger}_{q2} - \hat{a}^{\dagger}_{q1}  \hat{a}^{\dagger}_{q2} - \hat{a}_{q1}  \hat{a}_{q2} \right)
 \nonumber\\ 
&& + \sum_{k=1,2} \frac{2 g_k^2 (\alpha_{qk}+\alpha_{c})}{\Delta_k^2}  \hat{a}^{\dagger}_{qk}   \hat{a}_{qk}   \hat{a}^{\dagger}_{c} \hat{a}_{c} - \frac{1}{2}  \frac{g_1 g_2 }{\Delta_1 \Delta_2} \sum_{k=1,2} \alpha_{qk}
 \left( \hat{a}^{\dagger}_{qk} \hat{a}_{qk} \hat{a}_{qk} \hat{a}^{\dagger}_{q\bar{k}} +  H.c.   \right) \nonumber\\
 && +  \frac{g_1 g_2 \alpha_{c}}{\Delta_1 \Delta_2} \left(\hat{a}^{\dagger}_{q1} \hat{a}^{\dagger}_{q2} \hat{a}_{c} \hat{a}_{c}    + 2 \hat{a}^{\dagger}_{q1}  \hat{a}_{q2} \hat{a}^{\dagger}_{c} \hat{a}_{c}  +H.c.  \right)   +   \frac{1}{2}  \left(\frac{ g_1 g_2 }{\Delta_1 \Delta_2}\right)^2  \left(\alpha_{q1}+\alpha_{q2}+4\alpha_{c}\right)  \hat{a}^{\dagger}_{q1}   \hat{a}_{q1}  \hat{a}^{\dagger}_{q2} \hat{a}_{q2}  \nonumber\\
   && - \sum_{k=1,2} \left[ \frac{g_k \alpha_{qk}}{\Delta_k} \left( \hat{a}^{\dagger}_{qk} \hat{a}^{\dagger}_{qk} \hat{a}_{qk} \hat{a}_{c} + H.c. \right)   + \frac{g_k \alpha_c}{\Delta_k} \left( \hat{a}^{\dagger}_{qk} \hat{a}^{\dagger}_{c} \hat{a}_{c} \hat{a}_{c} +H.c.  \right)\right],
\end{eqnarray}
where the shifted qubit frequencies  are 
\begin{equation}
{\omega}'_{qk} = {\omega}_{qk} +  \frac{ g^2_k}{\Delta_k} - \frac{ g^2_k}{\Sigma_k} , ~ {\omega}'_{c} = {\omega}_{c} - \sum_{k=1,2} \left( \frac{ g^2_k}{\Delta_k} + \frac{ g^2_k}{\Sigma_k}  \right),
\end{equation}
and the shifted anharmonicity are 
\begin{equation}
\alpha'_{qk} = \alpha_{qk} \left(1 - 2\frac{g_k^2}{\Delta_k^2} \right),  ~\alpha'_{c} = \alpha_{c} \left[ 1-2 \left(\frac{g_1^2}{\Delta_1^2} + \frac{g_2^2}{\Delta_2^2} \right) \right] .
\end{equation}
Moreover, the effective coupling between the two computational qubits are 
\begin{equation}
{g}_{\rm eff} = {g}_{12} + \frac{g_1 g_2}{2} \left(\frac{1}{\Delta_1}+\frac{1}{\Delta_2}-\frac{1}{\Sigma_1}-\frac{1}{\Sigma_2} \right).
\end{equation}

Next, let us add some remarks to these new interaction terms of Eq.~\eqref{SW:second-order}.  First of all, the resonant terms $ \hat{a}^{\dagger}_{q1}   \hat{a}_{q1}   \hat{a}^{\dagger}_{c} \hat{a}_{c} $, $ \hat{a}^{\dagger}_{q2}   \hat{a}_{q2}   \hat{a}^{\dagger}_{c} \hat{a}_{c} $, $ \hat{a}^{\dagger}_{q1}   \hat{a}_{q2}   \hat{a}^{\dagger}_{c} \hat{a}_{c} $, and $ \hat{a}^{\dagger}_{q2}   \hat{a}_{q1}   \hat{a}^{\dagger}_{c} \hat{a}_{c} $ do not 
contribute at the end, because the coupler stays in ground sates all the time. Secondly, the last non-resonant terms generated from the first-order contributions 
indeed can be ignored with considering $\alpha_{\lambda}\ll \vert \Delta_{k} \vert$, $\lambda=q1, q2, c$, and $k=1,2$. However, they should be kept
 when the condition does not hold, e.g., $\alpha_{\lambda} \sim  \vert \Delta_{k} \vert$. 
 Later, we will see they will bring additional energy level shifts. 
 Thirdly, apart from these terms discussed, other interacting terms can be interpreted as parasitic couplings, which are discussed in the main text.  Finally, noted that during the whole derivation, the only approximated conditions we used are: $g_k/\vert \Delta_k \vert \ll1$ and  $(g_k/\sum_k)^2  \ll  (g_k/ \vert \Delta_k \vert)^2 $. 
 
 \subsection{2nd SWT}
 
The effective Hamiltonian after 1st SWT can be further simplified via performing a second SWT. The goal is to eliminate the last non-resonant term of Eq.~\eqref{Hamiltonian: 1st SW} .  
Introducing $\hat{s}_2$ with
\begin{eqnarray}
\hat{s}_2 &=&  - \frac{g_1 \alpha_{q1}}{\Delta_1 \left(\Delta'_{1} + \alpha'_{q1} \right) }  \left( \hat{a}^{\dagger}_{q1} \hat{a}^{\dagger}_{q1} \hat{a}_{q1} \hat{a}_{c} - \hat{a}^{\dagger}_{q1} \hat{a}_{q1} \hat{a}_{q1} \hat{a}^{\dagger}_{c} \right)   - \frac{g_2 \alpha_{q2}}{\Delta_2 \left(\Delta'_{2} + \alpha'_{q2} \right) }  \left( \hat{a}^{\dagger}_{q2} \hat{a}^{\dagger}_{q2} \hat{a}_{q2} \hat{a}_{c} - \hat{a}^{\dagger}_{q2} \hat{a}_{q2} \hat{a}_{q2} \hat{a}^{\dagger}_{c} \right) \nonumber\\
&& - \frac{g_1 \alpha_c}{\Delta_1 \left(\Delta'_{1} - \alpha'_c \right) } \left( \hat{a}^{\dagger}_{q1} \hat{a}^{\dagger}_{c} \hat{a}_{c} \hat{a}_{c} - \hat{a}_{q1}  \hat{a}^{\dagger}_{c} \hat{a}^{\dagger}_{c} \hat{a}_{c}  \right) - \frac{g_2 \alpha_c}{\Delta_2 \left(\Delta'_{2} - \alpha'_c \right) } \left( \hat{a}^{\dagger}_{q2} \hat{a}^{\dagger}_{c} \hat{a}_{c} \hat{a}_{c} - \hat{a}_{q2}  \hat{a}^{\dagger}_{c} \hat{a}^{\dagger}_{c} \hat{a}_{c}  \right),
\end{eqnarray}
where $\Delta'_{k}=\omega'_{qk}-\omega'_c$, $k=1,2$. Applying the 2nd SWT, i.e., the Hamiltonian is transformed to 
$
\hat{H}_{\rm eff}^2 = e^{\hat{s}_2} \hat{H}^1_{\rm eff} e^{- \hat{s}_2} = \hat{H}^1_{\rm eff}  + [\hat{s}_2, \hat{H}^1_{\rm eff} ] +  [\hat{s}_2, [\hat{s}_2, \hat{H}^1_{\rm eff}  ] ]/2! + \cdots,
$
where the 1st SWT result $\hat{H}_{\rm eff}^1$ was given in Eq.~\eqref{Hamiltonian: 1st SW}.
As did in 1st SWT, we have to compute the first-order contribution and the-second contribution, respectively.

Let us first compute the first-order contribution $[\hat{s}_2, \hat{H}^1_{\rm eff} ] $. Although $\hat{H}^1_{\rm eff} $ contains many terms, most of them can be neglected due to the smallness prefactors and only few of them contributes. Besides, $\hat{s}_2$ contains four terms, we compute them separately. At the end, we obtain

\begin{eqnarray}
[\hat{s}_2, \hat{H}^1_{\rm eff}] &\simeq &    \frac{g_1 \alpha_{q1}}{\Delta_1} \left( \hat{a}^{\dagger}_{q1} \hat{a}^{\dagger}_{q1} \hat{a}_{q1} \hat{a}_{c} + \hat{a}^{\dagger}_{q1} \hat{a}_{q1} \hat{a}_{q1} \hat{a}^{\dagger}_{c} \right)  + \frac{g_2 \alpha_{q2}}{\Delta_2} \left( \hat{a}^{\dagger}_{q2} \hat{a}^{\dagger}_{q2} \hat{a}_{q2} \hat{a}_{c} + \hat{a}^{\dagger}_{q2} \hat{a}_{q2} \hat{a}_{q2} \hat{a}^{\dagger}_{c} \right) - \frac{g_1 \alpha_c}{\Delta_1} \left( \hat{a}^{\dagger}_{q1} \hat{a}^{\dagger}_{c} \hat{a}_{c} \hat{a}_{c} +\hat{a}_{q1}  \hat{a}^{\dagger}_{c} \hat{a}^{\dagger}_{c} \hat{a}_{c}  \right)  \nonumber\\
&&  - \frac{g_2 \alpha_c}{\Delta_2} \left( \hat{a}^{\dagger}_{q2} \hat{a}^{\dagger}_{c} \hat{a}_{c} \hat{a}_{c} +\hat{a}_{q2}  \hat{a}^{\dagger}_{c} \hat{a}^{\dagger}_{c} \hat{a}_{c}  \right)  + 2   \frac{g_1^2 \alpha_{q1}^2}{\Delta_1^2 \left(\Delta'_{1} + \alpha'_{q1} \right) }   \hat{a}^{\dagger}_{q1} \hat{a}^{\dagger}_{q1} \hat{a}_{q1}  \hat{a}_{q1} + 2   \frac{g_2^2 \alpha_{q2}^2}{\Delta_2^2 \left(\Delta'_{2} + \alpha'_{q2} \right) }   \hat{a}^{\dagger}_{q2} \hat{a}^{\dagger}_{q2} \hat{a}_{q2}  \hat{a}_{q2}  \nonumber\\
&& -\left[ 2   \frac{g_1^2 \alpha_c^2}{\Delta_1^2 \left(\Delta'_{1} - \alpha'_c \right) } + 2   \frac{g_2^2 \alpha_c^2}{\Delta_2^2 \left(\Delta'_{2} - \alpha'_c \right) }  \right]  \hat{a}^{\dagger}_{c} \hat{a}^{\dagger}_{c} \hat{a}_{c}  \hat{a}_{c} . 
\end{eqnarray}
Similarly, the second order contribution is computed as
\begin{eqnarray}
\left[\hat{s}_2, [\hat{s}_2, \hat{H}^1_{\rm eff}] \right] &\simeq &    -2   \frac{g_1^2 \alpha_{q1}^2}{\Delta_1^2 \left(\Delta'_{1} + \alpha'_{q1} \right) }   \hat{a}^{\dagger}_{q1} \hat{a}^{\dagger}_{q1} \hat{a}_{q1}  \hat{a}_{q1} - 2   \frac{g_2^2 \alpha_{q2}^2}{\Delta_2^2 \left(\Delta'_{2} + \alpha'_{q2} \right) }   \hat{a}^{\dagger}_{q2} \hat{a}^{\dagger}_{q2} \hat{a}_{q2}  \hat{a}_{q2} \nonumber\\
&&+ \left[ 2   \frac{g_1^2 \alpha_c^2}{\Delta_1^2 \left(\Delta'_{1} - \alpha'_c \right) } + 2   \frac{g_2^2 \alpha_c^2}{\Delta_2^2 \left(\Delta'_{2} - \alpha'_c \right) }  \right]  \hat{a}^{\dagger}_{c} \hat{a}^{\dagger}_{c} \hat{a}_{c}  \hat{a}_{c} .
\end{eqnarray}
On top of the first SWT result, summing up all of these contributions,  we ultimately obtain
\begin{eqnarray} \label{Hamiltonian: 2nd SW}
  \hat{H}^2_{\rm eff} & \approx & \sum_{\lambda=q1,q2,c} \tilde{\omega}_{\lambda} \hat{a}^{\dagger}_{\lambda} \hat{a}_{\lambda} + \frac{\tilde{\alpha}_{\lambda}}{2} \hat{a}^{\dagger}_{\lambda} \hat{a}^{\dagger}_{\lambda} \hat{a}_{\lambda} \hat{a}_{\lambda} + {g}_{\rm eff} \left( \hat{a}^{\dagger}_{q1}  \hat{a}_{q2}  - \hat{a}^{\dagger}_{q1}  \hat{a}^{\dagger}_{q2} + H.c. \right), \nonumber\\ 
 &&  - \frac{1}{2}  \frac{g_1 g_2}{\Delta_1 \Delta_2}
 \left[ \alpha_{q1}( \hat{a}^{\dagger}_{q1} \hat{a}_{q1} \hat{a}_{q1} \hat{a}^{\dagger}_{q2} +  \hat{a}^{\dagger}_{q1} \hat{a}^{\dagger}_{q1} \hat{a}_{q1}  \hat{a}_{q2} ) + \alpha_{q2}
 ( \hat{a}^{\dagger}_{q2} \hat{a}_{q2} \hat{a}_{q2} \hat{a}^{\dagger}_{q1} +  \hat{a}^{\dagger}_{q2} \hat{a}^{\dagger}_{q2} \hat{a}_{q2}  \hat{a}_{q1} )  \right]\nonumber\\
 && + \frac{g_1 g_2 \alpha_{c}}{\Delta_1 \Delta_2} \left(\hat{a}^{\dagger}_{q1} \hat{a}^{\dagger}_{q2} \hat{a}_{c} \hat{a}_{c}  + \hat{a}_{q1} \hat{a}_{q2} \hat{a}^{\dagger}_{c} \hat{a}^{\dagger}_{c}  \right)  +   \frac{1}{2}  \left(\frac{ g_1 g_2 }{\Delta_1 \Delta_2}\right)^2  \left(\alpha_{q1}+\alpha_{q2}+4\alpha_{c}\right)  \hat{a}^{\dagger}_{q1}   \hat{a}_{q1}  \hat{a}^{\dagger}_{q2} \hat{a}_{q2},
\end{eqnarray}
where the qubits' frequency are unchanged, namely $\tilde{\omega}_{\lambda} = \omega'_{\lambda}$ , and shifted anharmonicity become 
\begin{eqnarray}
\tilde{\alpha}_{qk} &=& \alpha_{qk} \left(1 - 2 \frac{g_k^2}{\Delta_k^2} + 2 \frac{g_k^2}{\Delta_k^2} \frac{\alpha_{qk}}{\Delta'_k+\alpha'_{qk}}\right) \approx \alpha_{qk} \left[1 -2 \frac{g_k^2}{\Delta_k(\Delta_k + \alpha_{qk}) } \right] , \\  
 \tilde{\alpha}_{c} &=& \alpha_{c} \left[ 1- \sum_{k=1,2} \left( 2\frac{g_k^2}{\Delta_k^2}  +  2 \frac{g_k^2 \alpha_c}{\Delta_k^2 \left(\Delta'_{k} - \alpha'_c \right) } \right)  \right] \approx \alpha_{c} \left[  1- 2 \sum_{k=1,2} \frac{g_k^2}{\Delta_k (\Delta_k - \alpha_c) }  \right].
\end{eqnarray}
 Note that the additional approximate condition we used are: $\vert \alpha_{qk}/ (\Delta_k + \alpha_{qk}) \vert \leq 1$ and $\vert  \alpha_c/(\Delta_k - \alpha_c) \vert \leq 1$, $k=1,2$. Till now, we ultimately obtain the effective Hamiltonian, which was given in Eq.~\eqref{Hamiltonian: new representation} of the main text.

\section{Analytical expressions of ZZ coupling for different regime } 
\label{ZZ coupling: Analytical expression}
 
 In this Appendix, we show the derivation of the analytical ZZ coupling expressions $\zeta_{zz}$ for different regime discussed in Sec.~\ref{sec:Physical mechanisms of parasitic couplings}. For 
 easy reference, the system Hamiltonian [i.e., Eq.~\eqref{Hamiltonian: computational basis} in the main text] containing only the terms related to ZZ couplings is written down as follows.  
 \begin{eqnarray}
 {\hat{{H}}}'_{\rm eff} &=&   (\tilde{\omega}_{q1} + \tilde{\omega}_{q2} )\vert 101 \rangle \langle 101\vert  + (2\tilde{\omega}_{q1} + \tilde{\alpha}_{q1} )\vert 200 \rangle \langle 200\vert +(2\tilde{\omega}_{q2} + \tilde{\alpha}_{q2} ) \vert 002 \rangle \langle 002 \vert   +  (2\tilde{\omega}_{c} + \tilde{\alpha}_{c} )\vert 020 \rangle \langle 020 \vert    \nonumber\\
 &&+ \tilde{g}_{200}  \left(\vert 200 \rangle \langle 101 \vert + \vert 101 \rangle \langle 200 \vert \right)  + \tilde{g}_{002}  \left( \vert 002 \rangle \langle 101 \vert +  \vert 101 \rangle \langle 002 \vert  \right)  +  \tilde{g}_{020}  \left( \vert 020 \rangle \langle 101 \vert + \vert 101 \rangle \langle 020 \vert \right)  \nonumber\\
 &&  + \tilde{g}_{\rm cross-Kerr}   \left(\alpha_{q1}+\alpha_{q2}+4\alpha_{c}\right)  \vert 101 \rangle \langle 101 \vert ,
\end{eqnarray} 

where those irrelevant terms with ZZ coupling were neglected. The explicit expressions of the coupling strengths $ \tilde{g}_{200}$, $ \tilde{g}_{002}$, $ \tilde{g}_{020}$,   and  $\tilde{g}_{\rm cross-Kerr} $ were given in Eq.~\eqref{expressions: coupling strength}. With the help of this effective Hamiltonian, we are able to derive the analytical expression of ZZ couplings for different parameter regime. 

\subsection{Parasitic couplings due to high energy levels of computational qubits  }

When we pay specific attention to the effects of computational qubits' high energy levels, namely concentrating on the regime either $\tilde{\omega}_{q1} + \tilde{\omega}_{q2} \approx 2\tilde{\omega}_{q1} + \tilde{\alpha}_{q1}$ or $\tilde{\omega}_{q1} + \tilde{\omega}_{q2} \approx 2\tilde{\omega}_{q2} + \tilde{\alpha}_{q2}$, the resonant couplings between  $\vert 101 \rangle$ and  $\vert 200 \rangle$ (or $\vert 002 \rangle$) play a vital role.  As a consequence, the effective Hamiltonian in matrix form reduces to
\begin{eqnarray}
 \hat{{H}}^{200/002}_{\rm eff} \approx 
\begin{blockarray}{cccc}
& \vert 101 \rangle &   \vert 200 \rangle  &  \vert 002 \rangle  \cr
\begin{block}{c(ccc)}
  \vert 101 \rangle ~ & \tilde{\omega}_{q1} + \tilde{\omega}_{q2} &  \tilde{g}_{200}  &  \tilde{g}_{002} \cr
 \vert 200 \rangle ~& \tilde{g}_{200} &  2\tilde{\omega}_{q1} + \tilde{\alpha}_{q1} & 0   \cr
 \vert 002 \rangle ~ & \tilde{g}_{002} &  0 & 2\tilde{\omega}_{q2} + \tilde{\alpha}_{q2}  \cr
\end{block}
\end{blockarray} ~.
\end{eqnarray}

To compute the ZZ coupling strength $\zeta_{zz} = \tilde{\omega}_{101} -\tilde{\omega}_{100} -\tilde{\omega}_{001}$ ($\tilde{\omega}_{q1,c,q2}$ denotes the eigenenergy of the effective Hamiltonian), the key step is figuring out $\tilde{\omega}_{101} $ which are affected by the nearly resonant couplings with the states $\vert 200 \rangle$ or $\vert 002 \rangle$. In particular, the energy shift of $\tilde{\omega}_{101}$ is calculated by diagonalizing the matrix above. Ultimately, we obtain the resulting ZZ coupling expression, which is present in Eq.~\eqref{analytic:101 and 200} of the main text.

\subsection{Parasitic couplings due to high energy levels of the coupler}

Next, we turn to  consider the effects of the coupler's high energy levels, namely concentrating on the regime $\tilde{\omega}_{q1} + \tilde{\omega}_{q2} \approx 2\tilde{\omega}_{c} + \tilde{\alpha}_{c} $. 
Under this regime, we mainly concentrate on the resonant coupling between  $\vert 101 \rangle$ and $\vert 020 \rangle$.  As a consequence, the effective Hamiltonian becomes
\begin{eqnarray}
 \hat{{H}}^{020}_{\rm eff} \approx 
\begin{blockarray}{ccc}
& \vert 101 \rangle~ &   \vert 020 \rangle   \cr
\begin{block}{c(cc)}
  \vert 101 \rangle~ & \tilde{\omega}_{q1} + \tilde{\omega}_{q2} &  \tilde{g}_{020} \cr
 \vert 020 \rangle~ & \tilde{g}_{020}  & 2\tilde{\omega}_{c} + \tilde{\alpha}_{c}   \cr
\end{block}
\end{blockarray} ~.
\end{eqnarray}

As before, the critical step is to calculate the frequency shift of $\tilde{\omega}_{101} $. In particular, we figure it out via diagonalizing the equation above. Ultimately, we obtain the resulting ZZ coupling expression, which is present in Eq.~\eqref{analytic:101 and 020} of the main text.

\subsection{Parasitic couplings in dispersive regime}

In dispersive regime, we have to consider the couplings between  $\vert 101 \rangle$ and those states out of the computational space, including $\vert 200 \rangle$,  $\vert 002 \rangle$, $\vert 020 \rangle$.  As a consequence, the effective Hamiltonian in matrix form is given by
\begin{eqnarray}
 \hat{{H}}^{\rm dispersive}_{\rm eff} \approx 
\begin{blockarray}{ccccc}
& \vert 101 \rangle &   \vert 200 \rangle  &  \vert 002 \rangle & \vert 020 \rangle \cr
\begin{block}{c(cccc)}
  \vert 101 \rangle~ & \tilde{\omega}_{q1} + \tilde{\omega}_{q2} &  \tilde{g}_{200}  &  \tilde{g}_{002} &  \tilde{g}_{020} \cr
 \vert 200 \rangle~ & \tilde{g}_{200} &  2\tilde{\omega}_{q1} + \tilde{\alpha}_{q1}   & 0 & 0  \cr
 \vert 002 \rangle~ & \tilde{g}_{002} &  0 & 2\tilde{\omega}_{q2} + \tilde{\alpha}_{q2}    & 0 \cr
  \vert 020 \rangle~ &  \tilde{g}_{020} &  0 & 0 &2\tilde{\omega}_{c} + \tilde{\alpha}_{c}   \cr
\end{block}
\end{blockarray} ~~.
\end{eqnarray}

In the equation above, the coupling strengths satisfy: $\tilde{g}_{200} \ll \vert (\tilde{\omega}_{q1} + \tilde{\omega}_{q2})- (2\tilde{\omega}_{q1} + \tilde{\alpha}_{q1})  \vert$, $\tilde{g}_{020} \ll \vert (\tilde{\omega}_{q1} + \tilde{\omega}_{q2})- (2\tilde{\omega}_{c} + \tilde{\alpha}_{c})  \vert$, and $\tilde{g}_{002} \ll \vert (\tilde{\omega}_{q1} + \tilde{\omega}_{q2})- (2\tilde{\omega}_{q2} + \tilde{\alpha}_{q2})  \vert$. 
ZZ coupling contributed from each type of coupling can be solved independently via diagonalizing analytically the couplings between $\vert 101 \rangle$ and $\vert 200 \rangle$, $\vert 101 \rangle$ and $\vert 002 \rangle$, and $\vert 101 \rangle$ and $\vert 020 \rangle$, respectively. Ultimately, summing up different contributions, we obtain approximately
\begin{eqnarray}\label{analytic:general regime}
\zeta_{zz} & \approx &  \frac{\tilde{g}_{ 200 }^2}{(\tilde{\omega}_{q1} + \tilde{\omega}_{q2})- (2\tilde{\omega}_{q1} + \tilde{\alpha}_{q1})}  
 +\frac{\tilde{g}_{ 002 }^2}{(\tilde{\omega}_{q1} + \tilde{\omega}_{q2})- (2\tilde{\omega}_{q2} + \tilde{\alpha}_{q2})} 
 +  \frac{\tilde{g}_{ 020 }^2}{(\tilde{\omega}_{q1} + \tilde{\omega}_{q2})- (2\tilde{\omega}_{c} + \tilde{\alpha}_{c})}   + \zeta_{zz}^{\rm cross-Kerr}
 \nonumber\\
&\approx & -2  \left[\frac{(g_{12} + {g_1 g_2}/{\Delta_2})^2}{\alpha_{q1}+ \Delta_{12}} +\frac{(g_{12} + {g_1 g_2}/{\Delta_1} )^2}{\alpha_{q2}-  \Delta_{12}}  +  \frac{[g_1 g_2(1/\Delta_1 + 1/\Delta_2)]^2}{ \alpha_c - \Delta_1 - \Delta_2} -  \frac{2g_{12}g_1 g_2}{\Delta_1 \Delta_2} 
 \right],
\end{eqnarray}
where we used the approximated conditions $\tilde{\alpha}_{\lambda} \approx {\alpha}_{\lambda}$ ($\lambda=q1,c,q2$), $\tilde{\Delta}_{k} \approx {\Delta}_{k}$ ($k=1,2$), and $\Delta_{12} =\Delta_1 -\Delta_2$ is the frequency detuning between computational qubits. This result would be useful for exploring ZZ coupling characteristics. Considering $\Delta_1 = \Delta_2$ and rewriting the expression in terms of $g_{\rm eff}$ and $\tilde{g}_{12}$, the analytical result above reduces to Eq.~\eqref{zz coupling: analytic} of the main text. Note that  the analytical result of ZZ parasitic coupling valid for more general regimes comparing with previous investigations \cite{yan2018tunable, li2020tunable, zhao2020high,zhao2020switchable,mundada2019suppression};

In absence of the direct coupling $g_{12}$, the result above can be simplified. It becomes
\begin{eqnarray} \label{analytic:dispersive:general regime}
\zeta_{zz}  &=& - 2 g^2_1 g^2_2 \left[\frac{1}{\Delta^2_2} \frac{1}{\alpha_{q1} + \Delta_{12}}  + \frac{1}{\Delta^2_1} \frac{1}{\alpha_{q2} - \Delta_{12}}+ \left(\frac{1}{\Delta_1} + \frac{1}{\Delta_2} \right)^2 \frac{1}{\alpha_c - \Delta_1 - \Delta_2} \right].
\end{eqnarray}
Actually, a similar result was also referred in previous work \cite{zhao2020high,zhao2020suppression} which was obtained using perturbation analysis. Comparing with the previous method used, we give not only the origin for each term but also the clear physical mechanisms.  

As a further step, if we focus on the resonant case with $\omega_{q1}=\omega_{q2}$, namely $\Delta_{12}=0$, then the analytical expression reduces to a simple form [Eq.~\eqref{analytic:dispersive} of the main text]:
\begin{equation}
\zeta_{zz}  = -2  \frac{g_1^2 g_2^2}{\Delta^2} \left(\frac{1}{\alpha_{q1}} + \frac{1}{\alpha_{q2}}
+ \frac{4}{\alpha_{c}-2 \Delta} \right),
\end{equation}
where we take $\Delta_1=\Delta_2=\Delta$. This results can be used to explain the physical mechanism for the elimination of ZZ parasitic couplings, and more importantly trigger some novel parameter regions in which high-fidelity two-qubit gates are expected.

\end{widetext}

\bibliography{ref}

\begin{thebibliography}{65}%
\makeatletter
\providecommand \@ifxundefined [1]{%
 \@ifx{#1\undefined}
}%
\providecommand \@ifnum [1]{%
 \ifnum #1\expandafter \@firstoftwo
 \else \expandafter \@secondoftwo
 \fi
}%
\providecommand \@ifx [1]{%
 \ifx #1\expandafter \@firstoftwo
 \else \expandafter \@secondoftwo
 \fi
}%
\providecommand \natexlab [1]{#1}%
\providecommand \enquote  [1]{``#1''}%
\providecommand \bibnamefont  [1]{#1}%
\providecommand \bibfnamefont [1]{#1}%
\providecommand \citenamefont [1]{#1}%
\providecommand \href@noop [0]{\@secondoftwo}%
\providecommand \href [0]{\begingroup \@sanitize@url \@href}%
\providecommand \@href[1]{\@@startlink{#1}\@@href}%
\providecommand \@@href[1]{\endgroup#1\@@endlink}%
\providecommand \@sanitize@url [0]{\catcode `\\12\catcode `\$12\catcode
  `\&12\catcode `\#12\catcode `\^12\catcode `\_12\catcode `\%12\relax}%
\providecommand \@@startlink[1]{}%
\providecommand \@@endlink[0]{}%
\providecommand \url  [0]{\begingroup\@sanitize@url \@url }%
\providecommand \@url [1]{\endgroup\@href {#1}{\urlprefix }}%
\providecommand \urlprefix  [0]{URL }%
\providecommand \Eprint [0]{\href }%
\providecommand \doibase [0]{https://doi.org/}%
\providecommand \selectlanguage [0]{\@gobble}%
\providecommand \bibinfo  [0]{\@secondoftwo}%
\providecommand \bibfield  [0]{\@secondoftwo}%
\providecommand \translation [1]{[#1]}%
\providecommand \BibitemOpen [0]{}%
\providecommand \bibitemStop [0]{}%
\providecommand \bibitemNoStop [0]{.\EOS\space}%
\providecommand \EOS [0]{\spacefactor3000\relax}%
\providecommand \BibitemShut  [1]{\csname bibitem#1\endcsname}%
\let\auto@bib@innerbib\@empty
\bibitem [{\citenamefont {Gambetta}\ \emph {et~al.}(2017)\citenamefont
  {Gambetta}, \citenamefont {Chow},\ and\ \citenamefont
  {Steffen}}]{gambetta2017building}%
  \BibitemOpen
  \bibfield  {author} {\bibinfo {author} {\bibfnamefont {J.~M.}\ \bibnamefont
  {Gambetta}}, \bibinfo {author} {\bibfnamefont {J.~M.}\ \bibnamefont {Chow}},\
  and\ \bibinfo {author} {\bibfnamefont {M.}~\bibnamefont {Steffen}},\
  }\bibfield  {title} {\bibinfo {title} {Building logical qubits in a
  superconducting quantum computing system},\ }\href@noop {} {\bibfield
  {journal} {\bibinfo  {journal} {npj Quantum Information}\ }\textbf {\bibinfo
  {volume} {3}},\ \bibinfo {pages} {1} (\bibinfo {year} {2017})}\BibitemShut
  {NoStop}%
\bibitem [{\citenamefont {Barends}\ \emph {et~al.}(2016)\citenamefont
  {Barends}, \citenamefont {Shabani}, \citenamefont {Lamata}, \citenamefont
  {Kelly}, \citenamefont {Mezzacapo}, \citenamefont {Las~Heras}, \citenamefont
  {Babbush}, \citenamefont {Fowler}, \citenamefont {Campbell}, \citenamefont
  {Chen} \emph {et~al.}}]{barends2016digitized}%
  \BibitemOpen
  \bibfield  {author} {\bibinfo {author} {\bibfnamefont {R.}~\bibnamefont
  {Barends}}, \bibinfo {author} {\bibfnamefont {A.}~\bibnamefont {Shabani}},
  \bibinfo {author} {\bibfnamefont {L.}~\bibnamefont {Lamata}}, \bibinfo
  {author} {\bibfnamefont {J.}~\bibnamefont {Kelly}}, \bibinfo {author}
  {\bibfnamefont {A.}~\bibnamefont {Mezzacapo}}, \bibinfo {author}
  {\bibfnamefont {U.}~\bibnamefont {Las~Heras}}, \bibinfo {author}
  {\bibfnamefont {R.}~\bibnamefont {Babbush}}, \bibinfo {author} {\bibfnamefont
  {A.~G.}\ \bibnamefont {Fowler}}, \bibinfo {author} {\bibfnamefont
  {B.}~\bibnamefont {Campbell}}, \bibinfo {author} {\bibfnamefont
  {Y.}~\bibnamefont {Chen}}, \emph {et~al.},\ }\bibfield  {title} {\bibinfo
  {title} {Digitized adiabatic quantum computing with a superconducting
  circuit},\ }\href@noop {} {\bibfield  {journal} {\bibinfo  {journal}
  {Nature}\ }\textbf {\bibinfo {volume} {534}},\ \bibinfo {pages} {222}
  (\bibinfo {year} {2016})}\BibitemShut {NoStop}%
\bibitem [{\citenamefont {Wendin}(2017)}]{wendin2017quantum}%
  \BibitemOpen
  \bibfield  {author} {\bibinfo {author} {\bibfnamefont {G.}~\bibnamefont
  {Wendin}},\ }\bibfield  {title} {\bibinfo {title} {Quantum information
  processing with superconducting circuits: a review},\ }\href@noop {}
  {\bibfield  {journal} {\bibinfo  {journal} {Reports on Progress in Physics}\
  }\textbf {\bibinfo {volume} {80}},\ \bibinfo {pages} {106001} (\bibinfo
  {year} {2017})}\BibitemShut {NoStop}%
\bibitem [{\citenamefont {Arute}\ \emph {et~al.}(2019)\citenamefont {Arute},
  \citenamefont {Arya}, \citenamefont {Babbush}, \citenamefont {Bacon},
  \citenamefont {Bardin}, \citenamefont {Barends}, \citenamefont {Biswas},
  \citenamefont {Boixo}, \citenamefont {Brandao}, \citenamefont {Buell} \emph
  {et~al.}}]{arute2019quantum}%
  \BibitemOpen
  \bibfield  {author} {\bibinfo {author} {\bibfnamefont {F.}~\bibnamefont
  {Arute}}, \bibinfo {author} {\bibfnamefont {K.}~\bibnamefont {Arya}},
  \bibinfo {author} {\bibfnamefont {R.}~\bibnamefont {Babbush}}, \bibinfo
  {author} {\bibfnamefont {D.}~\bibnamefont {Bacon}}, \bibinfo {author}
  {\bibfnamefont {J.~C.}\ \bibnamefont {Bardin}}, \bibinfo {author}
  {\bibfnamefont {R.}~\bibnamefont {Barends}}, \bibinfo {author} {\bibfnamefont
  {R.}~\bibnamefont {Biswas}}, \bibinfo {author} {\bibfnamefont
  {S.}~\bibnamefont {Boixo}}, \bibinfo {author} {\bibfnamefont {F.~G.}\
  \bibnamefont {Brandao}}, \bibinfo {author} {\bibfnamefont {D.~A.}\
  \bibnamefont {Buell}}, \emph {et~al.},\ }\bibfield  {title} {\bibinfo {title}
  {Quantum supremacy using a programmable superconducting processor},\
  }\href@noop {} {\bibfield  {journal} {\bibinfo  {journal} {Nature}\ }\textbf
  {\bibinfo {volume} {574}},\ \bibinfo {pages} {505} (\bibinfo {year}
  {2019})}\BibitemShut {NoStop}%
\bibitem [{\citenamefont {Kjaergaard}\ \emph {et~al.}(2020)\citenamefont
  {Kjaergaard}, \citenamefont {Schwartz}, \citenamefont {Braum{\"u}ller},
  \citenamefont {Krantz}, \citenamefont {Wang}, \citenamefont {Gustavsson},\
  and\ \citenamefont {Oliver}}]{kjaergaard2020superconducting}%
  \BibitemOpen
  \bibfield  {author} {\bibinfo {author} {\bibfnamefont {M.}~\bibnamefont
  {Kjaergaard}}, \bibinfo {author} {\bibfnamefont {M.~E.}\ \bibnamefont
  {Schwartz}}, \bibinfo {author} {\bibfnamefont {J.}~\bibnamefont
  {Braum{\"u}ller}}, \bibinfo {author} {\bibfnamefont {P.}~\bibnamefont
  {Krantz}}, \bibinfo {author} {\bibfnamefont {J.~I.-J.}\ \bibnamefont {Wang}},
  \bibinfo {author} {\bibfnamefont {S.}~\bibnamefont {Gustavsson}},\ and\
  \bibinfo {author} {\bibfnamefont {W.~D.}\ \bibnamefont {Oliver}},\ }\bibfield
   {title} {\bibinfo {title} {Superconducting qubits: Current state of play},\
  }\href@noop {} {\bibfield  {journal} {\bibinfo  {journal} {Annual Review of
  Condensed Matter Physics}\ }\textbf {\bibinfo {volume} {11}},\ \bibinfo
  {pages} {369} (\bibinfo {year} {2020})}\BibitemShut {NoStop}%
\bibitem [{\citenamefont {Gong}\ \emph {et~al.}(2021)\citenamefont {Gong},
  \citenamefont {Wang}, \citenamefont {Zha}, \citenamefont {Chen},
  \citenamefont {Huang}, \citenamefont {Wu}, \citenamefont {Zhu}, \citenamefont
  {Zhao}, \citenamefont {Li}, \citenamefont {Guo} \emph
  {et~al.}}]{gong2021quantum}%
  \BibitemOpen
  \bibfield  {author} {\bibinfo {author} {\bibfnamefont {M.}~\bibnamefont
  {Gong}}, \bibinfo {author} {\bibfnamefont {S.}~\bibnamefont {Wang}}, \bibinfo
  {author} {\bibfnamefont {C.}~\bibnamefont {Zha}}, \bibinfo {author}
  {\bibfnamefont {M.-C.}\ \bibnamefont {Chen}}, \bibinfo {author}
  {\bibfnamefont {H.-L.}\ \bibnamefont {Huang}}, \bibinfo {author}
  {\bibfnamefont {Y.}~\bibnamefont {Wu}}, \bibinfo {author} {\bibfnamefont
  {Q.}~\bibnamefont {Zhu}}, \bibinfo {author} {\bibfnamefont {Y.}~\bibnamefont
  {Zhao}}, \bibinfo {author} {\bibfnamefont {S.}~\bibnamefont {Li}}, \bibinfo
  {author} {\bibfnamefont {S.}~\bibnamefont {Guo}}, \emph {et~al.},\ }\bibfield
   {title} {\bibinfo {title} {Quantum walks on a programmable two-dimensional
  62-qubit superconducting processor},\ }\href@noop {} {\bibfield  {journal}
  {\bibinfo  {journal} {arXiv:2102.02573}\ } (\bibinfo {year}
  {2021})}\BibitemShut {NoStop}%
\bibitem [{\citenamefont {Jurcevic}\ \emph {et~al.}(2021)\citenamefont
  {Jurcevic}, \citenamefont {Javadi-Abhari}, \citenamefont {Bishop},
  \citenamefont {Lauer}, \citenamefont {Bogorin}, \citenamefont {Brink},
  \citenamefont {Capelluto}, \citenamefont {G{\"u}nl{\"u}k}, \citenamefont
  {Itoko}, \citenamefont {Kanazawa} \emph
  {et~al.}}]{jurcevic2021demonstration}%
  \BibitemOpen
  \bibfield  {author} {\bibinfo {author} {\bibfnamefont {P.}~\bibnamefont
  {Jurcevic}}, \bibinfo {author} {\bibfnamefont {A.}~\bibnamefont
  {Javadi-Abhari}}, \bibinfo {author} {\bibfnamefont {L.~S.}\ \bibnamefont
  {Bishop}}, \bibinfo {author} {\bibfnamefont {I.}~\bibnamefont {Lauer}},
  \bibinfo {author} {\bibfnamefont {D.~F.}\ \bibnamefont {Bogorin}}, \bibinfo
  {author} {\bibfnamefont {M.}~\bibnamefont {Brink}}, \bibinfo {author}
  {\bibfnamefont {L.}~\bibnamefont {Capelluto}}, \bibinfo {author}
  {\bibfnamefont {O.}~\bibnamefont {G{\"u}nl{\"u}k}}, \bibinfo {author}
  {\bibfnamefont {T.}~\bibnamefont {Itoko}}, \bibinfo {author} {\bibfnamefont
  {N.}~\bibnamefont {Kanazawa}}, \emph {et~al.},\ }\bibfield  {title} {\bibinfo
  {title} {Demonstration of quantum volume 64 on a superconducting quantum
  computing system},\ }\href@noop {} {\bibfield  {journal} {\bibinfo  {journal}
  {Quantum Science and Technology}\ }\textbf {\bibinfo {volume} {6}},\ \bibinfo
  {pages} {025020} (\bibinfo {year} {2021})}\BibitemShut {NoStop}%
\bibitem [{\citenamefont {O’Malley}\ \emph {et~al.}(2016)\citenamefont
  {O’Malley}, \citenamefont {Babbush}, \citenamefont {Kivlichan},
  \citenamefont {Romero}, \citenamefont {McClean}, \citenamefont {Barends},
  \citenamefont {Kelly}, \citenamefont {Roushan}, \citenamefont {Tranter},
  \citenamefont {Ding} \emph {et~al.}}]{o2016scalable}%
  \BibitemOpen
  \bibfield  {author} {\bibinfo {author} {\bibfnamefont {P.~J.}\ \bibnamefont
  {O’Malley}}, \bibinfo {author} {\bibfnamefont {R.}~\bibnamefont {Babbush}},
  \bibinfo {author} {\bibfnamefont {I.~D.}\ \bibnamefont {Kivlichan}}, \bibinfo
  {author} {\bibfnamefont {J.}~\bibnamefont {Romero}}, \bibinfo {author}
  {\bibfnamefont {J.~R.}\ \bibnamefont {McClean}}, \bibinfo {author}
  {\bibfnamefont {R.}~\bibnamefont {Barends}}, \bibinfo {author} {\bibfnamefont
  {J.}~\bibnamefont {Kelly}}, \bibinfo {author} {\bibfnamefont
  {P.}~\bibnamefont {Roushan}}, \bibinfo {author} {\bibfnamefont
  {A.}~\bibnamefont {Tranter}}, \bibinfo {author} {\bibfnamefont
  {N.}~\bibnamefont {Ding}}, \emph {et~al.},\ }\bibfield  {title} {\bibinfo
  {title} {Scalable quantum simulation of molecular energies},\ }\href@noop {}
  {\bibfield  {journal} {\bibinfo  {journal} {Physical Review X}\ }\textbf
  {\bibinfo {volume} {6}},\ \bibinfo {pages} {031007} (\bibinfo {year}
  {2016})}\BibitemShut {NoStop}%
\bibitem [{\citenamefont {Kandala}\ \emph {et~al.}(2017)\citenamefont
  {Kandala}, \citenamefont {Mezzacapo}, \citenamefont {Temme}, \citenamefont
  {Takita}, \citenamefont {Brink}, \citenamefont {Chow},\ and\ \citenamefont
  {Gambetta}}]{kandala2017hardware}%
  \BibitemOpen
  \bibfield  {author} {\bibinfo {author} {\bibfnamefont {A.}~\bibnamefont
  {Kandala}}, \bibinfo {author} {\bibfnamefont {A.}~\bibnamefont {Mezzacapo}},
  \bibinfo {author} {\bibfnamefont {K.}~\bibnamefont {Temme}}, \bibinfo
  {author} {\bibfnamefont {M.}~\bibnamefont {Takita}}, \bibinfo {author}
  {\bibfnamefont {M.}~\bibnamefont {Brink}}, \bibinfo {author} {\bibfnamefont
  {J.~M.}\ \bibnamefont {Chow}},\ and\ \bibinfo {author} {\bibfnamefont
  {J.~M.}\ \bibnamefont {Gambetta}},\ }\bibfield  {title} {\bibinfo {title}
  {Hardware-efficient variational quantum eigensolver for small molecules and
  quantum magnets},\ }\href@noop {} {\bibfield  {journal} {\bibinfo  {journal}
  {Nature}\ }\textbf {\bibinfo {volume} {549}},\ \bibinfo {pages} {242}
  (\bibinfo {year} {2017})}\BibitemShut {NoStop}%
\bibitem [{\citenamefont {Havl{\'\i}{\v{c}}ek}\ \emph
  {et~al.}(2019)\citenamefont {Havl{\'\i}{\v{c}}ek}, \citenamefont
  {C{\'o}rcoles}, \citenamefont {Temme}, \citenamefont {Harrow}, \citenamefont
  {Kandala}, \citenamefont {Chow},\ and\ \citenamefont
  {Gambetta}}]{havlivcek2019supervised}%
  \BibitemOpen
  \bibfield  {author} {\bibinfo {author} {\bibfnamefont {V.}~\bibnamefont
  {Havl{\'\i}{\v{c}}ek}}, \bibinfo {author} {\bibfnamefont {A.~D.}\
  \bibnamefont {C{\'o}rcoles}}, \bibinfo {author} {\bibfnamefont
  {K.}~\bibnamefont {Temme}}, \bibinfo {author} {\bibfnamefont {A.~W.}\
  \bibnamefont {Harrow}}, \bibinfo {author} {\bibfnamefont {A.}~\bibnamefont
  {Kandala}}, \bibinfo {author} {\bibfnamefont {J.~M.}\ \bibnamefont {Chow}},\
  and\ \bibinfo {author} {\bibfnamefont {J.~M.}\ \bibnamefont {Gambetta}},\
  }\bibfield  {title} {\bibinfo {title} {Supervised learning with
  quantum-enhanced feature spaces},\ }\href@noop {} {\bibfield  {journal}
  {\bibinfo  {journal} {Nature}\ }\textbf {\bibinfo {volume} {567}},\ \bibinfo
  {pages} {209} (\bibinfo {year} {2019})}\BibitemShut {NoStop}%
\bibitem [{\citenamefont {Yordanov}\ \emph {et~al.}(2020)\citenamefont
  {Yordanov}, \citenamefont {Arvidsson-Shukur},\ and\ \citenamefont
  {Barnes}}]{yordanov2020efficient}%
  \BibitemOpen
  \bibfield  {author} {\bibinfo {author} {\bibfnamefont {Y.~S.}\ \bibnamefont
  {Yordanov}}, \bibinfo {author} {\bibfnamefont {D.~R.}\ \bibnamefont
  {Arvidsson-Shukur}},\ and\ \bibinfo {author} {\bibfnamefont {C.~H.}\
  \bibnamefont {Barnes}},\ }\bibfield  {title} {\bibinfo {title} {Efficient
  quantum circuits for quantum computational chemistry},\ }\href@noop {}
  {\bibfield  {journal} {\bibinfo  {journal} {Physical Review A}\ }\textbf
  {\bibinfo {volume} {102}},\ \bibinfo {pages} {062612} (\bibinfo {year}
  {2020})}\BibitemShut {NoStop}%
\bibitem [{\citenamefont {Harrigan}\ \emph {et~al.}(2021)\citenamefont
  {Harrigan}, \citenamefont {Sung}, \citenamefont {Neeley}, \citenamefont
  {Satzinger}, \citenamefont {Arute}, \citenamefont {Arya}, \citenamefont
  {Atalaya}, \citenamefont {Bardin}, \citenamefont {Barends}, \citenamefont
  {Boixo} \emph {et~al.}}]{harrigan2021quantum}%
  \BibitemOpen
  \bibfield  {author} {\bibinfo {author} {\bibfnamefont {M.~P.}\ \bibnamefont
  {Harrigan}}, \bibinfo {author} {\bibfnamefont {K.~J.}\ \bibnamefont {Sung}},
  \bibinfo {author} {\bibfnamefont {M.}~\bibnamefont {Neeley}}, \bibinfo
  {author} {\bibfnamefont {K.~J.}\ \bibnamefont {Satzinger}}, \bibinfo {author}
  {\bibfnamefont {F.}~\bibnamefont {Arute}}, \bibinfo {author} {\bibfnamefont
  {K.}~\bibnamefont {Arya}}, \bibinfo {author} {\bibfnamefont {J.}~\bibnamefont
  {Atalaya}}, \bibinfo {author} {\bibfnamefont {J.~C.}\ \bibnamefont {Bardin}},
  \bibinfo {author} {\bibfnamefont {R.}~\bibnamefont {Barends}}, \bibinfo
  {author} {\bibfnamefont {S.}~\bibnamefont {Boixo}}, \emph {et~al.},\
  }\bibfield  {title} {\bibinfo {title} {Quantum approximate optimization of
  non-planar graph problems on a planar superconducting processor},\
  }\href@noop {} {\bibfield  {journal} {\bibinfo  {journal} {Nature Physics}\
  }\textbf {\bibinfo {volume} {17}},\ \bibinfo {pages} {332} (\bibinfo {year}
  {2021})}\BibitemShut {NoStop}%
\bibitem [{\citenamefont {Barends}\ \emph {et~al.}(2013)\citenamefont
  {Barends}, \citenamefont {Kelly}, \citenamefont {Megrant}, \citenamefont
  {Sank}, \citenamefont {Jeffrey}, \citenamefont {Chen}, \citenamefont {Yin},
  \citenamefont {Chiaro}, \citenamefont {Mutus}, \citenamefont {Neill} \emph
  {et~al.}}]{barends2013coherent}%
  \BibitemOpen
  \bibfield  {author} {\bibinfo {author} {\bibfnamefont {R.}~\bibnamefont
  {Barends}}, \bibinfo {author} {\bibfnamefont {J.}~\bibnamefont {Kelly}},
  \bibinfo {author} {\bibfnamefont {A.}~\bibnamefont {Megrant}}, \bibinfo
  {author} {\bibfnamefont {D.}~\bibnamefont {Sank}}, \bibinfo {author}
  {\bibfnamefont {E.}~\bibnamefont {Jeffrey}}, \bibinfo {author} {\bibfnamefont
  {Y.}~\bibnamefont {Chen}}, \bibinfo {author} {\bibfnamefont {Y.}~\bibnamefont
  {Yin}}, \bibinfo {author} {\bibfnamefont {B.}~\bibnamefont {Chiaro}},
  \bibinfo {author} {\bibfnamefont {J.}~\bibnamefont {Mutus}}, \bibinfo
  {author} {\bibfnamefont {C.}~\bibnamefont {Neill}}, \emph {et~al.},\
  }\bibfield  {title} {\bibinfo {title} {Coherent josephson qubit suitable for
  scalable quantum integrated circuits},\ }\href@noop {} {\bibfield  {journal}
  {\bibinfo  {journal} {Physical Review Letters}\ }\textbf {\bibinfo {volume}
  {111}},\ \bibinfo {pages} {080502} (\bibinfo {year} {2013})}\BibitemShut
  {NoStop}%
\bibitem [{\citenamefont {Johnson}\ \emph {et~al.}(2011)\citenamefont
  {Johnson}, \citenamefont {Amin}, \citenamefont {Gildert}, \citenamefont
  {Lanting}, \citenamefont {Hamze}, \citenamefont {Dickson}, \citenamefont
  {Harris}, \citenamefont {Berkley}, \citenamefont {Johansson}, \citenamefont
  {Bunyk} \emph {et~al.}}]{johnson2011quantum}%
  \BibitemOpen
  \bibfield  {author} {\bibinfo {author} {\bibfnamefont {M.~W.}\ \bibnamefont
  {Johnson}}, \bibinfo {author} {\bibfnamefont {M.~H.}\ \bibnamefont {Amin}},
  \bibinfo {author} {\bibfnamefont {S.}~\bibnamefont {Gildert}}, \bibinfo
  {author} {\bibfnamefont {T.}~\bibnamefont {Lanting}}, \bibinfo {author}
  {\bibfnamefont {F.}~\bibnamefont {Hamze}}, \bibinfo {author} {\bibfnamefont
  {N.}~\bibnamefont {Dickson}}, \bibinfo {author} {\bibfnamefont
  {R.}~\bibnamefont {Harris}}, \bibinfo {author} {\bibfnamefont {A.~J.}\
  \bibnamefont {Berkley}}, \bibinfo {author} {\bibfnamefont {J.}~\bibnamefont
  {Johansson}}, \bibinfo {author} {\bibfnamefont {P.}~\bibnamefont {Bunyk}},
  \emph {et~al.},\ }\bibfield  {title} {\bibinfo {title} {Quantum annealing
  with manufactured spins},\ }\href@noop {} {\bibfield  {journal} {\bibinfo
  {journal} {Nature}\ }\textbf {\bibinfo {volume} {473}},\ \bibinfo {pages}
  {194} (\bibinfo {year} {2011})}\BibitemShut {NoStop}%
\bibitem [{\citenamefont {Niskanen}\ \emph {et~al.}(2007)\citenamefont
  {Niskanen}, \citenamefont {Harrabi}, \citenamefont {Yoshihara}, \citenamefont
  {Nakamura}, \citenamefont {Lloyd},\ and\ \citenamefont
  {Tsai}}]{niskanen2007quantum}%
  \BibitemOpen
  \bibfield  {author} {\bibinfo {author} {\bibfnamefont {A.}~\bibnamefont
  {Niskanen}}, \bibinfo {author} {\bibfnamefont {K.}~\bibnamefont {Harrabi}},
  \bibinfo {author} {\bibfnamefont {F.}~\bibnamefont {Yoshihara}}, \bibinfo
  {author} {\bibfnamefont {Y.}~\bibnamefont {Nakamura}}, \bibinfo {author}
  {\bibfnamefont {S.}~\bibnamefont {Lloyd}},\ and\ \bibinfo {author}
  {\bibfnamefont {J.~S.}\ \bibnamefont {Tsai}},\ }\bibfield  {title} {\bibinfo
  {title} {Quantum coherent tunable coupling of superconducting qubits},\
  }\href@noop {} {\bibfield  {journal} {\bibinfo  {journal} {Science}\ }\textbf
  {\bibinfo {volume} {316}},\ \bibinfo {pages} {723} (\bibinfo {year}
  {2007})}\BibitemShut {NoStop}%
\bibitem [{\citenamefont {Barends}\ \emph {et~al.}(2014)\citenamefont
  {Barends}, \citenamefont {Kelly}, \citenamefont {Megrant}, \citenamefont
  {Veitia}, \citenamefont {Sank}, \citenamefont {Jeffrey}, \citenamefont
  {White}, \citenamefont {Mutus}, \citenamefont {Fowler}, \citenamefont
  {Campbell} \emph {et~al.}}]{barends2014superconducting}%
  \BibitemOpen
  \bibfield  {author} {\bibinfo {author} {\bibfnamefont {R.}~\bibnamefont
  {Barends}}, \bibinfo {author} {\bibfnamefont {J.}~\bibnamefont {Kelly}},
  \bibinfo {author} {\bibfnamefont {A.}~\bibnamefont {Megrant}}, \bibinfo
  {author} {\bibfnamefont {A.}~\bibnamefont {Veitia}}, \bibinfo {author}
  {\bibfnamefont {D.}~\bibnamefont {Sank}}, \bibinfo {author} {\bibfnamefont
  {E.}~\bibnamefont {Jeffrey}}, \bibinfo {author} {\bibfnamefont {T.~C.}\
  \bibnamefont {White}}, \bibinfo {author} {\bibfnamefont {J.}~\bibnamefont
  {Mutus}}, \bibinfo {author} {\bibfnamefont {A.~G.}\ \bibnamefont {Fowler}},
  \bibinfo {author} {\bibfnamefont {B.}~\bibnamefont {Campbell}}, \emph
  {et~al.},\ }\bibfield  {title} {\bibinfo {title} {Superconducting quantum
  circuits at the surface code threshold for fault tolerance},\ }\href@noop {}
  {\bibfield  {journal} {\bibinfo  {journal} {Nature}\ }\textbf {\bibinfo
  {volume} {508}},\ \bibinfo {pages} {500} (\bibinfo {year}
  {2014})}\BibitemShut {NoStop}%
\bibitem [{\citenamefont {Kelly}\ \emph {et~al.}(2015)\citenamefont {Kelly},
  \citenamefont {Barends}, \citenamefont {Fowler}, \citenamefont {Megrant},
  \citenamefont {Jeffrey}, \citenamefont {White}, \citenamefont {Sank},
  \citenamefont {Mutus}, \citenamefont {Campbell}, \citenamefont {Chen} \emph
  {et~al.}}]{kelly2015state}%
  \BibitemOpen
  \bibfield  {author} {\bibinfo {author} {\bibfnamefont {J.}~\bibnamefont
  {Kelly}}, \bibinfo {author} {\bibfnamefont {R.}~\bibnamefont {Barends}},
  \bibinfo {author} {\bibfnamefont {A.~G.}\ \bibnamefont {Fowler}}, \bibinfo
  {author} {\bibfnamefont {A.}~\bibnamefont {Megrant}}, \bibinfo {author}
  {\bibfnamefont {E.}~\bibnamefont {Jeffrey}}, \bibinfo {author} {\bibfnamefont
  {T.~C.}\ \bibnamefont {White}}, \bibinfo {author} {\bibfnamefont
  {D.}~\bibnamefont {Sank}}, \bibinfo {author} {\bibfnamefont {J.~Y.}\
  \bibnamefont {Mutus}}, \bibinfo {author} {\bibfnamefont {B.}~\bibnamefont
  {Campbell}}, \bibinfo {author} {\bibfnamefont {Y.}~\bibnamefont {Chen}},
  \emph {et~al.},\ }\bibfield  {title} {\bibinfo {title} {State preservation by
  repetitive error detection in a superconducting quantum circuit},\
  }\href@noop {} {\bibfield  {journal} {\bibinfo  {journal} {Nature}\ }\textbf
  {\bibinfo {volume} {519}},\ \bibinfo {pages} {66} (\bibinfo {year}
  {2015})}\BibitemShut {NoStop}%
\bibitem [{\citenamefont {Majer}\ \emph {et~al.}(2007)\citenamefont {Majer},
  \citenamefont {Chow}, \citenamefont {Gambetta}, \citenamefont {Koch},
  \citenamefont {Johnson}, \citenamefont {Schreier}, \citenamefont {Frunzio},
  \citenamefont {Schuster}, \citenamefont {Houck}, \citenamefont {Wallraff}
  \emph {et~al.}}]{majer2007coupling}%
  \BibitemOpen
  \bibfield  {author} {\bibinfo {author} {\bibfnamefont {J.}~\bibnamefont
  {Majer}}, \bibinfo {author} {\bibfnamefont {J.}~\bibnamefont {Chow}},
  \bibinfo {author} {\bibfnamefont {J.}~\bibnamefont {Gambetta}}, \bibinfo
  {author} {\bibfnamefont {J.}~\bibnamefont {Koch}}, \bibinfo {author}
  {\bibfnamefont {B.}~\bibnamefont {Johnson}}, \bibinfo {author} {\bibfnamefont
  {J.}~\bibnamefont {Schreier}}, \bibinfo {author} {\bibfnamefont
  {L.}~\bibnamefont {Frunzio}}, \bibinfo {author} {\bibfnamefont
  {D.}~\bibnamefont {Schuster}}, \bibinfo {author} {\bibfnamefont {A.~A.}\
  \bibnamefont {Houck}}, \bibinfo {author} {\bibfnamefont {A.}~\bibnamefont
  {Wallraff}}, \emph {et~al.},\ }\bibfield  {title} {\bibinfo {title} {Coupling
  superconducting qubits via a cavity bus},\ }\href@noop {} {\bibfield
  {journal} {\bibinfo  {journal} {Nature}\ }\textbf {\bibinfo {volume} {449}},\
  \bibinfo {pages} {443} (\bibinfo {year} {2007})}\BibitemShut {NoStop}%
\bibitem [{\citenamefont {Chow}\ \emph {et~al.}(2014)\citenamefont {Chow},
  \citenamefont {Gambetta}, \citenamefont {Magesan}, \citenamefont {Abraham},
  \citenamefont {Cross}, \citenamefont {Johnson}, \citenamefont {Masluk},
  \citenamefont {Ryan}, \citenamefont {Smolin}, \citenamefont {Srinivasan}
  \emph {et~al.}}]{chow2014implementing}%
  \BibitemOpen
  \bibfield  {author} {\bibinfo {author} {\bibfnamefont {J.~M.}\ \bibnamefont
  {Chow}}, \bibinfo {author} {\bibfnamefont {J.~M.}\ \bibnamefont {Gambetta}},
  \bibinfo {author} {\bibfnamefont {E.}~\bibnamefont {Magesan}}, \bibinfo
  {author} {\bibfnamefont {D.~W.}\ \bibnamefont {Abraham}}, \bibinfo {author}
  {\bibfnamefont {A.~W.}\ \bibnamefont {Cross}}, \bibinfo {author}
  {\bibfnamefont {B.}~\bibnamefont {Johnson}}, \bibinfo {author} {\bibfnamefont
  {N.~A.}\ \bibnamefont {Masluk}}, \bibinfo {author} {\bibfnamefont {C.~A.}\
  \bibnamefont {Ryan}}, \bibinfo {author} {\bibfnamefont {J.~A.}\ \bibnamefont
  {Smolin}}, \bibinfo {author} {\bibfnamefont {S.~J.}\ \bibnamefont
  {Srinivasan}}, \emph {et~al.},\ }\bibfield  {title} {\bibinfo {title}
  {Implementing a strand of a scalable fault-tolerant quantum computing
  fabric},\ }\href@noop {} {\bibfield  {journal} {\bibinfo  {journal} {Nature
  communications}\ }\textbf {\bibinfo {volume} {5}},\ \bibinfo {pages} {4015}
  (\bibinfo {year} {2014})}\BibitemShut {NoStop}%
\bibitem [{\citenamefont {Song}\ \emph {et~al.}(2019)\citenamefont {Song},
  \citenamefont {Xu}, \citenamefont {Li}, \citenamefont {Zhang}, \citenamefont
  {Zhang}, \citenamefont {Liu}, \citenamefont {Guo}, \citenamefont {Wang},
  \citenamefont {Ren}, \citenamefont {Hao} \emph
  {et~al.}}]{song2019generation}%
  \BibitemOpen
  \bibfield  {author} {\bibinfo {author} {\bibfnamefont {C.}~\bibnamefont
  {Song}}, \bibinfo {author} {\bibfnamefont {K.}~\bibnamefont {Xu}}, \bibinfo
  {author} {\bibfnamefont {H.}~\bibnamefont {Li}}, \bibinfo {author}
  {\bibfnamefont {Y.-R.}\ \bibnamefont {Zhang}}, \bibinfo {author}
  {\bibfnamefont {X.}~\bibnamefont {Zhang}}, \bibinfo {author} {\bibfnamefont
  {W.}~\bibnamefont {Liu}}, \bibinfo {author} {\bibfnamefont {Q.}~\bibnamefont
  {Guo}}, \bibinfo {author} {\bibfnamefont {Z.}~\bibnamefont {Wang}}, \bibinfo
  {author} {\bibfnamefont {W.}~\bibnamefont {Ren}}, \bibinfo {author}
  {\bibfnamefont {J.}~\bibnamefont {Hao}}, \emph {et~al.},\ }\bibfield  {title}
  {\bibinfo {title} {Generation of multicomponent atomic schr{\"o}dinger cat
  states of up to 20 qubits},\ }\href@noop {} {\bibfield  {journal} {\bibinfo
  {journal} {Science}\ }\textbf {\bibinfo {volume} {365}},\ \bibinfo {pages}
  {574} (\bibinfo {year} {2019})}\BibitemShut {NoStop}%
\bibitem [{\citenamefont {Xu}\ \emph {et~al.}(2018)\citenamefont {Xu},
  \citenamefont {Chen}, \citenamefont {Zeng}, \citenamefont {Zhang},
  \citenamefont {Song}, \citenamefont {Liu}, \citenamefont {Guo}, \citenamefont
  {Zhang}, \citenamefont {Xu}, \citenamefont {Deng} \emph
  {et~al.}}]{xu2018emulating}%
  \BibitemOpen
  \bibfield  {author} {\bibinfo {author} {\bibfnamefont {K.}~\bibnamefont
  {Xu}}, \bibinfo {author} {\bibfnamefont {J.-J.}\ \bibnamefont {Chen}},
  \bibinfo {author} {\bibfnamefont {Y.}~\bibnamefont {Zeng}}, \bibinfo {author}
  {\bibfnamefont {Y.-R.}\ \bibnamefont {Zhang}}, \bibinfo {author}
  {\bibfnamefont {C.}~\bibnamefont {Song}}, \bibinfo {author} {\bibfnamefont
  {W.}~\bibnamefont {Liu}}, \bibinfo {author} {\bibfnamefont {Q.}~\bibnamefont
  {Guo}}, \bibinfo {author} {\bibfnamefont {P.}~\bibnamefont {Zhang}}, \bibinfo
  {author} {\bibfnamefont {D.}~\bibnamefont {Xu}}, \bibinfo {author}
  {\bibfnamefont {H.}~\bibnamefont {Deng}}, \emph {et~al.},\ }\bibfield
  {title} {\bibinfo {title} {Emulating many-body localization with a
  superconducting quantum processor},\ }\href@noop {} {\bibfield  {journal}
  {\bibinfo  {journal} {Physical Review Letters}\ }\textbf {\bibinfo {volume}
  {120}},\ \bibinfo {pages} {050507} (\bibinfo {year} {2018})}\BibitemShut
  {NoStop}%
\bibitem [{\citenamefont {Guo}\ \emph {et~al.}(2021)\citenamefont {Guo},
  \citenamefont {Cheng}, \citenamefont {Sun}, \citenamefont {Song},
  \citenamefont {Li}, \citenamefont {Wang}, \citenamefont {Ren}, \citenamefont
  {Dong}, \citenamefont {Zheng}, \citenamefont {Zhang} \emph
  {et~al.}}]{guo2021observation}%
  \BibitemOpen
  \bibfield  {author} {\bibinfo {author} {\bibfnamefont {Q.}~\bibnamefont
  {Guo}}, \bibinfo {author} {\bibfnamefont {C.}~\bibnamefont {Cheng}}, \bibinfo
  {author} {\bibfnamefont {Z.-H.}\ \bibnamefont {Sun}}, \bibinfo {author}
  {\bibfnamefont {Z.}~\bibnamefont {Song}}, \bibinfo {author} {\bibfnamefont
  {H.}~\bibnamefont {Li}}, \bibinfo {author} {\bibfnamefont {Z.}~\bibnamefont
  {Wang}}, \bibinfo {author} {\bibfnamefont {W.}~\bibnamefont {Ren}}, \bibinfo
  {author} {\bibfnamefont {H.}~\bibnamefont {Dong}}, \bibinfo {author}
  {\bibfnamefont {D.}~\bibnamefont {Zheng}}, \bibinfo {author} {\bibfnamefont
  {Y.-R.}\ \bibnamefont {Zhang}}, \emph {et~al.},\ }\bibfield  {title}
  {\bibinfo {title} {Observation of energy-resolved many-body localization},\
  }\href@noop {} {\bibfield  {journal} {\bibinfo  {journal} {Nature Physics}\
  }\textbf {\bibinfo {volume} {17}},\ \bibinfo {pages} {234} (\bibinfo {year}
  {2021})}\BibitemShut {NoStop}%
\bibitem [{\citenamefont {Chen}\ \emph {et~al.}(2014)\citenamefont {Chen},
  \citenamefont {Neill}, \citenamefont {Roushan}, \citenamefont {Leung},
  \citenamefont {Fang}, \citenamefont {Barends}, \citenamefont {Kelly},
  \citenamefont {Campbell}, \citenamefont {Chen}, \citenamefont {Chiaro} \emph
  {et~al.}}]{chen2014qubit}%
  \BibitemOpen
  \bibfield  {author} {\bibinfo {author} {\bibfnamefont {Y.}~\bibnamefont
  {Chen}}, \bibinfo {author} {\bibfnamefont {C.}~\bibnamefont {Neill}},
  \bibinfo {author} {\bibfnamefont {P.}~\bibnamefont {Roushan}}, \bibinfo
  {author} {\bibfnamefont {N.}~\bibnamefont {Leung}}, \bibinfo {author}
  {\bibfnamefont {M.}~\bibnamefont {Fang}}, \bibinfo {author} {\bibfnamefont
  {R.}~\bibnamefont {Barends}}, \bibinfo {author} {\bibfnamefont
  {J.}~\bibnamefont {Kelly}}, \bibinfo {author} {\bibfnamefont
  {B.}~\bibnamefont {Campbell}}, \bibinfo {author} {\bibfnamefont
  {Z.}~\bibnamefont {Chen}}, \bibinfo {author} {\bibfnamefont {B.}~\bibnamefont
  {Chiaro}}, \emph {et~al.},\ }\bibfield  {title} {\bibinfo {title} {Qubit
  architecture with high coherence and fast tunable coupling},\ }\href@noop {}
  {\bibfield  {journal} {\bibinfo  {journal} {Physical Review Letters}\
  }\textbf {\bibinfo {volume} {113}},\ \bibinfo {pages} {220502} (\bibinfo
  {year} {2014})}\BibitemShut {NoStop}%
\bibitem [{\citenamefont {Neill}\ \emph {et~al.}(2018)\citenamefont {Neill},
  \citenamefont {Roushan}, \citenamefont {Kechedzhi}, \citenamefont {Boixo},
  \citenamefont {Isakov}, \citenamefont {Smelyanskiy}, \citenamefont {Megrant},
  \citenamefont {Chiaro}, \citenamefont {Dunsworth}, \citenamefont {Arya} \emph
  {et~al.}}]{neill2018blueprint}%
  \BibitemOpen
  \bibfield  {author} {\bibinfo {author} {\bibfnamefont {C.}~\bibnamefont
  {Neill}}, \bibinfo {author} {\bibfnamefont {P.}~\bibnamefont {Roushan}},
  \bibinfo {author} {\bibfnamefont {K.}~\bibnamefont {Kechedzhi}}, \bibinfo
  {author} {\bibfnamefont {S.}~\bibnamefont {Boixo}}, \bibinfo {author}
  {\bibfnamefont {S.~V.}\ \bibnamefont {Isakov}}, \bibinfo {author}
  {\bibfnamefont {V.}~\bibnamefont {Smelyanskiy}}, \bibinfo {author}
  {\bibfnamefont {A.}~\bibnamefont {Megrant}}, \bibinfo {author} {\bibfnamefont
  {B.}~\bibnamefont {Chiaro}}, \bibinfo {author} {\bibfnamefont
  {A.}~\bibnamefont {Dunsworth}}, \bibinfo {author} {\bibfnamefont
  {K.}~\bibnamefont {Arya}}, \emph {et~al.},\ }\bibfield  {title} {\bibinfo
  {title} {A blueprint for demonstrating quantum supremacy with superconducting
  qubits},\ }\href@noop {} {\bibfield  {journal} {\bibinfo  {journal}
  {Science}\ }\textbf {\bibinfo {volume} {360}},\ \bibinfo {pages} {195}
  (\bibinfo {year} {2018})}\BibitemShut {NoStop}%
\bibitem [{\citenamefont {Yan}\ \emph {et~al.}(2018)\citenamefont {Yan},
  \citenamefont {Krantz}, \citenamefont {Sung}, \citenamefont {Kjaergaard},
  \citenamefont {Campbell}, \citenamefont {Orlando}, \citenamefont
  {Gustavsson},\ and\ \citenamefont {Oliver}}]{yan2018tunable}%
  \BibitemOpen
  \bibfield  {author} {\bibinfo {author} {\bibfnamefont {F.}~\bibnamefont
  {Yan}}, \bibinfo {author} {\bibfnamefont {P.}~\bibnamefont {Krantz}},
  \bibinfo {author} {\bibfnamefont {Y.}~\bibnamefont {Sung}}, \bibinfo {author}
  {\bibfnamefont {M.}~\bibnamefont {Kjaergaard}}, \bibinfo {author}
  {\bibfnamefont {D.~L.}\ \bibnamefont {Campbell}}, \bibinfo {author}
  {\bibfnamefont {T.~P.}\ \bibnamefont {Orlando}}, \bibinfo {author}
  {\bibfnamefont {S.}~\bibnamefont {Gustavsson}},\ and\ \bibinfo {author}
  {\bibfnamefont {W.~D.}\ \bibnamefont {Oliver}},\ }\bibfield  {title}
  {\bibinfo {title} {Tunable coupling scheme for implementing high-fidelity
  two-qubit gates},\ }\href@noop {} {\bibfield  {journal} {\bibinfo  {journal}
  {Physical Review Applied}\ }\textbf {\bibinfo {volume} {10}},\ \bibinfo
  {pages} {054062} (\bibinfo {year} {2018})}\BibitemShut {NoStop}%
\bibitem [{\citenamefont {Mundada}\ \emph {et~al.}(2019)\citenamefont
  {Mundada}, \citenamefont {Zhang}, \citenamefont {Hazard},\ and\ \citenamefont
  {Houck}}]{mundada2019suppression}%
  \BibitemOpen
  \bibfield  {author} {\bibinfo {author} {\bibfnamefont {P.}~\bibnamefont
  {Mundada}}, \bibinfo {author} {\bibfnamefont {G.}~\bibnamefont {Zhang}},
  \bibinfo {author} {\bibfnamefont {T.}~\bibnamefont {Hazard}},\ and\ \bibinfo
  {author} {\bibfnamefont {A.}~\bibnamefont {Houck}},\ }\bibfield  {title}
  {\bibinfo {title} {Suppression of qubit crosstalk in a tunable coupling
  superconducting circuit},\ }\href@noop {} {\bibfield  {journal} {\bibinfo
  {journal} {Physical Review Applied}\ }\textbf {\bibinfo {volume} {12}},\
  \bibinfo {pages} {054023} (\bibinfo {year} {2019})}\BibitemShut {NoStop}%
\bibitem [{\citenamefont {Li}\ \emph {et~al.}(2020)\citenamefont {Li},
  \citenamefont {Cai}, \citenamefont {Yan}, \citenamefont {Wang}, \citenamefont
  {Pan}, \citenamefont {Ma}, \citenamefont {Cai}, \citenamefont {Han},
  \citenamefont {Hua}, \citenamefont {Han} \emph {et~al.}}]{li2020tunable}%
  \BibitemOpen
  \bibfield  {author} {\bibinfo {author} {\bibfnamefont {X.}~\bibnamefont
  {Li}}, \bibinfo {author} {\bibfnamefont {T.}~\bibnamefont {Cai}}, \bibinfo
  {author} {\bibfnamefont {H.}~\bibnamefont {Yan}}, \bibinfo {author}
  {\bibfnamefont {Z.}~\bibnamefont {Wang}}, \bibinfo {author} {\bibfnamefont
  {X.}~\bibnamefont {Pan}}, \bibinfo {author} {\bibfnamefont {Y.}~\bibnamefont
  {Ma}}, \bibinfo {author} {\bibfnamefont {W.}~\bibnamefont {Cai}}, \bibinfo
  {author} {\bibfnamefont {J.}~\bibnamefont {Han}}, \bibinfo {author}
  {\bibfnamefont {Z.}~\bibnamefont {Hua}}, \bibinfo {author} {\bibfnamefont
  {X.}~\bibnamefont {Han}}, \emph {et~al.},\ }\bibfield  {title} {\bibinfo
  {title} {Tunable coupler for realizing a controlled-phase gate with
  dynamically decoupled regime in a superconducting circuit},\ }\href@noop {}
  {\bibfield  {journal} {\bibinfo  {journal} {Physical Review Applied}\
  }\textbf {\bibinfo {volume} {14}},\ \bibinfo {pages} {024070} (\bibinfo
  {year} {2020})}\BibitemShut {NoStop}%
\bibitem [{\citenamefont {Zhao}\ \emph
  {et~al.}(2020{\natexlab{a}})\citenamefont {Zhao}, \citenamefont {Xu},
  \citenamefont {Lan}, \citenamefont {Chu}, \citenamefont {Tan}, \citenamefont
  {Yu},\ and\ \citenamefont {Yu}}]{zhao2020high}%
  \BibitemOpen
  \bibfield  {author} {\bibinfo {author} {\bibfnamefont {P.}~\bibnamefont
  {Zhao}}, \bibinfo {author} {\bibfnamefont {P.}~\bibnamefont {Xu}}, \bibinfo
  {author} {\bibfnamefont {D.}~\bibnamefont {Lan}}, \bibinfo {author}
  {\bibfnamefont {J.}~\bibnamefont {Chu}}, \bibinfo {author} {\bibfnamefont
  {X.}~\bibnamefont {Tan}}, \bibinfo {author} {\bibfnamefont {H.}~\bibnamefont
  {Yu}},\ and\ \bibinfo {author} {\bibfnamefont {Y.}~\bibnamefont {Yu}},\
  }\bibfield  {title} {\bibinfo {title} {High-contrast zz interaction using
  superconducting qubits with opposite-sign anharmonicity},\ }\href@noop {}
  {\bibfield  {journal} {\bibinfo  {journal} {Physical Review Letters}\
  }\textbf {\bibinfo {volume} {125}},\ \bibinfo {pages} {200503} (\bibinfo
  {year} {2020}{\natexlab{a}})}\BibitemShut {NoStop}%
\bibitem [{\citenamefont {Xu}\ \emph {et~al.}(2020)\citenamefont {Xu},
  \citenamefont {Chu}, \citenamefont {Yuan}, \citenamefont {Qiu}, \citenamefont
  {Zhou}, \citenamefont {Zhang}, \citenamefont {Tan}, \citenamefont {Yu},
  \citenamefont {Liu}, \citenamefont {Li} \emph {et~al.}}]{xu2020high}%
  \BibitemOpen
  \bibfield  {author} {\bibinfo {author} {\bibfnamefont {Y.}~\bibnamefont
  {Xu}}, \bibinfo {author} {\bibfnamefont {J.}~\bibnamefont {Chu}}, \bibinfo
  {author} {\bibfnamefont {J.}~\bibnamefont {Yuan}}, \bibinfo {author}
  {\bibfnamefont {J.}~\bibnamefont {Qiu}}, \bibinfo {author} {\bibfnamefont
  {Y.}~\bibnamefont {Zhou}}, \bibinfo {author} {\bibfnamefont {L.}~\bibnamefont
  {Zhang}}, \bibinfo {author} {\bibfnamefont {X.}~\bibnamefont {Tan}}, \bibinfo
  {author} {\bibfnamefont {Y.}~\bibnamefont {Yu}}, \bibinfo {author}
  {\bibfnamefont {S.}~\bibnamefont {Liu}}, \bibinfo {author} {\bibfnamefont
  {J.}~\bibnamefont {Li}}, \emph {et~al.},\ }\bibfield  {title} {\bibinfo
  {title} {High-fidelity, high-scalability two-qubit gate scheme for
  superconducting qubits},\ }\href@noop {} {\bibfield  {journal} {\bibinfo
  {journal} {Physical Review Letters}\ }\textbf {\bibinfo {volume} {125}},\
  \bibinfo {pages} {240503} (\bibinfo {year} {2020})}\BibitemShut {NoStop}%
\bibitem [{\citenamefont {Zhao}\ \emph
  {et~al.}(2020{\natexlab{b}})\citenamefont {Zhao}, \citenamefont {Xu},
  \citenamefont {Lan}, \citenamefont {Tan}, \citenamefont {Yu},\ and\
  \citenamefont {Yu}}]{zhao2020switchable}%
  \BibitemOpen
  \bibfield  {author} {\bibinfo {author} {\bibfnamefont {P.}~\bibnamefont
  {Zhao}}, \bibinfo {author} {\bibfnamefont {P.}~\bibnamefont {Xu}}, \bibinfo
  {author} {\bibfnamefont {D.}~\bibnamefont {Lan}}, \bibinfo {author}
  {\bibfnamefont {X.}~\bibnamefont {Tan}}, \bibinfo {author} {\bibfnamefont
  {H.}~\bibnamefont {Yu}},\ and\ \bibinfo {author} {\bibfnamefont
  {Y.}~\bibnamefont {Yu}},\ }\bibfield  {title} {\bibinfo {title} {Switchable
  next-nearest-neighbor coupling for controlled two-qubit operations},\
  }\href@noop {} {\bibfield  {journal} {\bibinfo  {journal} {Physical Review
  Applied}\ }\textbf {\bibinfo {volume} {14}},\ \bibinfo {pages} {064016}
  (\bibinfo {year} {2020}{\natexlab{b}})}\BibitemShut {NoStop}%
\bibitem [{\citenamefont {Sung}\ \emph {et~al.}(2020)\citenamefont {Sung},
  \citenamefont {Ding}, \citenamefont {Braum{\"u}ller}, \citenamefont
  {Veps{\"a}l{\"a}inen}, \citenamefont {Kannan}, \citenamefont {Kjaergaard},
  \citenamefont {Greene}, \citenamefont {Samach}, \citenamefont {McNally},
  \citenamefont {Kim} \emph {et~al.}}]{sung2020realization}%
  \BibitemOpen
  \bibfield  {author} {\bibinfo {author} {\bibfnamefont {Y.}~\bibnamefont
  {Sung}}, \bibinfo {author} {\bibfnamefont {L.}~\bibnamefont {Ding}}, \bibinfo
  {author} {\bibfnamefont {J.}~\bibnamefont {Braum{\"u}ller}}, \bibinfo
  {author} {\bibfnamefont {A.}~\bibnamefont {Veps{\"a}l{\"a}inen}}, \bibinfo
  {author} {\bibfnamefont {B.}~\bibnamefont {Kannan}}, \bibinfo {author}
  {\bibfnamefont {M.}~\bibnamefont {Kjaergaard}}, \bibinfo {author}
  {\bibfnamefont {A.}~\bibnamefont {Greene}}, \bibinfo {author} {\bibfnamefont
  {G.~O.}\ \bibnamefont {Samach}}, \bibinfo {author} {\bibfnamefont
  {C.}~\bibnamefont {McNally}}, \bibinfo {author} {\bibfnamefont
  {D.}~\bibnamefont {Kim}}, \emph {et~al.},\ }\bibfield  {title} {\bibinfo
  {title} {Realization of high-fidelity cz and zz-free iswap gates with a
  tunable coupler},\ }\href@noop {} {\bibfield  {journal} {\bibinfo  {journal}
  {arXiv:2011.01261}\ } (\bibinfo {year} {2020})}\BibitemShut {NoStop}%
\bibitem [{\citenamefont {Collodo}\ \emph
  {et~al.}(2020{\natexlab{a}})\citenamefont {Collodo}, \citenamefont
  {Herrmann}, \citenamefont {Lacroix}, \citenamefont {Andersen}, \citenamefont
  {Remm}, \citenamefont {Lazar}, \citenamefont {Besse}, \citenamefont {Walter},
  \citenamefont {Wallraff},\ and\ \citenamefont
  {Eichler}}]{collodo2020implementation}%
  \BibitemOpen
  \bibfield  {author} {\bibinfo {author} {\bibfnamefont {M.~C.}\ \bibnamefont
  {Collodo}}, \bibinfo {author} {\bibfnamefont {J.}~\bibnamefont {Herrmann}},
  \bibinfo {author} {\bibfnamefont {N.}~\bibnamefont {Lacroix}}, \bibinfo
  {author} {\bibfnamefont {C.~K.}\ \bibnamefont {Andersen}}, \bibinfo {author}
  {\bibfnamefont {A.}~\bibnamefont {Remm}}, \bibinfo {author} {\bibfnamefont
  {S.}~\bibnamefont {Lazar}}, \bibinfo {author} {\bibfnamefont {J.-C.}\
  \bibnamefont {Besse}}, \bibinfo {author} {\bibfnamefont {T.}~\bibnamefont
  {Walter}}, \bibinfo {author} {\bibfnamefont {A.}~\bibnamefont {Wallraff}},\
  and\ \bibinfo {author} {\bibfnamefont {C.}~\bibnamefont {Eichler}},\
  }\bibfield  {title} {\bibinfo {title} {Implementation of conditional phase
  gates based on tunable z z interactions},\ }\href@noop {} {\bibfield
  {journal} {\bibinfo  {journal} {Physical Review Letters}\ }\textbf {\bibinfo
  {volume} {125}},\ \bibinfo {pages} {240502} (\bibinfo {year}
  {2020}{\natexlab{a}})}\BibitemShut {NoStop}%
\bibitem [{\citenamefont {Xu}\ and\ \citenamefont {Ansari}(2020)}]{xu2020zz}%
  \BibitemOpen
  \bibfield  {author} {\bibinfo {author} {\bibfnamefont {X.}~\bibnamefont
  {Xu}}\ and\ \bibinfo {author} {\bibfnamefont {M.}~\bibnamefont {Ansari}},\
  }\bibfield  {title} {\bibinfo {title} {zz freedom in two qubit gates},\
  }\href@noop {} {\bibfield  {journal} {\bibinfo  {journal} {arXiv:2009.00485}\
  } (\bibinfo {year} {2020})}\BibitemShut {NoStop}%
\bibitem [{\citenamefont {Zhao}\ \emph
  {et~al.}(2020{\natexlab{c}})\citenamefont {Zhao}, \citenamefont {Lan},
  \citenamefont {Xu}, \citenamefont {Xue}, \citenamefont {Blank}, \citenamefont
  {Tan}, \citenamefont {Yu},\ and\ \citenamefont {Yu}}]{zhao2020suppression}%
  \BibitemOpen
  \bibfield  {author} {\bibinfo {author} {\bibfnamefont {P.}~\bibnamefont
  {Zhao}}, \bibinfo {author} {\bibfnamefont {D.}~\bibnamefont {Lan}}, \bibinfo
  {author} {\bibfnamefont {P.}~\bibnamefont {Xu}}, \bibinfo {author}
  {\bibfnamefont {G.}~\bibnamefont {Xue}}, \bibinfo {author} {\bibfnamefont
  {M.}~\bibnamefont {Blank}}, \bibinfo {author} {\bibfnamefont
  {X.}~\bibnamefont {Tan}}, \bibinfo {author} {\bibfnamefont {H.}~\bibnamefont
  {Yu}},\ and\ \bibinfo {author} {\bibfnamefont {Y.}~\bibnamefont {Yu}},\
  }\bibfield  {title} {\bibinfo {title} {Suppression of static zz interaction
  in an all-transmon quantum processor},\ }\href@noop {} {\bibfield  {journal}
  {\bibinfo  {journal} {arXiv:2011.03976}\ } (\bibinfo {year}
  {2020}{\natexlab{c}})}\BibitemShut {NoStop}%
\bibitem [{\citenamefont {Han}\ \emph {et~al.}(2020)\citenamefont {Han},
  \citenamefont {Cai}, \citenamefont {Li}, \citenamefont {Wu}, \citenamefont
  {Ma}, \citenamefont {Ma}, \citenamefont {Wang}, \citenamefont {Zhang},
  \citenamefont {Song},\ and\ \citenamefont {Duan}}]{han2020error}%
  \BibitemOpen
  \bibfield  {author} {\bibinfo {author} {\bibfnamefont {X.}~\bibnamefont
  {Han}}, \bibinfo {author} {\bibfnamefont {T.}~\bibnamefont {Cai}}, \bibinfo
  {author} {\bibfnamefont {X.}~\bibnamefont {Li}}, \bibinfo {author}
  {\bibfnamefont {Y.}~\bibnamefont {Wu}}, \bibinfo {author} {\bibfnamefont
  {Y.}~\bibnamefont {Ma}}, \bibinfo {author} {\bibfnamefont {Y.}~\bibnamefont
  {Ma}}, \bibinfo {author} {\bibfnamefont {J.}~\bibnamefont {Wang}}, \bibinfo
  {author} {\bibfnamefont {H.}~\bibnamefont {Zhang}}, \bibinfo {author}
  {\bibfnamefont {Y.}~\bibnamefont {Song}},\ and\ \bibinfo {author}
  {\bibfnamefont {L.}~\bibnamefont {Duan}},\ }\bibfield  {title} {\bibinfo
  {title} {Error analysis in suppression of unwanted qubit interactions for a
  parametric gate in a tunable superconducting circuit},\ }\href@noop {}
  {\bibfield  {journal} {\bibinfo  {journal} {Physical Review A}\ }\textbf
  {\bibinfo {volume} {102}},\ \bibinfo {pages} {022619} (\bibinfo {year}
  {2020})}\BibitemShut {NoStop}%
\bibitem [{\citenamefont {Xu}\ \emph {et~al.}(2021)\citenamefont {Xu},
  \citenamefont {Liu}, \citenamefont {Li}, \citenamefont {Han}, \citenamefont
  {Zhang}, \citenamefont {Linghu}, \citenamefont {Li}, \citenamefont {Chen},
  \citenamefont {Yang}, \citenamefont {Wang} \emph
  {et~al.}}]{xu2021realization}%
  \BibitemOpen
  \bibfield  {author} {\bibinfo {author} {\bibfnamefont {H.}~\bibnamefont
  {Xu}}, \bibinfo {author} {\bibfnamefont {W.}~\bibnamefont {Liu}}, \bibinfo
  {author} {\bibfnamefont {Z.}~\bibnamefont {Li}}, \bibinfo {author}
  {\bibfnamefont {J.}~\bibnamefont {Han}}, \bibinfo {author} {\bibfnamefont
  {J.}~\bibnamefont {Zhang}}, \bibinfo {author} {\bibfnamefont
  {K.}~\bibnamefont {Linghu}}, \bibinfo {author} {\bibfnamefont
  {Y.}~\bibnamefont {Li}}, \bibinfo {author} {\bibfnamefont {M.}~\bibnamefont
  {Chen}}, \bibinfo {author} {\bibfnamefont {Z.}~\bibnamefont {Yang}}, \bibinfo
  {author} {\bibfnamefont {J.}~\bibnamefont {Wang}}, \emph {et~al.},\
  }\bibfield  {title} {\bibinfo {title} {Realization of adiabatic and diabatic
  cz gates in superconducting qubits coupled with a tunable coupler},\
  }\href@noop {} {\bibfield  {journal} {\bibinfo  {journal} {Chinese Physics
  B}\ }\textbf {\bibinfo {volume} {30}},\ \bibinfo {pages} {044212} (\bibinfo
  {year} {2021})}\BibitemShut {NoStop}%
\bibitem [{\citenamefont {Cai}\ \emph {et~al.}(2021)\citenamefont {Cai},
  \citenamefont {Han}, \citenamefont {Wu}, \citenamefont {Ma}, \citenamefont
  {Wang}, \citenamefont {Zhang}, \citenamefont {Wang}, \citenamefont {Song},\
  and\ \citenamefont {Duan}}]{cai2021perturbation}%
  \BibitemOpen
  \bibfield  {author} {\bibinfo {author} {\bibfnamefont {T.}~\bibnamefont
  {Cai}}, \bibinfo {author} {\bibfnamefont {X.}~\bibnamefont {Han}}, \bibinfo
  {author} {\bibfnamefont {Y.}~\bibnamefont {Wu}}, \bibinfo {author}
  {\bibfnamefont {Y.}~\bibnamefont {Ma}}, \bibinfo {author} {\bibfnamefont
  {J.}~\bibnamefont {Wang}}, \bibinfo {author} {\bibfnamefont {H.}~\bibnamefont
  {Zhang}}, \bibinfo {author} {\bibfnamefont {H.}~\bibnamefont {Wang}},
  \bibinfo {author} {\bibfnamefont {Y.}~\bibnamefont {Song}},\ and\ \bibinfo
  {author} {\bibfnamefont {L.}~\bibnamefont {Duan}},\ }\bibfield  {title}
  {\bibinfo {title} {Perturbation impact of spectators on a cross-resonance
  gate in a tunable coupling superconducting circuit},\ }\href@noop {}
  {\bibfield  {journal} {\bibinfo  {journal} {arXiv:2101.01854}\ } (\bibinfo
  {year} {2021})}\BibitemShut {NoStop}%
\bibitem [{\citenamefont {Sete}\ \emph
  {et~al.}(2021{\natexlab{a}})\citenamefont {Sete}, \citenamefont {Didier},
  \citenamefont {Chen}, \citenamefont {Kulshreshtha}, \citenamefont {Manenti},\
  and\ \citenamefont {Poletto}}]{sete2021parametric}%
  \BibitemOpen
  \bibfield  {author} {\bibinfo {author} {\bibfnamefont {E.~A.}\ \bibnamefont
  {Sete}}, \bibinfo {author} {\bibfnamefont {N.}~\bibnamefont {Didier}},
  \bibinfo {author} {\bibfnamefont {A.~Q.}\ \bibnamefont {Chen}}, \bibinfo
  {author} {\bibfnamefont {S.}~\bibnamefont {Kulshreshtha}}, \bibinfo {author}
  {\bibfnamefont {R.}~\bibnamefont {Manenti}},\ and\ \bibinfo {author}
  {\bibfnamefont {S.}~\bibnamefont {Poletto}},\ }\bibfield  {title} {\bibinfo
  {title} {Parametric-resonance entangling gates with a tunable coupler},\
  }\href@noop {} {\bibfield  {journal} {\bibinfo  {journal} {arXiv:2104.03511}\
  } (\bibinfo {year} {2021}{\natexlab{a}})}\BibitemShut {NoStop}%
\bibitem [{\citenamefont {Campbell}\ \emph {et~al.}(2017)\citenamefont
  {Campbell}, \citenamefont {Terhal},\ and\ \citenamefont
  {Vuillot}}]{campbell2017roads}%
  \BibitemOpen
  \bibfield  {author} {\bibinfo {author} {\bibfnamefont {E.~T.}\ \bibnamefont
  {Campbell}}, \bibinfo {author} {\bibfnamefont {B.~M.}\ \bibnamefont
  {Terhal}},\ and\ \bibinfo {author} {\bibfnamefont {C.}~\bibnamefont
  {Vuillot}},\ }\bibfield  {title} {\bibinfo {title} {Roads towards
  fault-tolerant universal quantum computation},\ }\href@noop {} {\bibfield
  {journal} {\bibinfo  {journal} {Nature}\ }\textbf {\bibinfo {volume} {549}},\
  \bibinfo {pages} {172} (\bibinfo {year} {2017})}\BibitemShut {NoStop}%
\bibitem [{\citenamefont {Barends}\ \emph {et~al.}(2019)\citenamefont
  {Barends}, \citenamefont {Quintana}, \citenamefont {Petukhov}, \citenamefont
  {Chen}, \citenamefont {Kafri}, \citenamefont {Kechedzhi}, \citenamefont
  {Collins}, \citenamefont {Naaman}, \citenamefont {Boixo}, \citenamefont
  {Arute} \emph {et~al.}}]{barends2019diabatic}%
  \BibitemOpen
  \bibfield  {author} {\bibinfo {author} {\bibfnamefont {R.}~\bibnamefont
  {Barends}}, \bibinfo {author} {\bibfnamefont {C.}~\bibnamefont {Quintana}},
  \bibinfo {author} {\bibfnamefont {A.}~\bibnamefont {Petukhov}}, \bibinfo
  {author} {\bibfnamefont {Y.}~\bibnamefont {Chen}}, \bibinfo {author}
  {\bibfnamefont {D.}~\bibnamefont {Kafri}}, \bibinfo {author} {\bibfnamefont
  {K.}~\bibnamefont {Kechedzhi}}, \bibinfo {author} {\bibfnamefont
  {R.}~\bibnamefont {Collins}}, \bibinfo {author} {\bibfnamefont
  {O.}~\bibnamefont {Naaman}}, \bibinfo {author} {\bibfnamefont
  {S.}~\bibnamefont {Boixo}}, \bibinfo {author} {\bibfnamefont
  {F.}~\bibnamefont {Arute}}, \emph {et~al.},\ }\bibfield  {title} {\bibinfo
  {title} {Diabatic gates for frequency-tunable superconducting qubits},\
  }\href@noop {} {\bibfield  {journal} {\bibinfo  {journal} {Physical Review
  Letters}\ }\textbf {\bibinfo {volume} {123}},\ \bibinfo {pages} {210501}
  (\bibinfo {year} {2019})}\BibitemShut {NoStop}%
\bibitem [{\citenamefont {Foxen}\ \emph {et~al.}(2020)\citenamefont {Foxen},
  \citenamefont {Neill}, \citenamefont {Dunsworth}, \citenamefont {Roushan},
  \citenamefont {Chiaro}, \citenamefont {Megrant}, \citenamefont {Kelly},
  \citenamefont {Chen}, \citenamefont {Satzinger}, \citenamefont {Barends}
  \emph {et~al.}}]{foxen2020demonstrating}%
  \BibitemOpen
  \bibfield  {author} {\bibinfo {author} {\bibfnamefont {B.}~\bibnamefont
  {Foxen}}, \bibinfo {author} {\bibfnamefont {C.}~\bibnamefont {Neill}},
  \bibinfo {author} {\bibfnamefont {A.}~\bibnamefont {Dunsworth}}, \bibinfo
  {author} {\bibfnamefont {P.}~\bibnamefont {Roushan}}, \bibinfo {author}
  {\bibfnamefont {B.}~\bibnamefont {Chiaro}}, \bibinfo {author} {\bibfnamefont
  {A.}~\bibnamefont {Megrant}}, \bibinfo {author} {\bibfnamefont
  {J.}~\bibnamefont {Kelly}}, \bibinfo {author} {\bibfnamefont
  {Z.}~\bibnamefont {Chen}}, \bibinfo {author} {\bibfnamefont {K.}~\bibnamefont
  {Satzinger}}, \bibinfo {author} {\bibfnamefont {R.}~\bibnamefont {Barends}},
  \emph {et~al.},\ }\bibfield  {title} {\bibinfo {title} {Demonstrating a
  continuous set of two-qubit gates for near-term quantum algorithms},\
  }\href@noop {} {\bibfield  {journal} {\bibinfo  {journal} {Physical Review
  Letters}\ }\textbf {\bibinfo {volume} {125}},\ \bibinfo {pages} {120504}
  (\bibinfo {year} {2020})}\BibitemShut {NoStop}%
\bibitem [{\citenamefont {Ku}\ \emph {et~al.}(2020)\citenamefont {Ku},
  \citenamefont {Xu}, \citenamefont {Brink}, \citenamefont {McKay},
  \citenamefont {Hertzberg}, \citenamefont {Ansari},\ and\ \citenamefont
  {Plourde}}]{ku2020suppression}%
  \BibitemOpen
  \bibfield  {author} {\bibinfo {author} {\bibfnamefont {J.}~\bibnamefont
  {Ku}}, \bibinfo {author} {\bibfnamefont {X.}~\bibnamefont {Xu}}, \bibinfo
  {author} {\bibfnamefont {M.}~\bibnamefont {Brink}}, \bibinfo {author}
  {\bibfnamefont {D.~C.}\ \bibnamefont {McKay}}, \bibinfo {author}
  {\bibfnamefont {J.~B.}\ \bibnamefont {Hertzberg}}, \bibinfo {author}
  {\bibfnamefont {M.~H.}\ \bibnamefont {Ansari}},\ and\ \bibinfo {author}
  {\bibfnamefont {B.}~\bibnamefont {Plourde}},\ }\bibfield  {title} {\bibinfo
  {title} {Suppression of unwanted z z interactions in a hybrid two-qubit
  system},\ }\href@noop {} {\bibfield  {journal} {\bibinfo  {journal} {Physical
  Review Letters}\ }\textbf {\bibinfo {volume} {125}},\ \bibinfo {pages}
  {200504} (\bibinfo {year} {2020})}\BibitemShut {NoStop}%
\bibitem [{\citenamefont {Bravyi}\ \emph {et~al.}(2011)\citenamefont {Bravyi},
  \citenamefont {DiVincenzo},\ and\ \citenamefont
  {Loss}}]{bravyi2011schrieffer}%
  \BibitemOpen
  \bibfield  {author} {\bibinfo {author} {\bibfnamefont {S.}~\bibnamefont
  {Bravyi}}, \bibinfo {author} {\bibfnamefont {D.~P.}\ \bibnamefont
  {DiVincenzo}},\ and\ \bibinfo {author} {\bibfnamefont {D.}~\bibnamefont
  {Loss}},\ }\bibfield  {title} {\bibinfo {title} {Schrieffer--wolff
  transformation for quantum many-body systems},\ }\href@noop {} {\bibfield
  {journal} {\bibinfo  {journal} {Annals of Physics}\ }\textbf {\bibinfo
  {volume} {326}},\ \bibinfo {pages} {2793} (\bibinfo {year}
  {2011})}\BibitemShut {NoStop}%
\bibitem [{\citenamefont {Kovacic}\ and\ \citenamefont
  {Brennan}(2011)}]{kovacic2011duffing}%
  \BibitemOpen
  \bibfield  {author} {\bibinfo {author} {\bibfnamefont {I.}~\bibnamefont
  {Kovacic}}\ and\ \bibinfo {author} {\bibfnamefont {M.~J.}\ \bibnamefont
  {Brennan}},\ }\href@noop {} {\emph {\bibinfo {title} {The Duffing equation:
  nonlinear oscillators and their behaviour}}}\ (\bibinfo  {publisher} {John
  Wiley \& Sons},\ \bibinfo {year} {2011})\BibitemShut {NoStop}%
\bibitem [{\citenamefont {Krantz}\ \emph {et~al.}(2019)\citenamefont {Krantz},
  \citenamefont {Kjaergaard}, \citenamefont {Yan}, \citenamefont {Orlando},
  \citenamefont {Gustavsson},\ and\ \citenamefont
  {Oliver}}]{krantz2019quantum}%
  \BibitemOpen
  \bibfield  {author} {\bibinfo {author} {\bibfnamefont {P.}~\bibnamefont
  {Krantz}}, \bibinfo {author} {\bibfnamefont {M.}~\bibnamefont {Kjaergaard}},
  \bibinfo {author} {\bibfnamefont {F.}~\bibnamefont {Yan}}, \bibinfo {author}
  {\bibfnamefont {T.~P.}\ \bibnamefont {Orlando}}, \bibinfo {author}
  {\bibfnamefont {S.}~\bibnamefont {Gustavsson}},\ and\ \bibinfo {author}
  {\bibfnamefont {W.~D.}\ \bibnamefont {Oliver}},\ }\bibfield  {title}
  {\bibinfo {title} {A quantum engineer's guide to superconducting qubits},\
  }\href@noop {} {\bibfield  {journal} {\bibinfo  {journal} {Applied Physics
  Reviews}\ }\textbf {\bibinfo {volume} {6}},\ \bibinfo {pages} {021318}
  (\bibinfo {year} {2019})}\BibitemShut {NoStop}%
\bibitem [{\citenamefont {McKay}\ \emph {et~al.}(2019)\citenamefont {McKay},
  \citenamefont {Sheldon}, \citenamefont {Smolin}, \citenamefont {Chow},\ and\
  \citenamefont {Gambetta}}]{mckay2019three}%
  \BibitemOpen
  \bibfield  {author} {\bibinfo {author} {\bibfnamefont {D.~C.}\ \bibnamefont
  {McKay}}, \bibinfo {author} {\bibfnamefont {S.}~\bibnamefont {Sheldon}},
  \bibinfo {author} {\bibfnamefont {J.~A.}\ \bibnamefont {Smolin}}, \bibinfo
  {author} {\bibfnamefont {J.~M.}\ \bibnamefont {Chow}},\ and\ \bibinfo
  {author} {\bibfnamefont {J.~M.}\ \bibnamefont {Gambetta}},\ }\bibfield
  {title} {\bibinfo {title} {Three-qubit randomized benchmarking},\ }\href@noop
  {} {\bibfield  {journal} {\bibinfo  {journal} {Physical review letters}\
  }\textbf {\bibinfo {volume} {122}},\ \bibinfo {pages} {200502} (\bibinfo
  {year} {2019})}\BibitemShut {NoStop}%
\bibitem [{\citenamefont {Sheldon}\ \emph {et~al.}(2016)\citenamefont
  {Sheldon}, \citenamefont {Magesan}, \citenamefont {Chow},\ and\ \citenamefont
  {Gambetta}}]{sheldon2016procedure}%
  \BibitemOpen
  \bibfield  {author} {\bibinfo {author} {\bibfnamefont {S.}~\bibnamefont
  {Sheldon}}, \bibinfo {author} {\bibfnamefont {E.}~\bibnamefont {Magesan}},
  \bibinfo {author} {\bibfnamefont {J.~M.}\ \bibnamefont {Chow}},\ and\
  \bibinfo {author} {\bibfnamefont {J.~M.}\ \bibnamefont {Gambetta}},\
  }\bibfield  {title} {\bibinfo {title} {Procedure for systematically tuning up
  cross-talk in the cross-resonance gate},\ }\href@noop {} {\bibfield
  {journal} {\bibinfo  {journal} {Physical Review A}\ }\textbf {\bibinfo
  {volume} {93}},\ \bibinfo {pages} {060302} (\bibinfo {year}
  {2016})}\BibitemShut {NoStop}%
\bibitem [{\citenamefont {McKay}\ \emph {et~al.}(2016)\citenamefont {McKay},
  \citenamefont {Filipp}, \citenamefont {Mezzacapo}, \citenamefont {Magesan},
  \citenamefont {Chow},\ and\ \citenamefont {Gambetta}}]{mckay2016universal}%
  \BibitemOpen
  \bibfield  {author} {\bibinfo {author} {\bibfnamefont {D.~C.}\ \bibnamefont
  {McKay}}, \bibinfo {author} {\bibfnamefont {S.}~\bibnamefont {Filipp}},
  \bibinfo {author} {\bibfnamefont {A.}~\bibnamefont {Mezzacapo}}, \bibinfo
  {author} {\bibfnamefont {E.}~\bibnamefont {Magesan}}, \bibinfo {author}
  {\bibfnamefont {J.~M.}\ \bibnamefont {Chow}},\ and\ \bibinfo {author}
  {\bibfnamefont {J.~M.}\ \bibnamefont {Gambetta}},\ }\bibfield  {title}
  {\bibinfo {title} {Universal gate for fixed-frequency qubits via a tunable
  bus},\ }\href@noop {} {\bibfield  {journal} {\bibinfo  {journal} {Physical
  Review Applied}\ }\textbf {\bibinfo {volume} {6}},\ \bibinfo {pages} {064007}
  (\bibinfo {year} {2016})}\BibitemShut {NoStop}%
\bibitem [{\citenamefont {Magesan}\ and\ \citenamefont
  {Gambetta}(2020)}]{magesan2020effective}%
  \BibitemOpen
  \bibfield  {author} {\bibinfo {author} {\bibfnamefont {E.}~\bibnamefont
  {Magesan}}\ and\ \bibinfo {author} {\bibfnamefont {J.~M.}\ \bibnamefont
  {Gambetta}},\ }\bibfield  {title} {\bibinfo {title} {Effective hamiltonian
  models of the cross-resonance gate},\ }\href@noop {} {\bibfield  {journal}
  {\bibinfo  {journal} {Physical Review A}\ }\textbf {\bibinfo {volume}
  {101}},\ \bibinfo {pages} {052308} (\bibinfo {year} {2020})}\BibitemShut
  {NoStop}%
\bibitem [{\citenamefont {Takita}\ \emph {et~al.}(2016)\citenamefont {Takita},
  \citenamefont {C{\'o}rcoles}, \citenamefont {Magesan}, \citenamefont {Abdo},
  \citenamefont {Brink}, \citenamefont {Cross}, \citenamefont {Chow},\ and\
  \citenamefont {Gambetta}}]{takita2016demonstration}%
  \BibitemOpen
  \bibfield  {author} {\bibinfo {author} {\bibfnamefont {M.}~\bibnamefont
  {Takita}}, \bibinfo {author} {\bibfnamefont {A.~D.}\ \bibnamefont
  {C{\'o}rcoles}}, \bibinfo {author} {\bibfnamefont {E.}~\bibnamefont
  {Magesan}}, \bibinfo {author} {\bibfnamefont {B.}~\bibnamefont {Abdo}},
  \bibinfo {author} {\bibfnamefont {M.}~\bibnamefont {Brink}}, \bibinfo
  {author} {\bibfnamefont {A.}~\bibnamefont {Cross}}, \bibinfo {author}
  {\bibfnamefont {J.~M.}\ \bibnamefont {Chow}},\ and\ \bibinfo {author}
  {\bibfnamefont {J.~M.}\ \bibnamefont {Gambetta}},\ }\bibfield  {title}
  {\bibinfo {title} {Demonstration of weight-four parity measurements in the
  surface code architecture},\ }\href@noop {} {\bibfield  {journal} {\bibinfo
  {journal} {Physical review letters}\ }\textbf {\bibinfo {volume} {117}},\
  \bibinfo {pages} {210505} (\bibinfo {year} {2016})}\BibitemShut {NoStop}%
\bibitem [{\citenamefont {Winick}\ \emph {et~al.}(2020)\citenamefont {Winick},
  \citenamefont {Wallman},\ and\ \citenamefont
  {Emerson}}]{winick2020simulating}%
  \BibitemOpen
  \bibfield  {author} {\bibinfo {author} {\bibfnamefont {A.}~\bibnamefont
  {Winick}}, \bibinfo {author} {\bibfnamefont {J.~J.}\ \bibnamefont
  {Wallman}},\ and\ \bibinfo {author} {\bibfnamefont {J.}~\bibnamefont
  {Emerson}},\ }\bibfield  {title} {\bibinfo {title} {Simulating and mitigating
  crosstalk},\ }\href@noop {} {\bibfield  {journal} {\bibinfo  {journal}
  {arXiv:2006.09596}\ } (\bibinfo {year} {2020})}\BibitemShut {NoStop}%
\bibitem [{\citenamefont {Collodo}\ \emph
  {et~al.}(2020{\natexlab{b}})\citenamefont {Collodo}, \citenamefont
  {Herrmann}, \citenamefont {Lacroix}, \citenamefont {Andersen}, \citenamefont
  {Remm}, \citenamefont {Lazar}, \citenamefont {Besse}, \citenamefont {Walter},
  \citenamefont {Wallraff},\ and\ \citenamefont
  {Eichler}}]{PhysRevLett.125.240502}%
  \BibitemOpen
  \bibfield  {author} {\bibinfo {author} {\bibfnamefont {M.~C.}\ \bibnamefont
  {Collodo}}, \bibinfo {author} {\bibfnamefont {J.}~\bibnamefont {Herrmann}},
  \bibinfo {author} {\bibfnamefont {N.}~\bibnamefont {Lacroix}}, \bibinfo
  {author} {\bibfnamefont {C.~K.}\ \bibnamefont {Andersen}}, \bibinfo {author}
  {\bibfnamefont {A.}~\bibnamefont {Remm}}, \bibinfo {author} {\bibfnamefont
  {S.}~\bibnamefont {Lazar}}, \bibinfo {author} {\bibfnamefont {J.-C.}\
  \bibnamefont {Besse}}, \bibinfo {author} {\bibfnamefont {T.}~\bibnamefont
  {Walter}}, \bibinfo {author} {\bibfnamefont {A.}~\bibnamefont {Wallraff}},\
  and\ \bibinfo {author} {\bibfnamefont {C.}~\bibnamefont {Eichler}},\
  }\bibfield  {title} {\bibinfo {title} {Implementation of conditional phase
  gates based on tunable $zz$ interactions},\ }\href@noop {} {\bibfield
  {journal} {\bibinfo  {journal} {Physical Review Letters}\ }\textbf {\bibinfo
  {volume} {125}},\ \bibinfo {pages} {240502} (\bibinfo {year}
  {2020}{\natexlab{b}})}\BibitemShut {NoStop}%
\bibitem [{\citenamefont {Koch}\ \emph {et~al.}(2007)\citenamefont {Koch},
  \citenamefont {Terri}, \citenamefont {Gambetta}, \citenamefont {Houck},
  \citenamefont {Schuster}, \citenamefont {Majer}, \citenamefont {Blais},
  \citenamefont {Devoret}, \citenamefont {Girvin},\ and\ \citenamefont
  {Schoelkopf}}]{koch2007charge}%
  \BibitemOpen
  \bibfield  {author} {\bibinfo {author} {\bibfnamefont {J.}~\bibnamefont
  {Koch}}, \bibinfo {author} {\bibfnamefont {M.~Y.}\ \bibnamefont {Terri}},
  \bibinfo {author} {\bibfnamefont {J.}~\bibnamefont {Gambetta}}, \bibinfo
  {author} {\bibfnamefont {A.~A.}\ \bibnamefont {Houck}}, \bibinfo {author}
  {\bibfnamefont {D.}~\bibnamefont {Schuster}}, \bibinfo {author}
  {\bibfnamefont {J.}~\bibnamefont {Majer}}, \bibinfo {author} {\bibfnamefont
  {A.}~\bibnamefont {Blais}}, \bibinfo {author} {\bibfnamefont {M.~H.}\
  \bibnamefont {Devoret}}, \bibinfo {author} {\bibfnamefont {S.~M.}\
  \bibnamefont {Girvin}},\ and\ \bibinfo {author} {\bibfnamefont {R.~J.}\
  \bibnamefont {Schoelkopf}},\ }\bibfield  {title} {\bibinfo {title}
  {Charge-insensitive qubit design derived from the cooper pair box},\
  }\href@noop {} {\bibfield  {journal} {\bibinfo  {journal} {Physical Review
  A}\ }\textbf {\bibinfo {volume} {76}},\ \bibinfo {pages} {042319} (\bibinfo
  {year} {2007})}\BibitemShut {NoStop}%
\bibitem [{\citenamefont {Steffen}\ \emph {et~al.}(2010)\citenamefont
  {Steffen}, \citenamefont {Kumar}, \citenamefont {DiVincenzo}, \citenamefont
  {Rozen}, \citenamefont {Keefe}, \citenamefont {Rothwell},\ and\ \citenamefont
  {Ketchen}}]{steffen2010high}%
  \BibitemOpen
  \bibfield  {author} {\bibinfo {author} {\bibfnamefont {M.}~\bibnamefont
  {Steffen}}, \bibinfo {author} {\bibfnamefont {S.}~\bibnamefont {Kumar}},
  \bibinfo {author} {\bibfnamefont {D.~P.}\ \bibnamefont {DiVincenzo}},
  \bibinfo {author} {\bibfnamefont {J.}~\bibnamefont {Rozen}}, \bibinfo
  {author} {\bibfnamefont {G.~A.}\ \bibnamefont {Keefe}}, \bibinfo {author}
  {\bibfnamefont {M.~B.}\ \bibnamefont {Rothwell}},\ and\ \bibinfo {author}
  {\bibfnamefont {M.~B.}\ \bibnamefont {Ketchen}},\ }\bibfield  {title}
  {\bibinfo {title} {High-coherence hybrid superconducting qubit},\ }\href@noop
  {} {\bibfield  {journal} {\bibinfo  {journal} {Physical Review Letters}\
  }\textbf {\bibinfo {volume} {105}},\ \bibinfo {pages} {100502} (\bibinfo
  {year} {2010})}\BibitemShut {NoStop}%
\bibitem [{\citenamefont {Chow}\ \emph {et~al.}(2011)\citenamefont {Chow},
  \citenamefont {C{\'o}rcoles}, \citenamefont {Gambetta}, \citenamefont
  {Rigetti}, \citenamefont {Johnson}, \citenamefont {Smolin}, \citenamefont
  {Rozen}, \citenamefont {Keefe}, \citenamefont {Rothwell}, \citenamefont
  {Ketchen} \emph {et~al.}}]{chow2011simple}%
  \BibitemOpen
  \bibfield  {author} {\bibinfo {author} {\bibfnamefont {J.~M.}\ \bibnamefont
  {Chow}}, \bibinfo {author} {\bibfnamefont {A.}~\bibnamefont {C{\'o}rcoles}},
  \bibinfo {author} {\bibfnamefont {J.~M.}\ \bibnamefont {Gambetta}}, \bibinfo
  {author} {\bibfnamefont {C.}~\bibnamefont {Rigetti}}, \bibinfo {author}
  {\bibfnamefont {B.}~\bibnamefont {Johnson}}, \bibinfo {author} {\bibfnamefont
  {J.~A.}\ \bibnamefont {Smolin}}, \bibinfo {author} {\bibfnamefont
  {J.}~\bibnamefont {Rozen}}, \bibinfo {author} {\bibfnamefont {G.~A.}\
  \bibnamefont {Keefe}}, \bibinfo {author} {\bibfnamefont {M.~B.}\ \bibnamefont
  {Rothwell}}, \bibinfo {author} {\bibfnamefont {M.~B.}\ \bibnamefont
  {Ketchen}}, \emph {et~al.},\ }\bibfield  {title} {\bibinfo {title} {Simple
  all-microwave entangling gate for fixed-frequency superconducting qubits},\
  }\href@noop {} {\bibfield  {journal} {\bibinfo  {journal} {Physical review
  letters}\ }\textbf {\bibinfo {volume} {107}},\ \bibinfo {pages} {080502}
  (\bibinfo {year} {2011})}\BibitemShut {NoStop}%
\bibitem [{\citenamefont {Yan}\ \emph {et~al.}(2016)\citenamefont {Yan},
  \citenamefont {Gustavsson}, \citenamefont {Kamal}, \citenamefont {Birenbaum},
  \citenamefont {Sears}, \citenamefont {Hover}, \citenamefont {Gudmundsen},
  \citenamefont {Rosenberg}, \citenamefont {Samach}, \citenamefont {Weber}
  \emph {et~al.}}]{yan2016flux}%
  \BibitemOpen
  \bibfield  {author} {\bibinfo {author} {\bibfnamefont {F.}~\bibnamefont
  {Yan}}, \bibinfo {author} {\bibfnamefont {S.}~\bibnamefont {Gustavsson}},
  \bibinfo {author} {\bibfnamefont {A.}~\bibnamefont {Kamal}}, \bibinfo
  {author} {\bibfnamefont {J.}~\bibnamefont {Birenbaum}}, \bibinfo {author}
  {\bibfnamefont {A.~P.}\ \bibnamefont {Sears}}, \bibinfo {author}
  {\bibfnamefont {D.}~\bibnamefont {Hover}}, \bibinfo {author} {\bibfnamefont
  {T.~J.}\ \bibnamefont {Gudmundsen}}, \bibinfo {author} {\bibfnamefont
  {D.}~\bibnamefont {Rosenberg}}, \bibinfo {author} {\bibfnamefont
  {G.}~\bibnamefont {Samach}}, \bibinfo {author} {\bibfnamefont
  {S.}~\bibnamefont {Weber}}, \emph {et~al.},\ }\bibfield  {title} {\bibinfo
  {title} {The flux qubit revisited to enhance coherence and reproducibility},\
  }\href@noop {} {\bibfield  {journal} {\bibinfo  {journal} {Nature
  communications}\ }\textbf {\bibinfo {volume} {7}},\ \bibinfo {pages} {1}
  (\bibinfo {year} {2016})}\BibitemShut {NoStop}%
\bibitem [{\citenamefont {Nielsen}\ and\ \citenamefont
  {Chuang}(2010)}]{nielsen2002quantum}%
  \BibitemOpen
  \bibfield  {author} {\bibinfo {author} {\bibfnamefont {M.~A.}\ \bibnamefont
  {Nielsen}}\ and\ \bibinfo {author} {\bibfnamefont {I.~L.}\ \bibnamefont
  {Chuang}},\ }\href@noop {} {\emph {\bibinfo {title} {Quantum computation and
  quantum information: 10th Anniversary Edition}}}\ (\bibinfo  {publisher}
  {Cambridge University Press},\ \bibinfo {year} {2010})\BibitemShut {NoStop}%
\bibitem [{\citenamefont {Jozsa}(1994)}]{jozsa1994fidelity}%
  \BibitemOpen
  \bibfield  {author} {\bibinfo {author} {\bibfnamefont {R.}~\bibnamefont
  {Jozsa}},\ }\bibfield  {title} {\bibinfo {title} {Fidelity for mixed quantum
  states},\ }\href@noop {} {\bibfield  {journal} {\bibinfo  {journal} {Journal
  of Modern Optics}\ }\textbf {\bibinfo {volume} {41}},\ \bibinfo {pages}
  {2315} (\bibinfo {year} {1994})}\BibitemShut {NoStop}%
\bibitem [{\citenamefont {Liang}\ \emph {et~al.}(2019)\citenamefont {Liang},
  \citenamefont {Yeh}, \citenamefont {Mendon{\c{c}}a}, \citenamefont {Teh},
  \citenamefont {Reid},\ and\ \citenamefont {Drummond}}]{liang2019quantum}%
  \BibitemOpen
  \bibfield  {author} {\bibinfo {author} {\bibfnamefont {Y.-C.}\ \bibnamefont
  {Liang}}, \bibinfo {author} {\bibfnamefont {Y.-H.}\ \bibnamefont {Yeh}},
  \bibinfo {author} {\bibfnamefont {P.~E.}\ \bibnamefont {Mendon{\c{c}}a}},
  \bibinfo {author} {\bibfnamefont {R.~Y.}\ \bibnamefont {Teh}}, \bibinfo
  {author} {\bibfnamefont {M.~D.}\ \bibnamefont {Reid}},\ and\ \bibinfo
  {author} {\bibfnamefont {P.~D.}\ \bibnamefont {Drummond}},\ }\bibfield
  {title} {\bibinfo {title} {Quantum fidelity measures for mixed states},\
  }\href@noop {} {\bibfield  {journal} {\bibinfo  {journal} {Reports on
  Progress in Physics}\ }\textbf {\bibinfo {volume} {82}},\ \bibinfo {pages}
  {076001} (\bibinfo {year} {2019})}\BibitemShut {NoStop}%
\bibitem [{\citenamefont {Place}\ \emph {et~al.}(2021)\citenamefont {Place},
  \citenamefont {Rodgers}, \citenamefont {Mundada}, \citenamefont {Smitham},
  \citenamefont {Fitzpatrick}, \citenamefont {Leng}, \citenamefont {Premkumar},
  \citenamefont {Bryon}, \citenamefont {Vrajitoarea}, \citenamefont {Sussman}
  \emph {et~al.}}]{place2021new}%
  \BibitemOpen
  \bibfield  {author} {\bibinfo {author} {\bibfnamefont {A.~P.}\ \bibnamefont
  {Place}}, \bibinfo {author} {\bibfnamefont {L.~V.}\ \bibnamefont {Rodgers}},
  \bibinfo {author} {\bibfnamefont {P.}~\bibnamefont {Mundada}}, \bibinfo
  {author} {\bibfnamefont {B.~M.}\ \bibnamefont {Smitham}}, \bibinfo {author}
  {\bibfnamefont {M.}~\bibnamefont {Fitzpatrick}}, \bibinfo {author}
  {\bibfnamefont {Z.}~\bibnamefont {Leng}}, \bibinfo {author} {\bibfnamefont
  {A.}~\bibnamefont {Premkumar}}, \bibinfo {author} {\bibfnamefont
  {J.}~\bibnamefont {Bryon}}, \bibinfo {author} {\bibfnamefont
  {A.}~\bibnamefont {Vrajitoarea}}, \bibinfo {author} {\bibfnamefont
  {S.}~\bibnamefont {Sussman}}, \emph {et~al.},\ }\bibfield  {title} {\bibinfo
  {title} {New material platform for superconducting transmon qubits with
  coherence times exceeding 0.3 milliseconds},\ }\href@noop {} {\bibfield
  {journal} {\bibinfo  {journal} {Nature communications}\ }\textbf {\bibinfo
  {volume} {12}},\ \bibinfo {pages} {1} (\bibinfo {year} {2021})}\BibitemShut
  {NoStop}%
\bibitem [{\citenamefont {Abdurakhimov}\ \emph {et~al.}(2019)\citenamefont
  {Abdurakhimov}, \citenamefont {Mahboob}, \citenamefont {Toida}, \citenamefont
  {Kakuyanagi},\ and\ \citenamefont {Saito}}]{abdurakhimov2019long}%
  \BibitemOpen
  \bibfield  {author} {\bibinfo {author} {\bibfnamefont {L.~V.}\ \bibnamefont
  {Abdurakhimov}}, \bibinfo {author} {\bibfnamefont {I.}~\bibnamefont
  {Mahboob}}, \bibinfo {author} {\bibfnamefont {H.}~\bibnamefont {Toida}},
  \bibinfo {author} {\bibfnamefont {K.}~\bibnamefont {Kakuyanagi}},\ and\
  \bibinfo {author} {\bibfnamefont {S.}~\bibnamefont {Saito}},\ }\bibfield
  {title} {\bibinfo {title} {A long-lived capacitively shunted flux qubit
  embedded in a 3d cavity},\ }\href@noop {} {\bibfield  {journal} {\bibinfo
  {journal} {Applied Physics Letters}\ }\textbf {\bibinfo {volume} {115}},\
  \bibinfo {pages} {262601} (\bibinfo {year} {2019})}\BibitemShut {NoStop}%
\bibitem [{\citenamefont {Jakob}\ and\ \citenamefont
  {Stenholm}(2003)}]{jakob2003variational}%
  \BibitemOpen
  \bibfield  {author} {\bibinfo {author} {\bibfnamefont {M.}~\bibnamefont
  {Jakob}}\ and\ \bibinfo {author} {\bibfnamefont {S.}~\bibnamefont
  {Stenholm}},\ }\bibfield  {title} {\bibinfo {title} {Variational functions in
  driven open quantum systems},\ }\href@noop {} {\bibfield  {journal} {\bibinfo
   {journal} {Physical Review A}\ }\textbf {\bibinfo {volume} {67}},\ \bibinfo
  {pages} {032111} (\bibinfo {year} {2003})}\BibitemShut {NoStop}%
\bibitem [{\citenamefont {Mascarenhas}\ \emph {et~al.}(2015)\citenamefont
  {Mascarenhas}, \citenamefont {Flayac},\ and\ \citenamefont
  {Savona}}]{mascarenhas2015matrix}%
  \BibitemOpen
  \bibfield  {author} {\bibinfo {author} {\bibfnamefont {E.}~\bibnamefont
  {Mascarenhas}}, \bibinfo {author} {\bibfnamefont {H.}~\bibnamefont
  {Flayac}},\ and\ \bibinfo {author} {\bibfnamefont {V.}~\bibnamefont
  {Savona}},\ }\bibfield  {title} {\bibinfo {title} {Matrix-product-operator
  approach to the nonequilibrium steady state of driven-dissipative quantum
  arrays},\ }\href@noop {} {\bibfield  {journal} {\bibinfo  {journal} {Physical
  Review A}\ }\textbf {\bibinfo {volume} {92}},\ \bibinfo {pages} {022116}
  (\bibinfo {year} {2015})}\BibitemShut {NoStop}%
\bibitem [{\citenamefont {Stehlik}\ \emph {et~al.}(2021)\citenamefont
  {Stehlik}, \citenamefont {Zajac}, \citenamefont {Underwood}, \citenamefont
  {Phung}, \citenamefont {Blair}, \citenamefont {Carnevale}, \citenamefont
  {Klaus}, \citenamefont {Keefe}, \citenamefont {Carniol}, \citenamefont
  {Kumph} \emph {et~al.}}]{stehlik2021tunable}%
  \BibitemOpen
  \bibfield  {author} {\bibinfo {author} {\bibfnamefont {J.}~\bibnamefont
  {Stehlik}}, \bibinfo {author} {\bibfnamefont {D.}~\bibnamefont {Zajac}},
  \bibinfo {author} {\bibfnamefont {D.}~\bibnamefont {Underwood}}, \bibinfo
  {author} {\bibfnamefont {T.}~\bibnamefont {Phung}}, \bibinfo {author}
  {\bibfnamefont {J.}~\bibnamefont {Blair}}, \bibinfo {author} {\bibfnamefont
  {S.}~\bibnamefont {Carnevale}}, \bibinfo {author} {\bibfnamefont
  {D.}~\bibnamefont {Klaus}}, \bibinfo {author} {\bibfnamefont
  {G.}~\bibnamefont {Keefe}}, \bibinfo {author} {\bibfnamefont
  {A.}~\bibnamefont {Carniol}}, \bibinfo {author} {\bibfnamefont
  {M.}~\bibnamefont {Kumph}}, \emph {et~al.},\ }\bibfield  {title} {\bibinfo
  {title} {Tunable coupling architecture for fixed-frequency transmons},\
  }\href@noop {} {\bibfield  {journal} {\bibinfo  {journal} {arXiv:2101.07746}\
  } (\bibinfo {year} {2021})}\BibitemShut {NoStop}%
\bibitem [{\citenamefont {Sete}\ \emph
  {et~al.}(2021{\natexlab{b}})\citenamefont {Sete}, \citenamefont {Chen},
  \citenamefont {Manenti}, \citenamefont {Kulshreshtha},\ and\ \citenamefont
  {Poletto}}]{sete2021floating}%
  \BibitemOpen
  \bibfield  {author} {\bibinfo {author} {\bibfnamefont {E.~A.}\ \bibnamefont
  {Sete}}, \bibinfo {author} {\bibfnamefont {A.~Q.}\ \bibnamefont {Chen}},
  \bibinfo {author} {\bibfnamefont {R.}~\bibnamefont {Manenti}}, \bibinfo
  {author} {\bibfnamefont {S.}~\bibnamefont {Kulshreshtha}},\ and\ \bibinfo
  {author} {\bibfnamefont {S.}~\bibnamefont {Poletto}},\ }\bibfield  {title}
  {\bibinfo {title} {Floating tunable coupler for scalable quantum computing
  architectures},\ }\href@noop {} {\bibfield  {journal} {\bibinfo  {journal}
  {arXiv:2103.07030}\ } (\bibinfo {year} {2021}{\natexlab{b}})}\BibitemShut
  {NoStop}%
\end{thebibliography}%


%

\end{document}